\documentclass[preprint,aps,nofootinbib,11pt]{revtex4-1}
\usepackage{geometry}                % See geometry.pdf to learn the layout options. There are lots.
\usepackage{amsmath,amssymb,amsfonts,dcolumn,color,graphicx,graphics,latexsym,placeins,epsfig,tikz,
afterpage}
\usetikzlibrary{arrows,shapes}
\usepackage{subfigure,rotating,bm,mathrsfs}
\usepackage[title]{appendix}
\usepackage{psfrag}

\newcommand{\be}{\begin{equation}}
\newcommand{\ee}{\end{equation}}
\newcommand{\bear}{\begin{eqnarray}}
\newcommand{\eear}{\end{eqnarray}}
\newcommand{\ba}{\begin{array}}
\newcommand{\ea}{\end{array}}

\usepackage[pagebackref=false, colorlinks=true]{hyperref}
\hypersetup{linkcolor=red,         % color of internal links
                  citecolor=blue,        % color of links to bibliography
                  filecolor=magenta,      % color of file links
                  urlcolor=blue}          % color of the url
\begin{document}

\title{\Large Homogeneous spacetime with shear viscosity}

\author{Inyong Cho}
\email{iycho@seoultech.ac.kr}
\author{Rajibul Shaikh}
\email{lrajibulsk@gmail.com}
\affiliation{Institute of Convergence Fundamental Studies, School of Natural Sciences, College of Liberal Arts, \\ Seoul National University of Science and Technology, Seoul 01811, Korea}

\begin{abstract}
We study the homogeneous and anisotropic evolution of Bianchi type-I 
spacetime driven by perfect fluid with shear viscosity. We obtain exact 
solutions by considering the simplest form of the equation of state wherein 
the pressure and the shear stress are proportional to the energy density 
individually. A special case of our general solutions represent Bianchi 
type-VII cosmology. We analyse the singularity structure of the solutions 
and its connection with various energy conditions. We find that the 
initial singularity can be removed only for the Bianchi type-VII. We also 
analyse the late-time behaviour of the solutions and find that, compared 
to the usual Friedmann universe, the spacetime expands less rapidly and 
the energy density drops faster.

\end{abstract}

\maketitle

\section{Introduction}

In relativistic cosmology, people introduce perfect fluid to 
model a homogeneous and isotropic universe \citep{MTW}. It is, however, 
plausible to consider that the Universe in the beginning might have had 
anisotropy which had died out in the course of evolution resulting in 
the present Universe. There is also a debate that anisotropies may have 
hung around and not dissipated \citep{Aluri}. More realistic models, 
therefore, include 
dissipative processes which are caused by bulk and shear viscosities. 
Misner has argued that the neutrino viscosity may reduce the shear 
anisotropy present in the early universe \citep{Misner:1967uu}. Later, 
However, Stewart \citep{Stewart1968,Stewart1969} and Doroshkevich 
{\em et. al.} \citep{Doro1968} pointed out that Misner's findings were 
based on the assumption that the anisotropy in the early universe is 
small. Stewart argued that, for large anisotropy, the viscous effects 
of neutrinos are inefficient in reducing the shear completely 
\citep{Stewart1968}.

Shear viscosity usually manifests as anisotropy. In this work, we 
shall consider both diagonal and off-diagonal components of the 
energy-momentum tensor modelled by shear viscosity. It is quite 
natural to consider the shear viscosity in the realistic astrophysical 
situations involving fluid plasma. It is also 
natural to consider it in cosmology; however, as the current 
universe exhibits very small anisotropy, people often ignore it 
simply for convenience. In cosmological perturbation theories, 
an anisotropic tensor field which represents the shear viscosity 
inevitably arises in theories higher than linear order 
\citep{Noh:2004bc,Hwang:2007ni,Hwang:2017oxa}.

In Eckart's theory the energy-momentum tensor of fluid is given by 
\citep{Eckart:1940te,MTW,Pimentel}
\begin{equation}\label{eq:IF}
T^{\mu\nu}=\rho u^\mu u^\nu+(p-\xi\Theta)h^{\mu\nu}+q^\mu u^\nu+q^\nu u^\mu -2\eta \sigma^{\mu\nu}.
\end{equation}
Here, $\xi$ and $\eta$ are the bulk and the shear viscosity 
coefficients, $q^\mu$ is the heat-flux four-vector, $u^\mu$ is 
the four-velocity of fluid, $h_{\mu\nu}=g_{\mu\nu}+u_\mu u_\nu$ is 
the projection tensor, $\Theta=u^{\mu}_{;\mu}$ is the expansion, and 
$\sigma_{\mu\nu}=(u_{\mu;\delta}h^\delta_\nu+u_{\nu;\delta}h^\delta_\mu)/2-\Theta h_{\mu\nu}/3$ 
is the symmetric traceless shear tensor. The above energy-momentum 
consists of three components; the perfect fluid component, 
$T^{\mu\nu}_{\rm pf}=\rho u^\mu u^\nu+p h^{\mu\nu}$ makes , the 
heat-flux component, $T^{\mu\nu}_{\rm heat}=q^\mu u^\nu+q^\nu u^\mu$, 
and the viscosity component, $T^{\mu\nu}_{\rm visc}=-\xi\Theta h^{\mu\nu}-2\eta \sigma^{\mu\nu}$. 
It is to be noted that the shear tensor contributes to  both the 
spatial diagonal and off-diagonal components of the energy-momentum 
tensor in general while the bulk viscosity contributes to the 
diagonal components only.

Anisotropic cosmological models with both bulk and shear viscosity has 
been studied in the literature in Refs.~\citep{VC1,VC1-1,VC3,VC2,VC4,
VC5,VC6,VC7}, although exact solutions are not much known. Most of the 
available exact solutions are those with non-zero bulk viscosity; exact 
solutions with non-zero shear viscosity are very few. Besides the 
equation of state between the pressure and the energy density, most of 
these works assume a `second equation of state' associated with the 
{\it viscosity coefficient} $\eta$. Belinski and Khalatnikov 
\citep{VC1-1} studied Bianchi type-I cosmology with both bulk and shear 
viscosities. In the asymptotic limits of both small and large $\rho$, 
they considered $\eta\propto \rho^A$. They showed that the viscosities 
were not capable of removing the cosmological singularity. They found 
that the curvature invariants diverge at the singularity while the 
energy density vanishes. Similar results were obtained for some cases 
in Ref. \cite{VC3} in which the authors studied Bianchi type-I 
cosmology with stiff matter ($p=\rho$) with $\eta\propto \rho^A$. 
Anisotropic cosmology with $\eta\propto \rho$ was studied in Ref. 
\cite{VC2}, although a complete exact solution was not obtained. 
Banerjee et. al. \citep{VC5} considered the shear to be proportional to 
the expansion, i.e., $\sigma^{\mu\nu}\sigma_{\mu\nu}\propto \Theta^2$, 
and studied exact cosmological solutions of several Bianchi types. 
Anisotropic cosmological models with $\eta\propto \rho^A$ and 
$\eta\propto H$, where $H$ is the average Hubble parameter, were 
studied in Ref. \citep{VC6,VC7}. It is to be noted that $\eta$ does not 
have any specific form of equation of state. Most of the works in the 
literature utilize certain simplifying assumptions for the equation of 
state to get the exact solutions. In this work, we shall impose the 
equation of state for viscosity directly to the off-diagonal stress 
term $\eta \sigma^{\mu\nu}$. In particular, we consider that this 
off-diagonal term is proportional to the energy density.

In Ref. \cite{Cho:2022rgs}, we studied anisotropic cosmology and obtained 
exact solutions by considering an energy-momentum tensor with all the 
spatial off-diagonal terms to be the same. We considered two equations of 
states $p=\alpha\rho$ and $\sigma=T^i_{j\; \rm shear\; visc} (i\neq j)=\beta\rho$. 
We found that the spacetime expands less rapidly at late times than the 
usual Friedmann universe as the energy density drops faster due to the 
transfer to the shear stress. We found also that the initial big-bang 
singularity can be removed in the parameter region $1+\alpha+2\beta\leq 0$. 
In the present work, we study anisotropic cosmology with an energy-momentum 
tensor containing only one non-zero off-diagonal stress. If we perform the 
diagonalization of the set-up, the present case corresponds to the Bianchi 
type-I which has anisotropy along all three spatial directions (the previous 
work is Bianchi type-VII which is a special case of Bianchi type-I). We 
obtain exact solution by considering the same equations of state as in 
the previous work. We find that the initial singularity always exists in 
the present case. 

The paper is organized as follows. In Sec. \ref{sec:model}, we introduce our 
model and obtain the exact solutions of the Einstein's field equation. In Sec. 
\ref{sec:solution}, we analyze the early-time behaviour of the solutions, 
the energy density and the Kretschmann scalar, and discuss the effect of the 
shear stress on the singularity structure of the solution. We also discuss the 
connection of the singularity structure with various energy conditions. In Sec. 
\ref{sec:latetime}, we analyze the late-time behaviour of the solutions and 
compare with that of the Friedmann universe. In Sec. \ref{sec:decceleration}, 
we study the evolution of the deceleration and the anisotropy parameters, and compare 
with those of the Friedmann universe. Finally, we conclude in Sec. \ref{sec:conclusion}.

\section{Model and field equations}
\label{sec:model}
In Ref. \cite{Cho:2022rgs}, we studied anisotropic cosmology and obtained exact 
solutions by considering an energy-momentum tensor with all the spatial off-diagonal 
terms to be the same. In this work, we consider a fluid whose energy-momentum tensor 
has only one spatial off-diagonal component and is given by
\begin{equation}\label{eq:PF}
T^\mu_\nu=
\begin{bmatrix}
   \; -\rho \;\;\;\; &  0 \;\;\;\;\;  &  0 \;\;\;\;\;  &  0  \;\;\;\; \\
      0   &  p_1+\sigma_1  & \sigma & 0 \\
      0   & \sigma & p_1+\sigma_1 & 0 \\
      0   & 0 & 0 &  p_2-2\sigma_1 
\end{bmatrix},
\end{equation}
where the energy density $\rho$, the pressures $p_1$ and $p_2$, and the 
stresses $\sigma$ and $\sigma_1$ depend only on time $t$. Later, we set 
$p_1=p_2$. The above energy-momentum tensor can solves the Einstein equations 
consistently if we consider a metric of following ansatz:
\begin{align} 
ds^2 &= -dt^2+a^2(t)\left(dx^2+dy^2\right)+2b(t) dxdy +e^2(t) dz^2 \label{metricab}\\
&=-dt^2+\frac{1}{2}\left[c^2(t)+d^2(t)\right] \left(dx^2+dy^2\right)
+\left[d^2(t)-c^2(t)\right]dxdy+e^2(t)dz^2 \label{metriccd},
\end{align}
where we rewrote the metric in the last step by defining 
$a^2=(c^2+d^2)/2$ and $b=(d^2-c^2)/2$ as it is easier to solve the field 
equations using the metric in Eq. \eqref{metriccd} (see Ref. 
\citep{Cho:2022rgs}). The three-volume density is given by 
${\cal V}_3 =(a^4-b^2)^{1/2}e= cde$. For the co-moving fluid four-velocity 
$u^\mu=(u^t,u^x,u^y,u^z)=(1,0,0,0)$, the expansion and the nonzero components 
of the shear tensor become $\Theta=\left(\dot{c}/c+\dot{d}/d+\dot{e}/e\right)$, 
$\sigma^x_y=\sigma^y_x=-(\dot{c}/c-\dot{d}/d)/2$ and 
$\sigma^x_x=\sigma^y_y=-\sigma^z_z/2=(\dot{c}/c+\dot{d}/d-2\dot{e}/e)/6$. 
It is to be noted that, the energy-momentum tensor \eqref{eq:PF} takes the 
form \eqref{eq:IF} if we set $\eta\neq 0$, $\xi=0$, $q^\mu=0$, 
$\sigma=\eta(\dot{c}/c-\dot{d}/d)$, 
$\sigma_1=-\eta(\dot{c}/c+\dot{d}/d-2\dot{e}/e)/3$ and $p_1=p_2=p$.

We now consider the isotropic pressure $p_1=p_2=p$. The Einstein field equations 
require $e^2=c^{1+3n}d^{1-3n}$, where $n$ is a constant. This 
gives $\sigma_1=n\sigma$.

One can transform both the metric and the energy-momentum tensor to diagonal 
form by suitable coordinate transformations. The coordinate transformations,
\begin{align}\label{CT}
X = \frac{1}{\sqrt{2}} (x+y),\quad
Y = \frac{1}{\sqrt{2}} (x-y),\quad
Z = z,
\end{align}
transforms the metric \eqref{metriccd} to the Bianchi type I metric,
\begin{align}\label{metricBianchi} 
ds^2 = -dt^2+d^2(t)dX^2+c^2(t)dY^2 + c^{1+3n}(t)d^{1-3n}(t)dZ^2 .
\end{align}
Similarly, the energy-momentum tensor \eqref{eq:PF} is transformed to
\begin{equation}\label{EMTdiag}
T^\mu_\nu=
\begin{bmatrix}
   \; -\rho \;\;\;\; &  0 \;\;\;\;\;  &  0 \;\;\;\;\;  &  0  \;\;\;\; \\
      0   &  p+(n+1)\sigma  & 0 & 0 \\
      0   & 0 & p+(n-1)\sigma & 0 \\
      0   & 0 & 0 &  p-2n\sigma 
\end{bmatrix}.
\end{equation}
The Einstein's equations in the diagonal form are equivalent to the 
ones in the off-diagonal form. It is to be noted that, for 
$n=\pm 1/3$, the above set up reduces to Bianchi type-VII as two of the 
scale factors in the metric \eqref{metricBianchi} and two of the spatial 
diagonal terms in the energy-momentum \eqref{EMTdiag} become equal. This 
special case has been studied in great detail in Ref. \cite{Cho:2022rgs}.

The components of the Einstein's equation $G^\mu_\nu= T^\mu_\nu$ ($8\pi G=1$) yield
\begin{align}
(1+3n)\frac{\dot{c}^2}{c^2}+4\frac{\dot{c}}{c}\frac{\dot{d}}{d}+(1-3n)\frac{\dot{d}^2}{d^2} &= 2\rho, \label{eq:FE1}\\
2(1+n)\frac{\ddot{c}}{c}+2(1-n)\frac{\ddot{d}}{d}+n(1+3n)\frac{\dot{c}^2}{c^2}+2(1-3n^2)\frac{\dot{c}}{c}\frac{\dot{d}}{d}-n(1-3n)\frac{\dot{d}^2}{d^2} &= -2p, \label{eq:FE2}\\
2\frac{\ddot{c}}{c}-2\frac{\ddot{d}}{d}+(1+3n)\frac{\dot{c}^2}{c^2}-6n\frac{\dot{c}}{c}\frac{\dot{d}}{d}-(1-3n)\frac{\dot{d}^2}{d^2} &= -4\sigma. \label{eq:FE3}
\end{align}
We, therefore, have three equations for the five unknowns $c$, $d$, $\rho$, 
$p$ and $\sigma$. We need two more equations to consistently solve the field 
equations. For that, we consider following two equations of state in the 
same way as in Ref. \cite{Cho:2022rgs},
\be\label{eoss}
p=\alpha\rho, \qquad \sigma = \beta\rho.
\ee
With this, we manipulate the field equations and integrate them 
(see Appendix \ref{appendixA}). After integrating once, the field 
equations with new variables $u=c^{\bar{n}_1}d^{\bar{n}_2}$ and 
$v=c^{n_1} d^{n_2}$ reduce to [see Eqs. \eqref{eq:Udot2_app}, 
\eqref{eq:Vdot2_app} and \eqref{eq:rho_app}]
\begin{equation}
\frac{\dot{u}}{u}=\frac{A_1}{u^{\sqrt{3}/4}v^{\sqrt{3}/4-s_2}},
\label{eq:Udot2}
\end{equation}
\begin{equation}
\frac{\dot{v}}{v}=\frac{A_2}{u^{\sqrt{3}/4-s_1}v^{\sqrt{3}/4}},
\label{eq:Vdot2}
\end{equation}
\begin{equation}
\rho=\frac{A_1A_2/4}{u^{\sqrt{3}/2-s_1}v^{\sqrt{3}/2-s_2}},
\label{eq:rho}
\end{equation}
where $A_1$ and $A_2$ are integration constants  and 
\begin{equation}
s_1=\frac{1}{4}\left[\sqrt{3}(1-\alpha)-2\sqrt{1+3n^2}\beta\right], \quad
s_2=\frac{1}{4}\left[\sqrt{3}(1-\alpha)+2\sqrt{1+3n^2}\beta\right],
\end{equation}
\begin{equation}
n_1=\sqrt{3}(1+n)+\sqrt{1+3n^2}, \quad n_2=\sqrt{3}(1-n)-\sqrt{1+3n^2},
\end{equation}
\begin{equation}
\bar{n}_1=\sqrt{3}(1+n)-\sqrt{1+3n^2}, \quad \bar{n}_2=\sqrt{3}(1-n)+\sqrt{1+3n^2}.
\end{equation}
From Eqs. \eqref{eq:Udot2} and \eqref{eq:Vdot2}, we obtain for later use,
\begin{equation}
u^{s_1-1}\dot{u}=\frac{A_1}{A_2}v^{s_2-1}\dot{v}.
\label{eq:UVdot}
\end{equation}
The solution to this equation are classified by $s_1$ and $s_2$.

\vspace{12pt}
\noindent
\underline{{\bf Class G}: General Class}
\vspace{12pt}

For $s_1\neq 0$ and $s_2\neq 0$, the solution to Eq. \eqref{eq:UVdot} is 
given by
\begin{equation}
u=\left( \frac{A_1 s_1}{A_2 s_2} v^{s_2}+A_{3}\right)^{1/{s_1}},
\label{eq:U_gen}
\end{equation}
where $A_{3}$ is an integration constant. Then the solution to 
Eq. \eqref{eq:Vdot2} is given by an integral form,
\begin{equation}
t=\frac{1}{A2}\int_{v_0}^v v^{\sqrt{3}/4-1} \left( \frac{A_1 s_1}{A_2 s_2} v^{s_2}+A_{3}\right)^{\frac{\sqrt{3}}{4s_1}-1} dv,
\label{eq:t_gen}
\end{equation}
where $v_0=v(t=0)$. Note that these solutions are not valid for 
$s_1=0$, or $s_2=0$.

\vspace{12pt}
\noindent
\underline{{\bf Class A}: $\sqrt{3}(1-\alpha)+2\sqrt{1+3n^2}\beta=0$ 
($\alpha\neq 1$ and $\beta\neq 0$)}\footnote{If $\alpha=1$ and $\beta=0$, 
it is the stress-free case. It is dealt with in Class C.}
\vspace{12pt}

For this case, $s_1=\sqrt{3}(1-\alpha)/2$ and $s_2=0$. Integrating 
Eq. \eqref{eq:UVdot}, we obtain
\begin{equation}
u=\left[\frac{\sqrt{3}(1-\alpha)A_1}{2 A_2} \log v+A_{3A}\right]^{\frac{2}{\sqrt{3}(1-\alpha)}},
\label{eq:U_SC1}
\end{equation}
where $A_{3A}$ is an integration constant. From Eq. \eqref{eq:Vdot2}, we obtain
\begin{equation}
t=\frac{1}{A2}\int_{v_0}^v v^{\sqrt{3}/4-1} \left[\frac{\sqrt{3}(1-\alpha)A_1}{2A_2} \log v+A_{3A}\right]^{\frac{2\alpha-1}{2(1-\alpha)}} dv.
\label{eq:t_SC1}
\end{equation}

\vspace{12pt}
\noindent
\underline{{\bf Class B}: $\sqrt{3}(1-\alpha)-2\sqrt{1+3n^2}\beta=0$ ($\alpha\neq 1$ and $\beta\neq 0$)}
\vspace{12pt}

For this case, $s_1=0$ and $s_2=\sqrt{3}(1-\alpha)/2$. Integrating Eq. \eqref{eq:UVdot}, we obtain
\begin{equation}
u=A_{3B}\exp\left[\frac{2A_1}{\sqrt{3}(1-\alpha)A_2} v^{\sqrt{3}(1-\alpha)/2}\right],
\label{eq:U_SC2}
\end{equation}
where $A_{3B}$ is an integration constant. From Eq. \eqref{eq:Vdot2}, we obtain
\begin{equation}
t=\frac{1}{A2}\int_{v_0}^v v^{\sqrt{3}/4-1} \left\{ A_{3B}\exp\left[\frac{2A_1}{\sqrt{3}(1-\alpha)A_2} v^{\sqrt{3}(1-\alpha)/2}\right]\right\}^{\frac{\sqrt{3}}{4}} dv.
\label{eq:t_SC2}
\end{equation}

\vspace{12pt}
\noindent
\underline{{\bf Class C}: $\beta=0$}
\vspace{12pt}

The off-diagonal stress terms vanish, i.e., $T^i_j(i\neq j)=0$ in this 
case. The usual Friedmann universe belongs to this class. 
Integrating Eq. \eqref{eq:UVdot}, we obtain
\begin{equation}
u=\left\{
  \begin{array}{lr}
    A_{3C} v^{A_1/A_2} & (\alpha =1)\\
  \left[\frac{A_1}{A_2 }v^{\sqrt{3}(1-\alpha)/4}+A_{3C}\right]^{\frac{4}{\sqrt{3}(1-\alpha)}} & (\alpha \neq 1)
  \end{array}
\right.,
\label{eq:U_SC3}
\end{equation}
where $A_{3C}$ is an integration constant. From \eqref{eq:Vdot2}, we get
\begin{equation}
A_2 t+A_4=\left\{
  \begin{array}{lr}
  \frac{4A_2 (A_{3C})^{\sqrt{3}/4}}{\sqrt{3}(A_1+A_2)} v^{\sqrt{3}(A_1+A_2)/(4A_2)} & (\alpha =1)\\
  \int v^{\sqrt{3}/4-1} \left[\frac{A_1}{A_2}v^{\sqrt{3}(1-\alpha)/4}+A_{3C}\right]^{\frac{\alpha}{1-\alpha}} dv & (\alpha \neq 1)
  \end{array}
\right.,
\label{eq:t_SC3}
\end{equation}
where $A_4$ is an integration constant. In terms of $u$ and $v$, $c$ and $d$ are 
given by $c=u^{-n_2/(n_1\bar{n}_2-\bar{n}_1 n_2)} v^{\bar{n}_2/(n_1\bar{n}_2-\bar{n}_1 n_2)}$ 
and $d=u^{n_1/(n_1\bar{n}_2-\bar{n}_1 n_2)} v^{-\bar{n}_1/(n_1\bar{n}_2-\bar{n}_1 n_2)}$. 
Note that $c\neq d$ as $u\neq v$ in general. One can check, $c=d=e$ if $u=v$. 
Therefore, although the off-diagonal stress $\sigma$ vanishes in this case, 
the shear tensor $\sigma^i_j$ does not in general. We, thus, have following 
two sub-classes.

{\bf (i) FRW universe}: 
The usual Friedmann-Robertson-Walker universe represents shear-free cosmology. 
The shear tensor vanishes identically only when $c=d=e$, i.e., $u=v$ all the time. 
This can be achieved by setting the conditions $c=d$ and $\dot{c}=\dot{d}$, i.e., 
$u=v$ and $\dot{u}=\dot{v}$ at a 
certain time, say at $t=0$. We apply these conditions on Eq.~\eqref{eq:U_SC3} 
and find that $A_2=A_1$, and $A_{3C}=1$ for $\alpha=1$ and $A_{3C}=0$ for 
$\alpha \neq 1$. Therefore, from Eq.~\eqref{eq:t_SC3}, we obtain
\begin{equation}
u=v=\left\{
  \begin{array}{lr}
    e^{\left(A_2 t+A_4\right)} & (\alpha =-1) \\
  \left[\frac{\sqrt{3}(1+\alpha)}{4}\left(A_2 t+A_4\right)\right]^{\frac{4}{\sqrt{3}(1+\alpha)}} &(\alpha \neq -1)
  \end{array}
\right..
\label{eq:UV_SC3}
\end{equation}
Using above equation, we have
\begin{equation}
c=d=\left\{
  \begin{array}{lr}
    e^{\left(A_2 t+A_4\right)/(2\sqrt{3})} & (\alpha =-1) \\
  \left[\frac{\sqrt{3}(1+\alpha)}{4}\left(A_2 t+A_4\right)\right]^{\frac{2}{3(1+\alpha)}} &(\alpha \neq -1)
  \end{array}
\right..
\label{eq:c_SC3}
\end{equation}
If we further impose $v(0)=v_0=1$, i.e., $c(0)=c_0=1$ without loss of 
generality, we get $A_4=0$ for $\alpha=-1$ and $A_4=4/[\sqrt{3}(1+\alpha)]$ 
for $\alpha \neq -1$.

{\bf (ii) Non-FRW universe}:
However, if we do not impose the conditions $c=d$ and $\dot{c}=\dot{d}$, i.e., 
$u=v$ and $\dot{u}=\dot{v}$ at $t=0$, the shear tensor is nonzero 
($\sigma^i_j \neq 0$), although the off-diagonal stress terms vanish 
($T^i_j(i\neq j)=0$). The solution in this sub-class is given by 
Eqs. \eqref{eq:U_SC3} and \eqref{eq:t_SC3}. We shall not pay much attention 
to this case in what follows.

\vspace{12pt}
\noindent
\underline{{\bf Kasner type}}
\vspace{12pt}

There exists a vacuum solution of Kasner type. If we set either $A_1=0$ 
or $A_2=0$, then $\rho=0$ (and hence $p=\sigma=0$), but the Ktetshmann 
scalar is non-vanishing. Setting $A_1=0$, we get from Eqs. \eqref{eq:Udot2} 
and \eqref{eq:Vdot2} $u=B_1$ and $v=(B_2 t+B_3)^{4/\sqrt{3}}$, where 
$B_i$'s are constants. Therefore, the three scale factors vary as 
$a_1\equiv c\sim t^{p_1}$, $a_2\equiv d\sim t^{p_2}$ and 
$a_3\equiv e\sim t^{p_3}$, where we have set $B_3=0$ and
\begin{equation}
p_1=\frac{\sqrt{3}(1-n)+\sqrt{1+3n^2}}{3\sqrt{1+3n^2}}, \quad p_2=-\frac{\sqrt{3}(1+n)-\sqrt{1+3n^2}}{3\sqrt{1+3n^2}}, \quad p_3=\frac{2\sqrt{3}n+\sqrt{1+3n^2}}{3\sqrt{1+3n^2}}.
\nonumber
\end{equation}
It is to be noted that the sum of the Kasner exponents $p_i$ as well 
as the sum of their squares are equal to 1. We get a similar results 
when we set $A_2=0$ and $A_1\neq 0$. 

%%%%%%%%%%%%%%%%%%%%%%%%%%%%%%%%%%%%%%%%%%%%%%%%%%%%%%%%%%%%%%%%%%%%%%%%%%%%%%%%%%%%%

\section{Solutions and Effect of stress}
\label{sec:solution}

In this section, we study the behaviour of the solutions
and discuss the effect of the stress $\sigma$. We integrate 
the $v$-solutions in different class and obtain $u$ from 
Eq.~\eqref{eq:U_gen} [Eqs.~\eqref{eq:U_SC1}, \eqref{eq:U_SC2}, 
and \eqref{eq:U_SC3} for special cases], and $\rho$ from 
Eq.~\eqref{eq:rho}.

We determine the four integrations constants $v_0$ and three 
$A_i$'s by imposing initial conditions. We keep the same initial 
conditions across all the classes. Since we want the solutions 
to reduce to the Friedmann universe ($c=d=e$, i.e., $u=v$) for 
$\beta=0$, we impose the same initial conditions as imposed for 
the FRW case in Sec.~\ref{sec:model}, i.e., we impose $v(0)=u(0)=1$ 
and $\dot{v}(0)=\dot{u}(0)$. This implies $v_0=1$. This gives 
$A_2=A_1$ from Eq.~\eqref{eq:UVdot}, and $A_{3i}$ is obtained from 
Eq.~\eqref{eq:U_gen} [Eqs.~\eqref{eq:U_SC1}, \eqref{eq:U_SC2}, and 
\eqref{eq:U_SC3} for the special cases]. The remaining constant $A_1$ 
can be determined from Eq.~\eqref{eq:rho} by imposing the values of 
$\rho(0)$.

We integrate the $v$-solution in both directions in $t$ from $t=0$. 
The time-evolution of the metric functions $c$, $d$ and $e$, the 
three-volume density ${\cal V}_3$, the energy density $\rho$, and 
the Kretschmann scalar ${\cal K}$ for $n=0,\pm 1$ and 
$\alpha, \beta = 0, \pm 1/3,\pm 1$ is shown in Figs.~\ref{fig1}, 
\ref{fig1_n1} and \ref{fig1_nm1}. The green plots represent the 
usual Friedmann case.

\subsection{Metric functions $c$, $d$ and $e$, and three-volume density ${\cal V}_3$}

For the Friedmann universe the metric functions become $c=d=e=a$. 
For $\alpha > -1$, the scale factor $a(t)$ evolves as
\be
a(t) = a_0 (t-t_s)^{2/[3(1+\alpha)]},
\ee
where $t_s$ is the moment of big bang (the moment when ${\cal V}_3$ 
vanishes in our numerical calculations). For $\alpha = -1$, $a(t)$ 
becomes that of de Sitter space with $t_s \to -\infty$.

\begin{figure}[]
\centering
\subfigure{\includegraphics[scale=0.44]{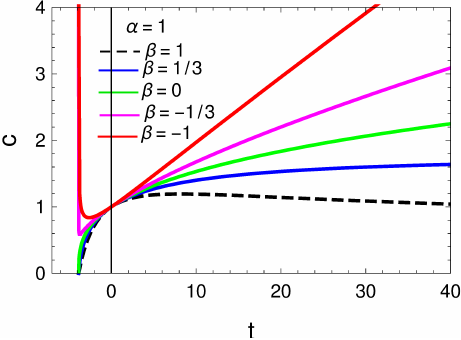}}
\subfigure{\includegraphics[scale=0.44]{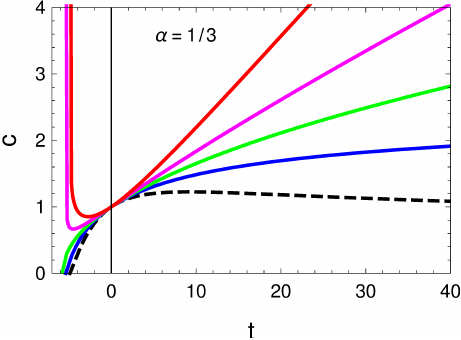}}
\subfigure{\includegraphics[scale=0.44]{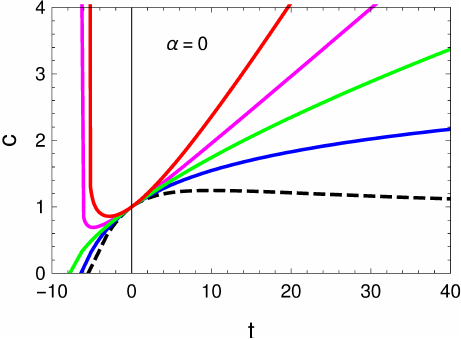}}
\subfigure{\includegraphics[scale=0.44]{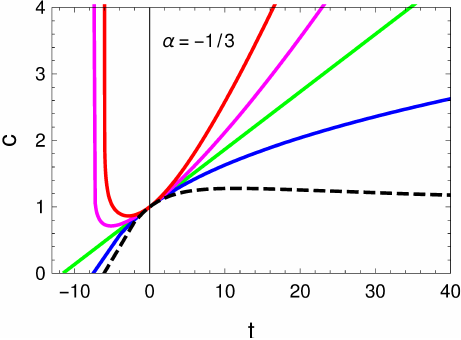}}
\subfigure{\includegraphics[scale=0.44]{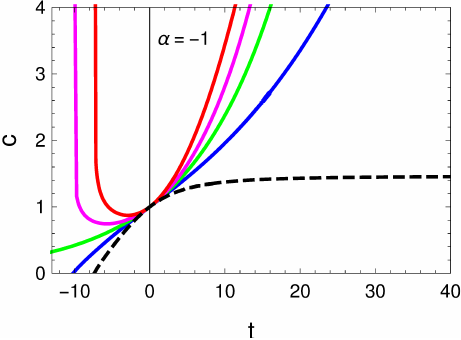}}
\subfigure{\includegraphics[scale=0.44]{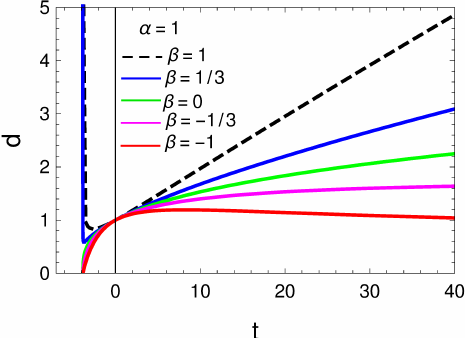}}
\subfigure{\includegraphics[scale=0.44]{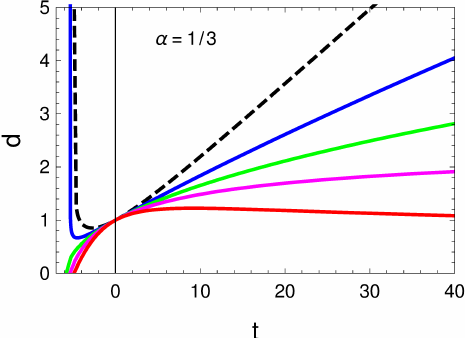}}
\subfigure{\includegraphics[scale=0.44]{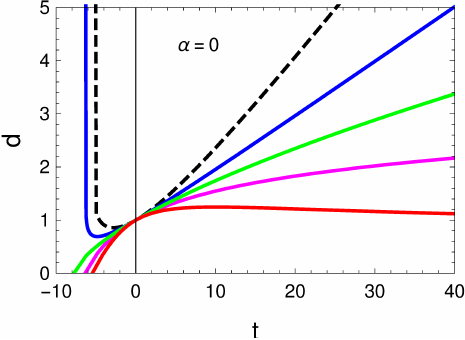}}
\subfigure{\includegraphics[scale=0.44]{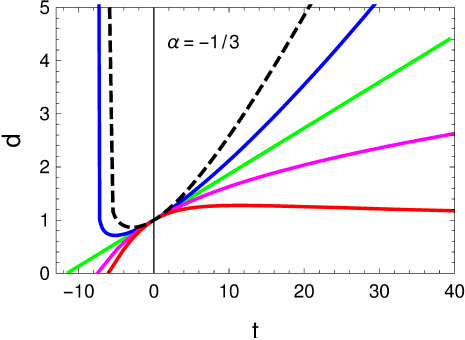}}
\subfigure{\includegraphics[scale=0.44]{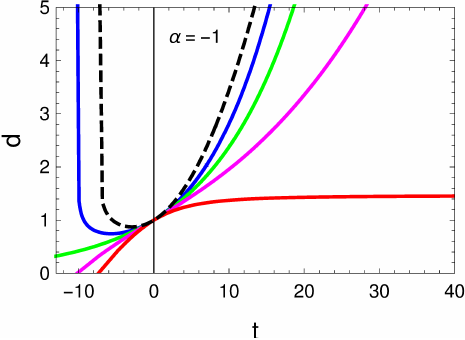}}
\subfigure{\includegraphics[scale=0.46]{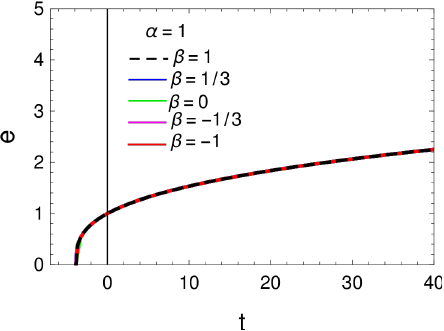}}
\subfigure{\includegraphics[scale=0.46]{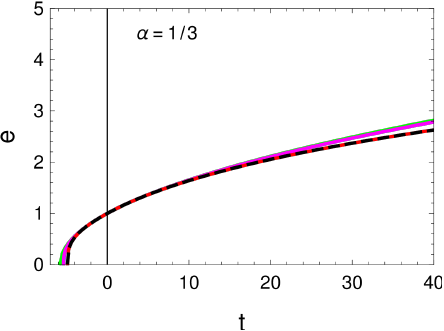}}
\subfigure{\includegraphics[scale=0.46]{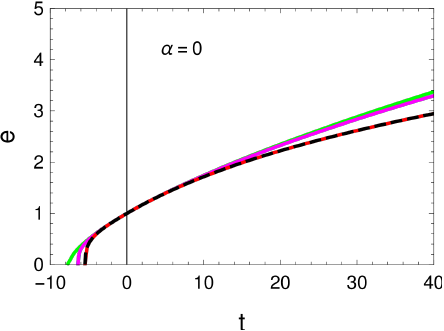}}
\subfigure{\includegraphics[scale=0.46]{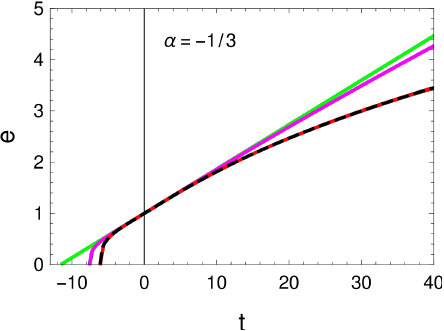}}
\subfigure{\includegraphics[scale=0.46]{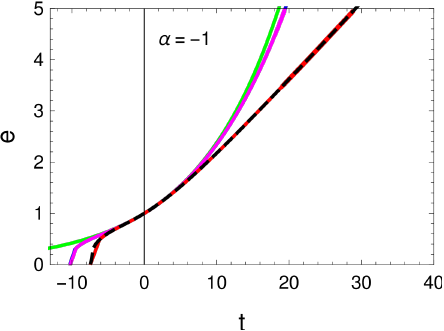}}
\subfigure{\includegraphics[scale=0.143]{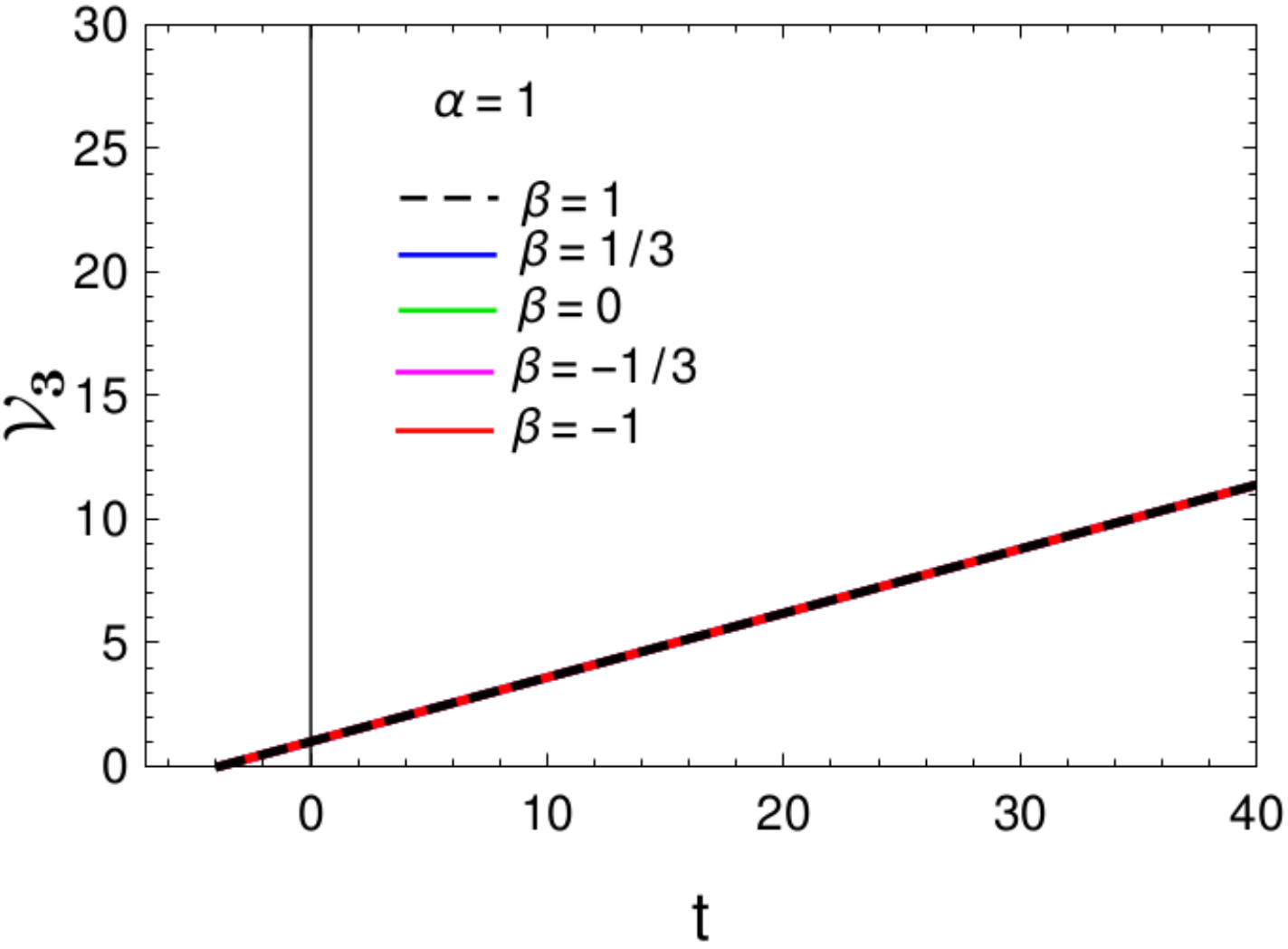}}
\subfigure{\includegraphics[scale=0.143]{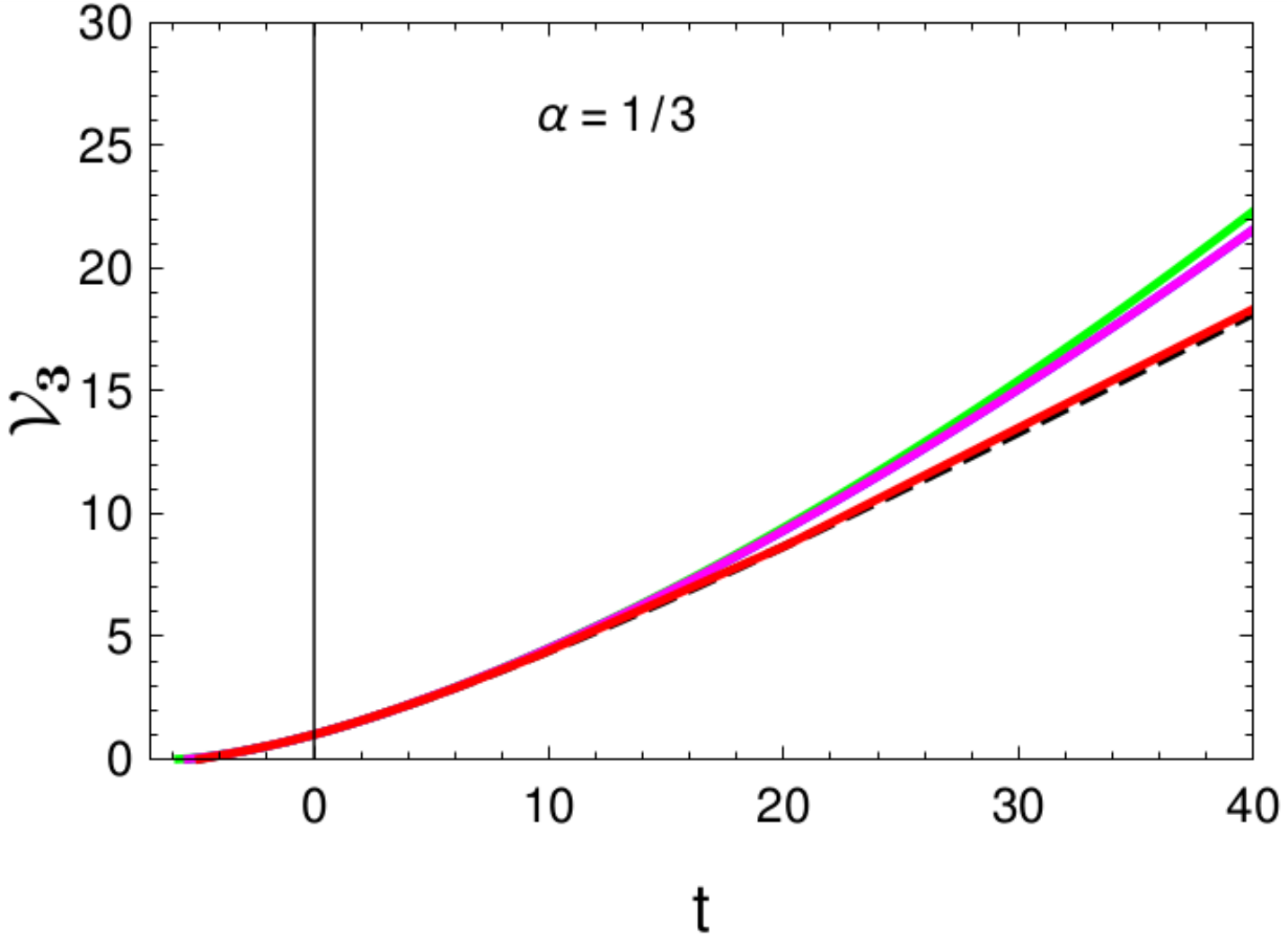}}
\subfigure{\includegraphics[scale=0.143]{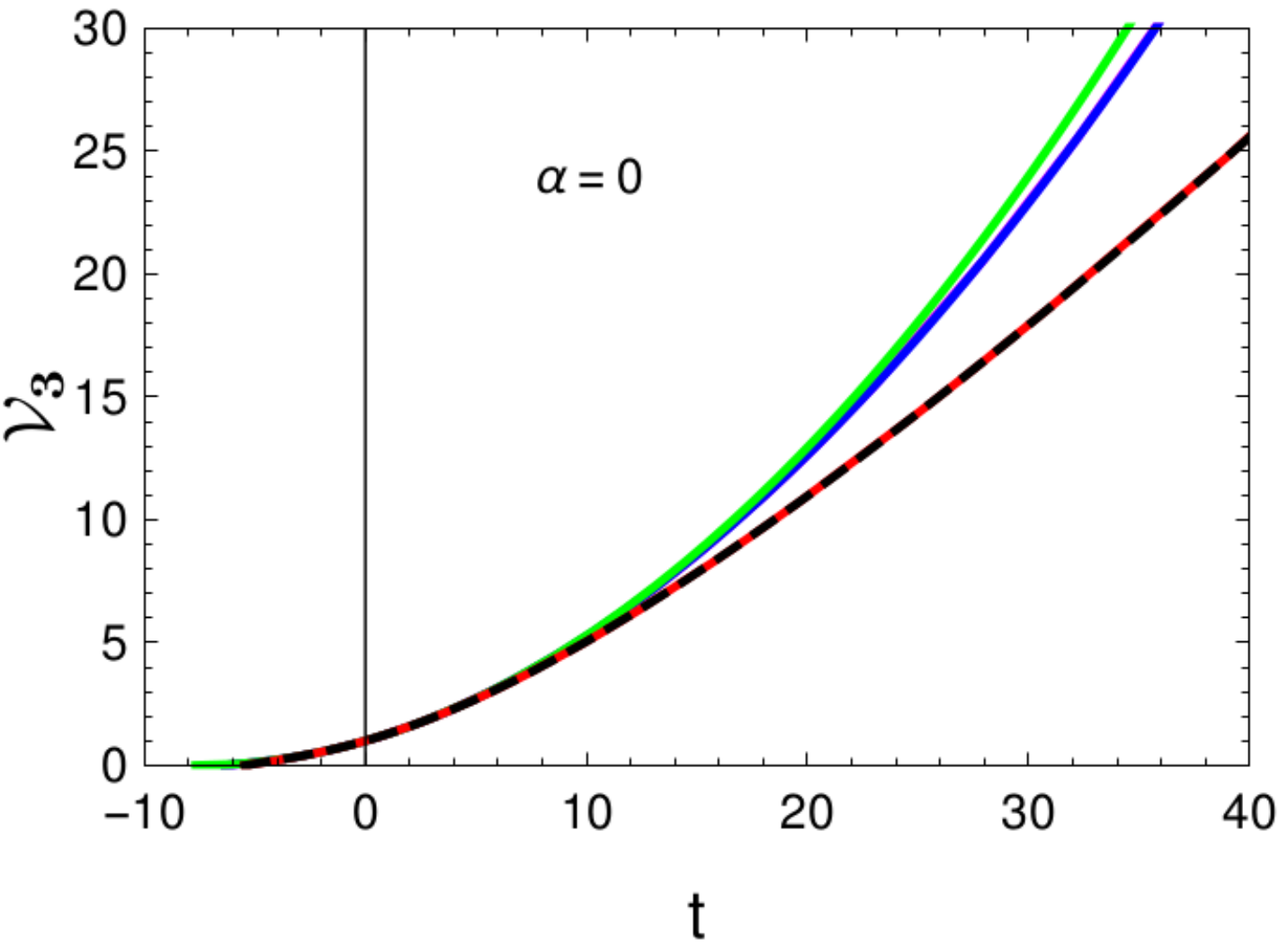}}
\subfigure{\includegraphics[scale=0.143]{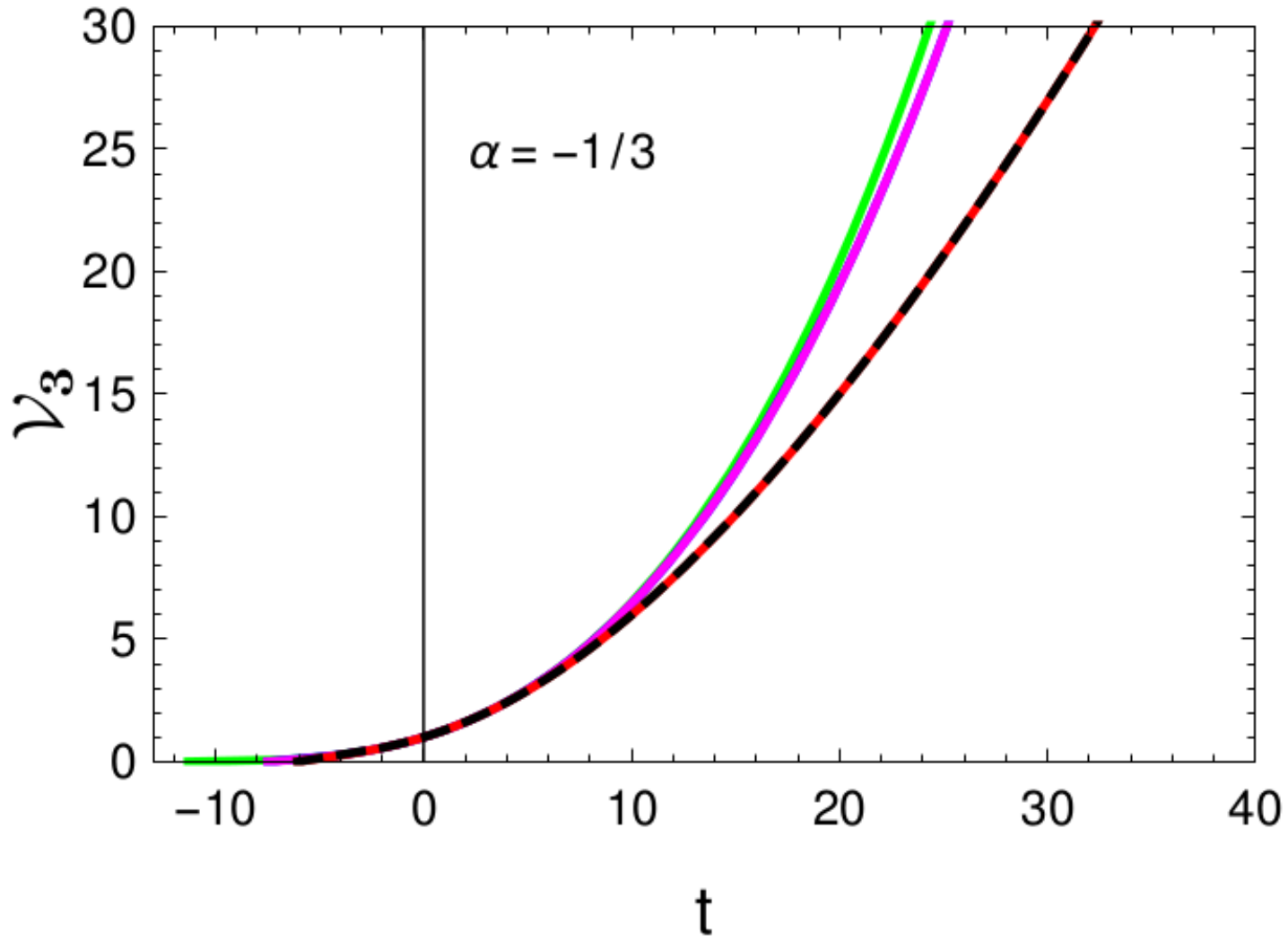}}
\subfigure{\includegraphics[scale=0.143]{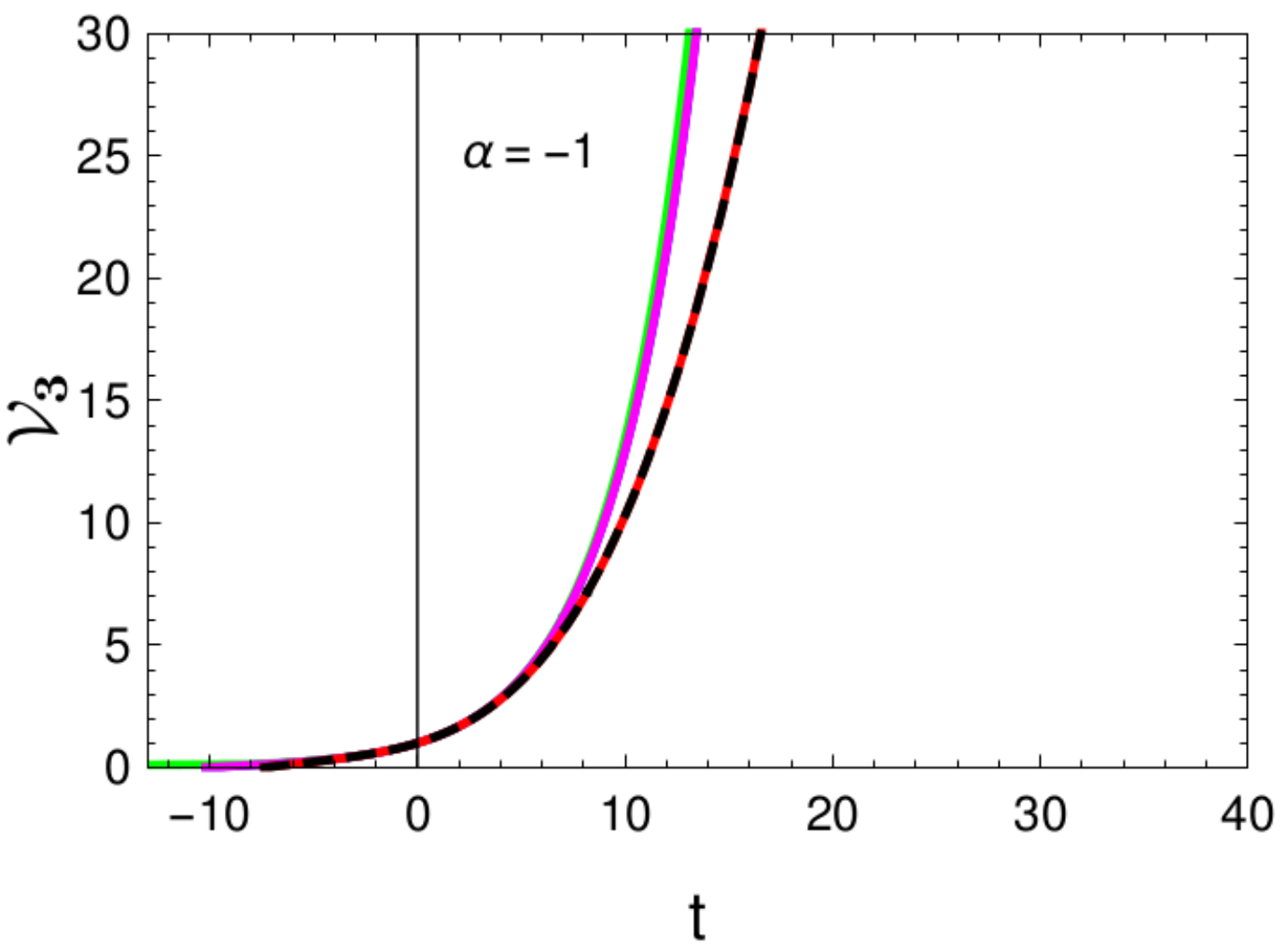}}
\subfigure{\includegraphics[scale=0.465]{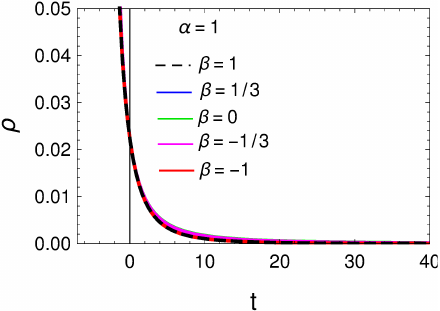}}
\subfigure{\includegraphics[scale=0.465]{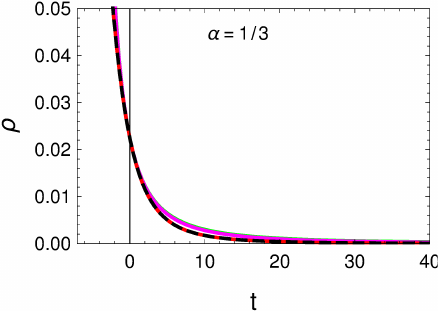}}
\subfigure{\includegraphics[scale=0.465]{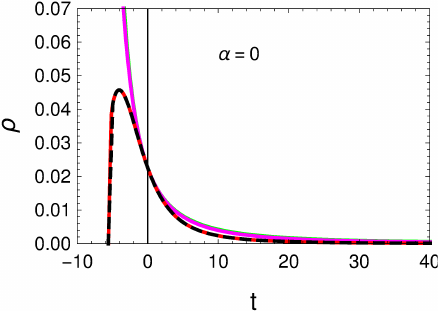}}
\subfigure{\includegraphics[scale=0.465]{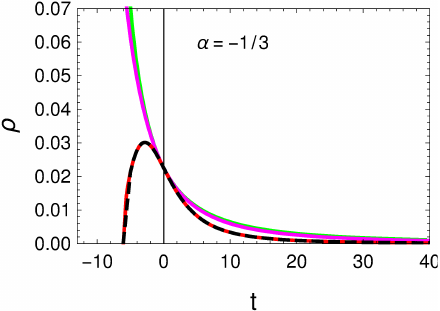}}
\subfigure{\includegraphics[scale=0.465]{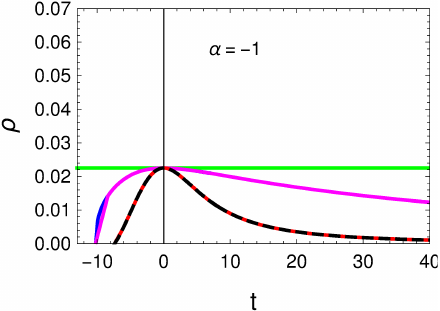}}
\subfigure{\includegraphics[scale=0.141]{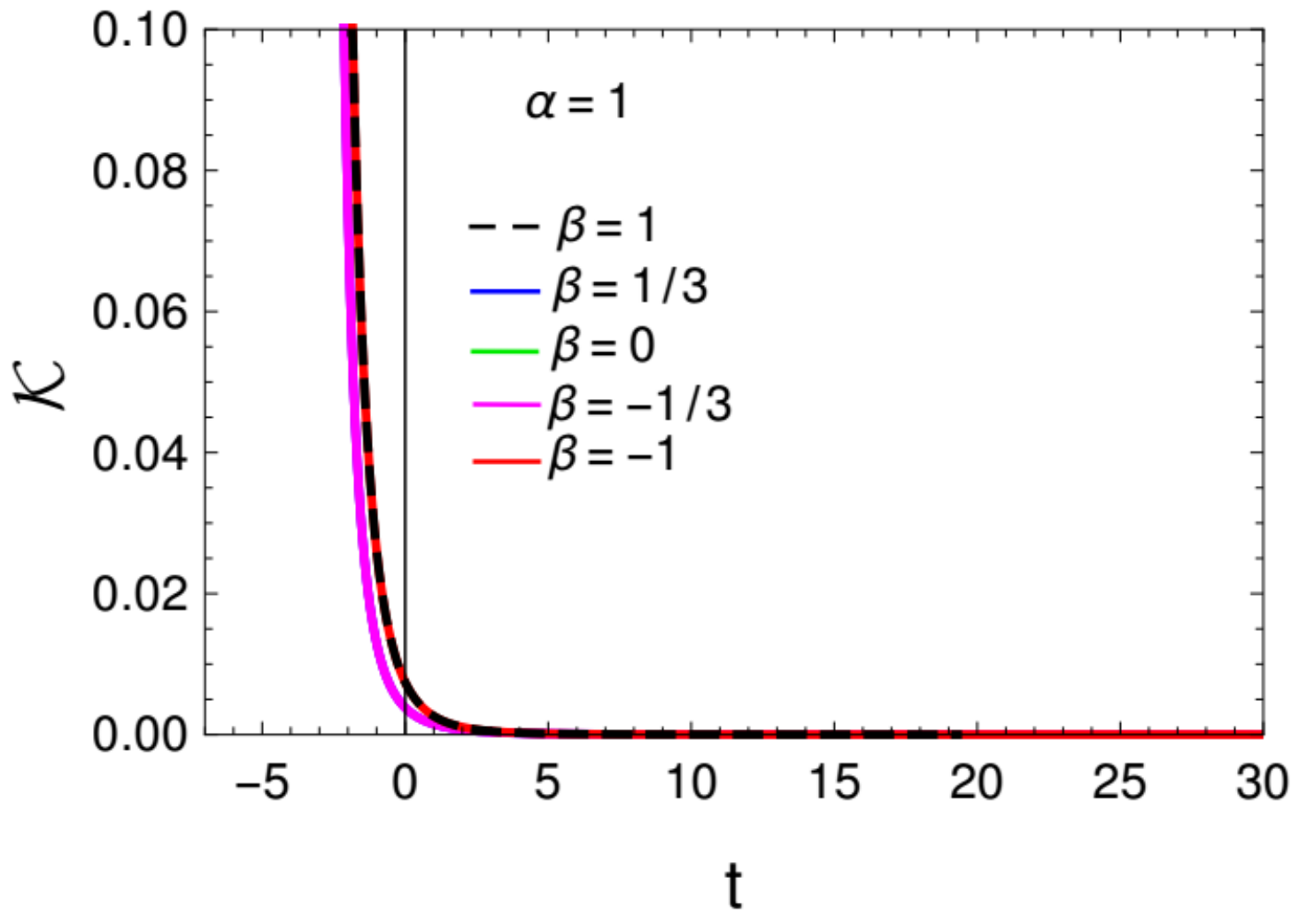}}
\subfigure{\includegraphics[scale=0.141]{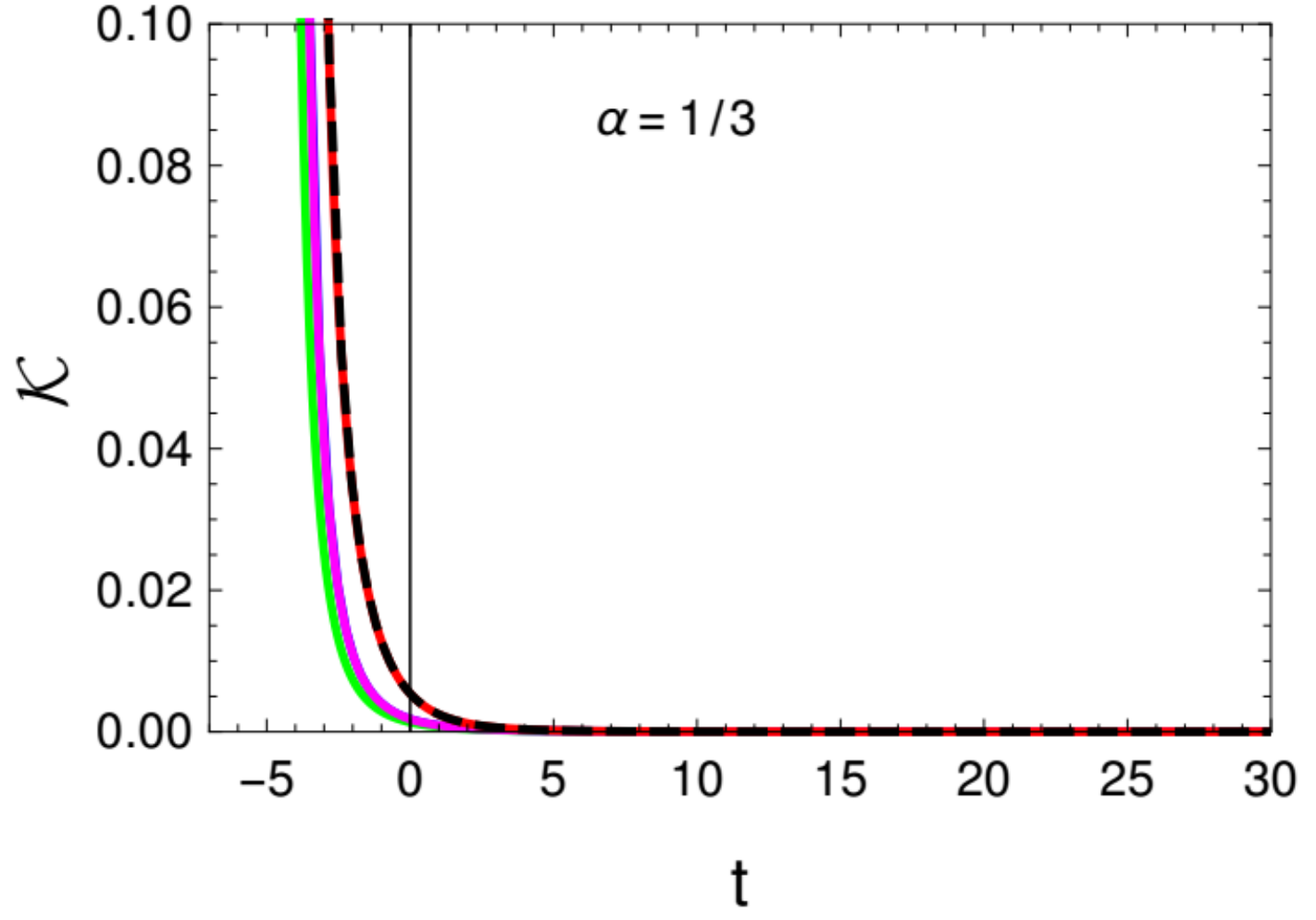}}
\subfigure{\includegraphics[scale=0.141]{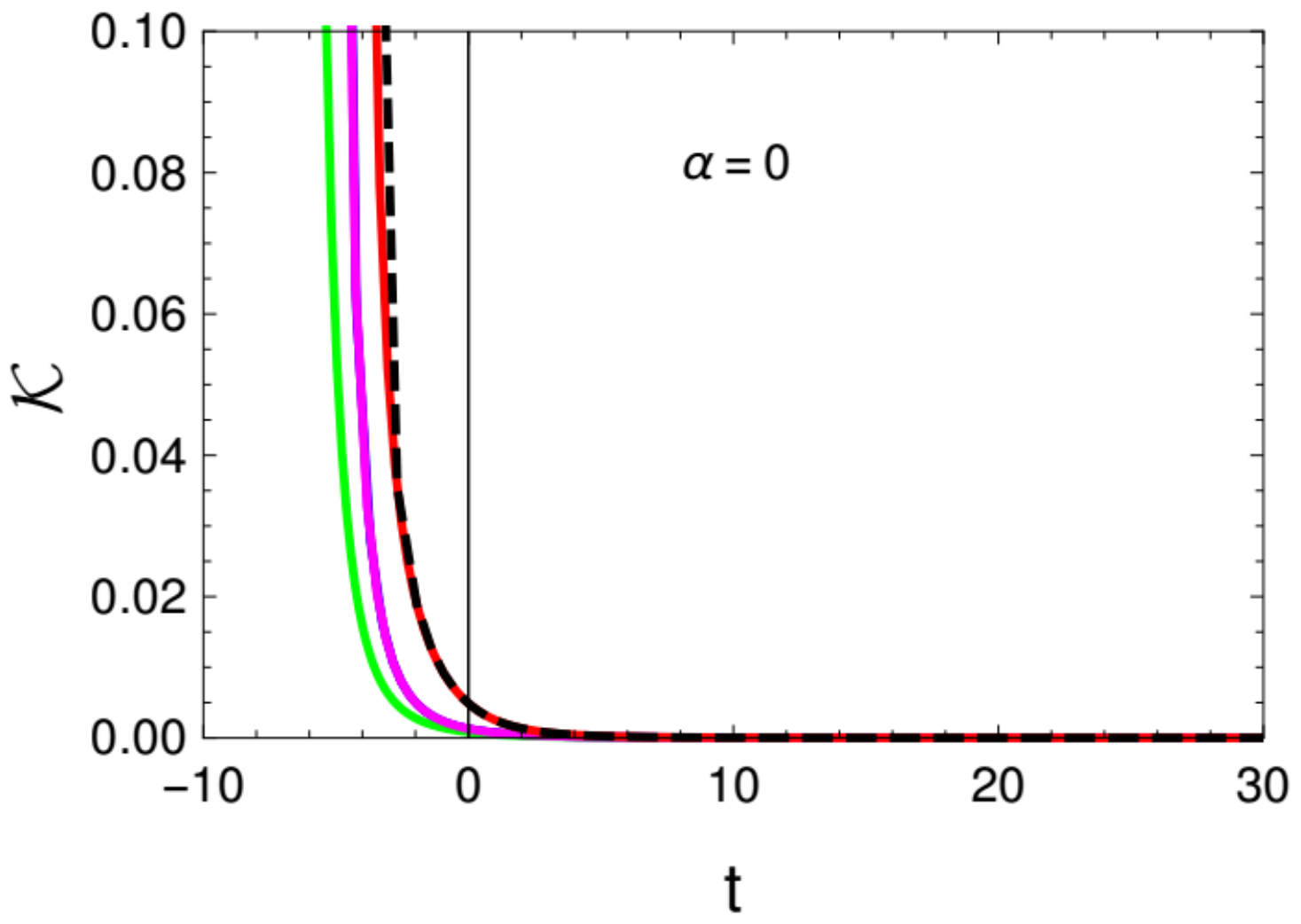}}
\subfigure{\includegraphics[scale=0.141]{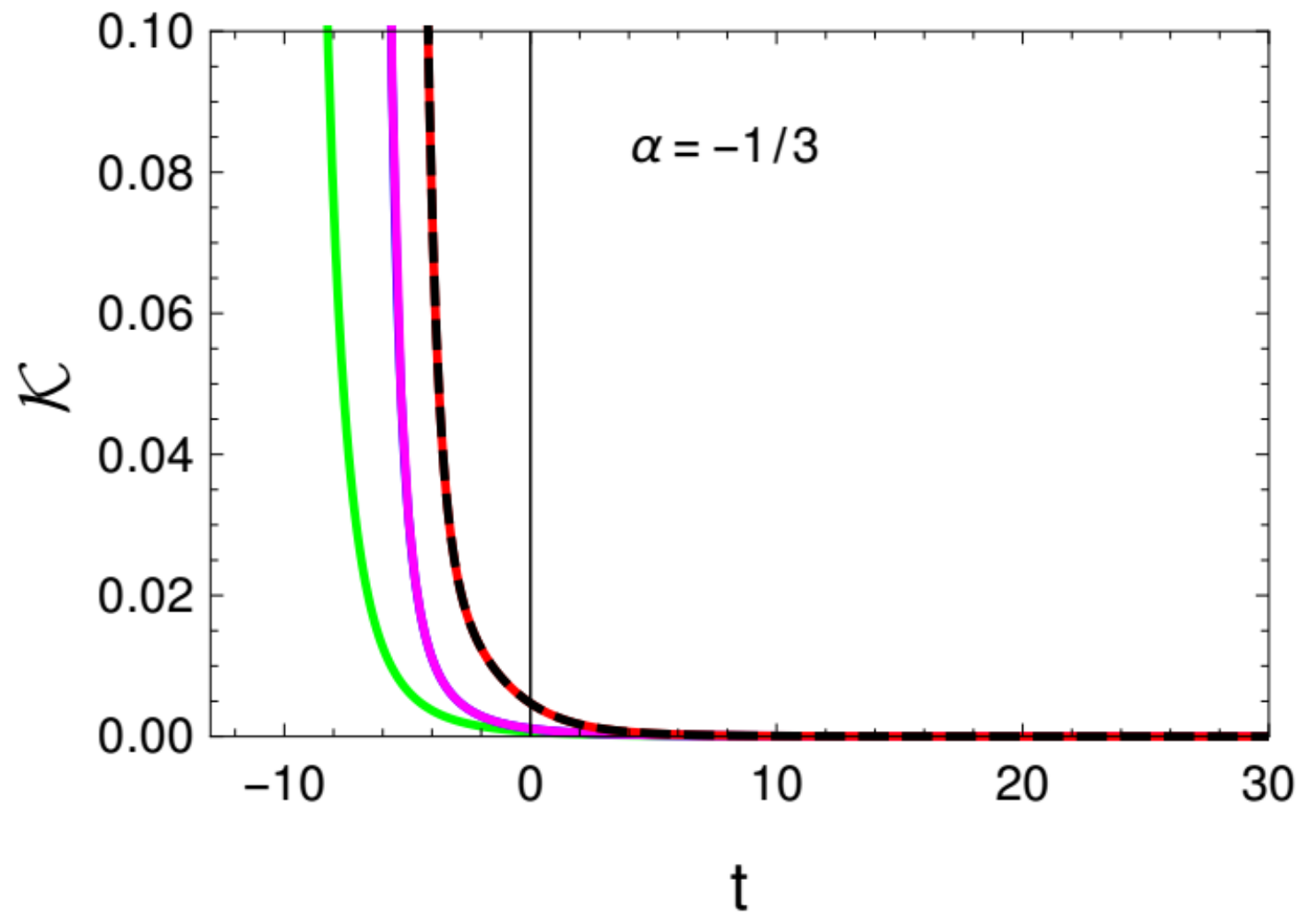}}
\subfigure{\includegraphics[scale=0.141]{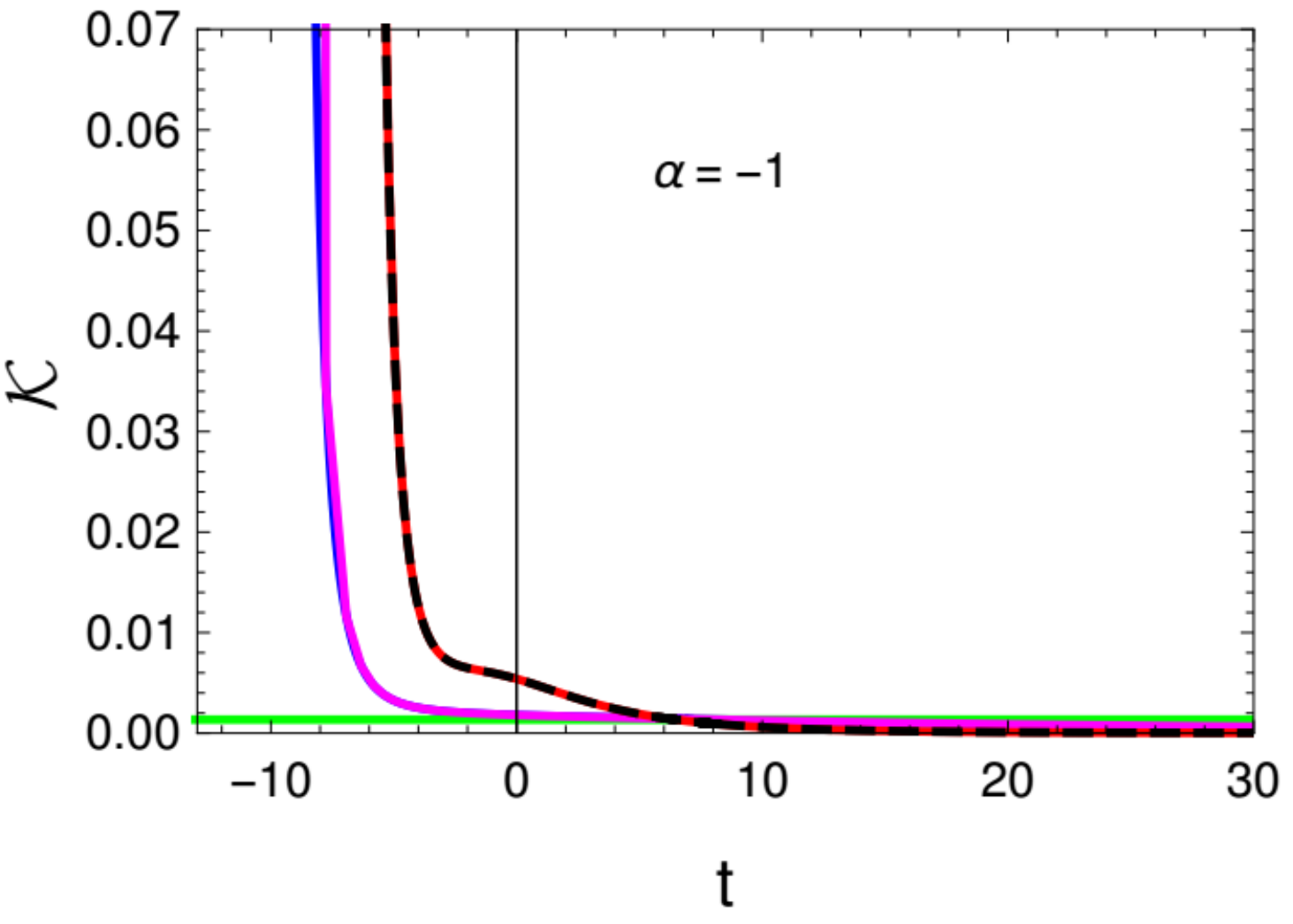}}
\caption{Plots of $c$, $d$, $e$, ${\cal V}_3$, $\rho$ and $\mathcal{K}$ 
for $\alpha, \beta = 0, \pm 1/3,\pm 1$ and $n=0$.
Each column is for the same value of $\alpha$. 
Values of $\beta$ are distinguished by color.}
\label{fig1}
\end{figure}

\begin{figure}[]
\centering
\subfigure{\includegraphics[scale=0.45]{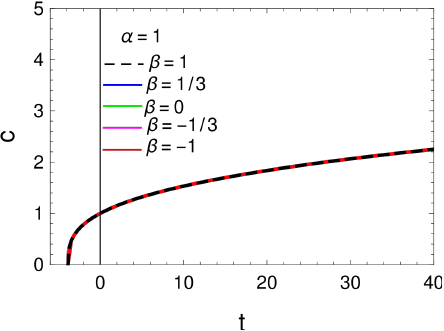}}
\subfigure{\includegraphics[scale=0.45]{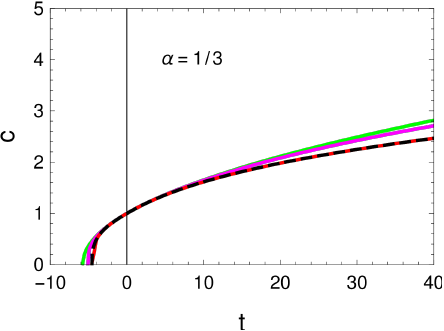}}
\subfigure{\includegraphics[scale=0.45]{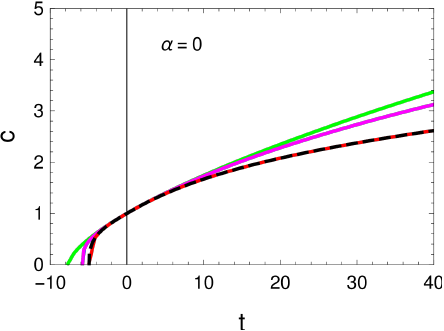}}
\subfigure{\includegraphics[scale=0.45]{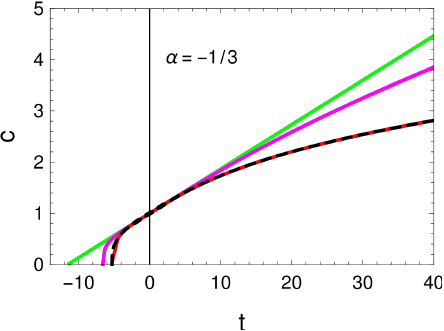}}
\subfigure{\includegraphics[scale=0.45]{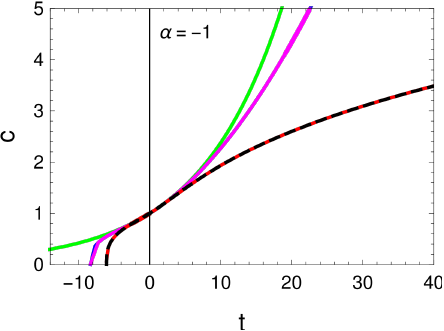}}
\subfigure{\includegraphics[scale=0.45]{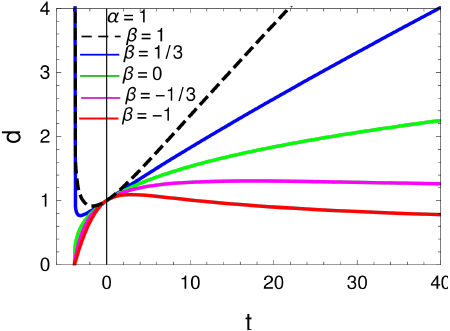}}
\subfigure{\includegraphics[scale=0.45]{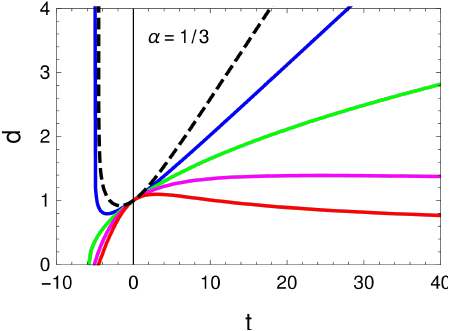}}
\subfigure{\includegraphics[scale=0.45]{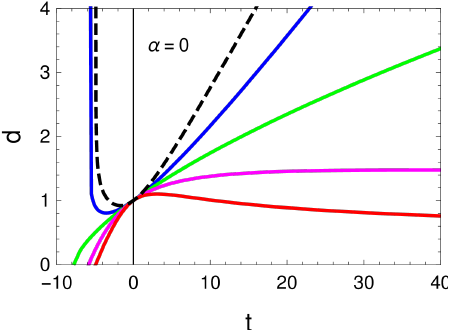}}
\subfigure{\includegraphics[scale=0.45]{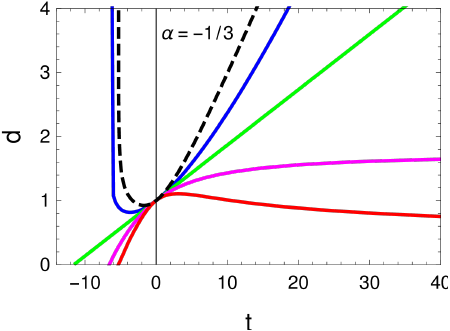}}
\subfigure{\includegraphics[scale=0.45]{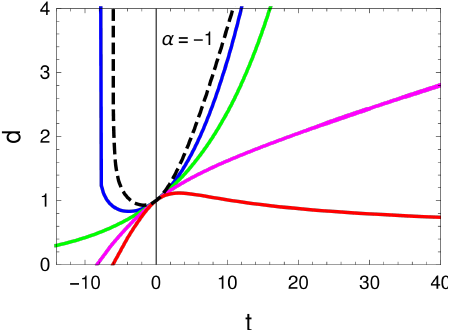}}
\subfigure{\includegraphics[scale=0.46]{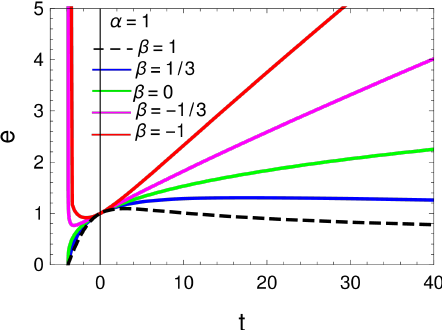}}
\subfigure{\includegraphics[scale=0.46]{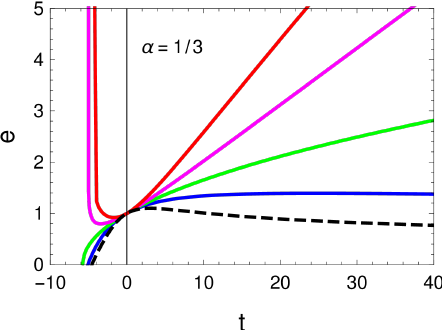}}
\subfigure{\includegraphics[scale=0.46]{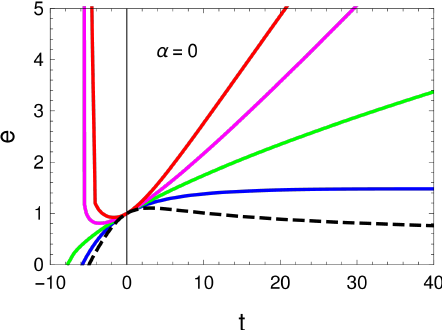}}
\subfigure{\includegraphics[scale=0.46]{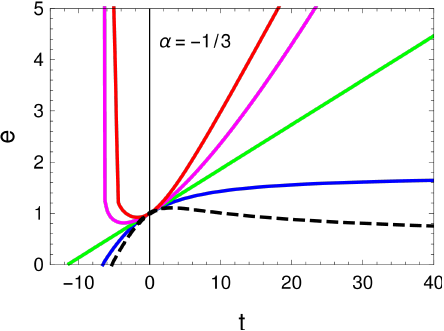}}
\subfigure{\includegraphics[scale=0.46]{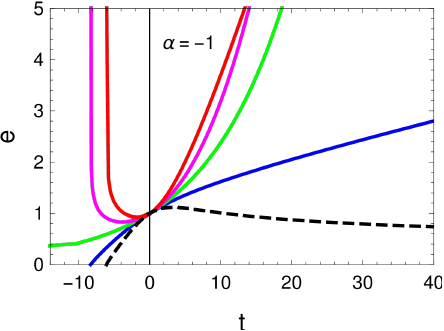}}
\subfigure{\includegraphics[scale=0.19]{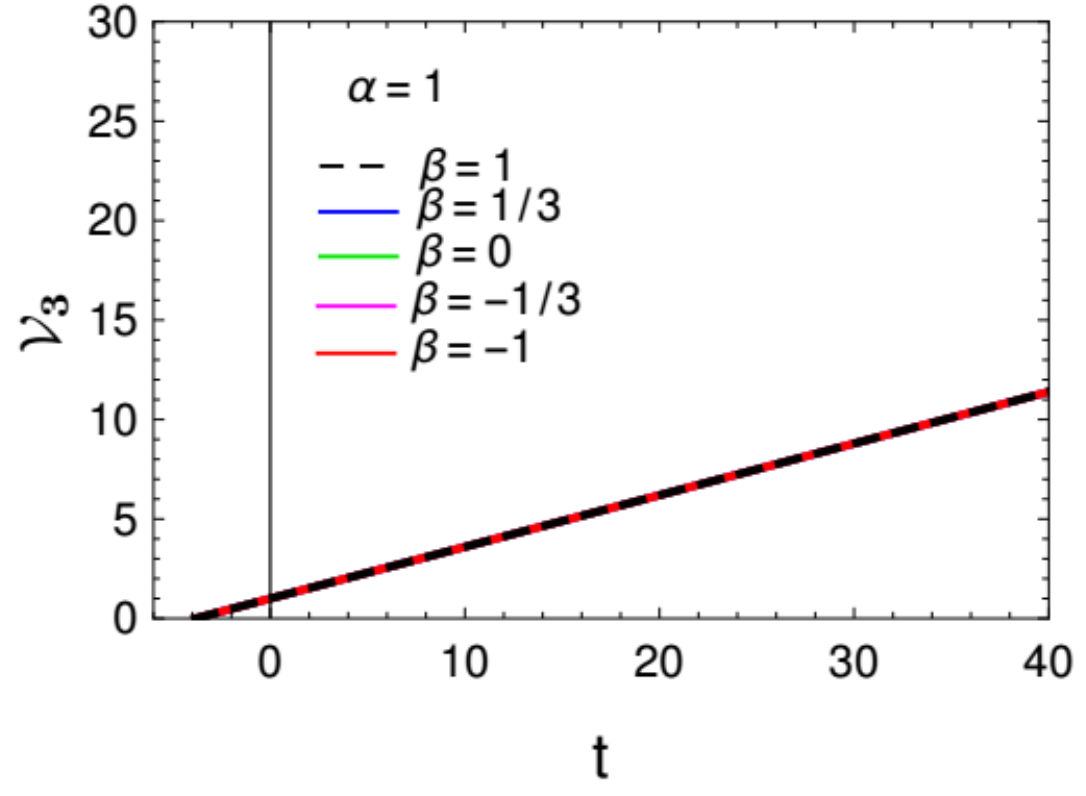}}
\subfigure{\includegraphics[scale=0.19]{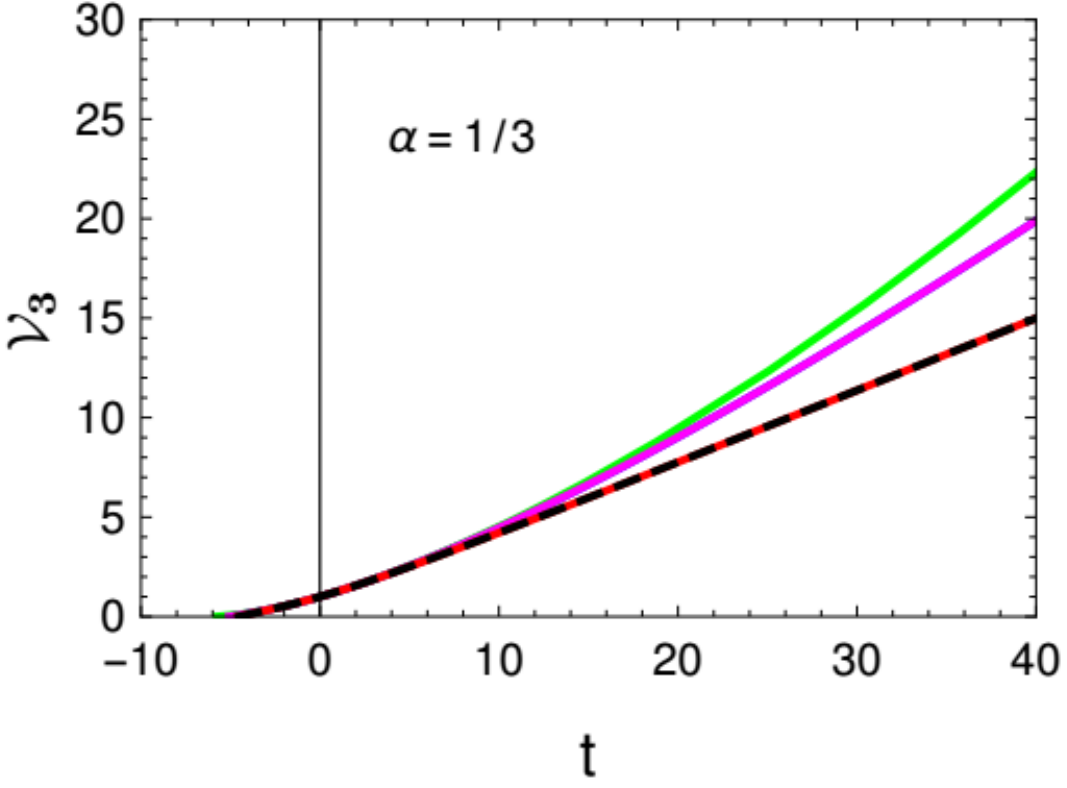}}
\subfigure{\includegraphics[scale=0.19]{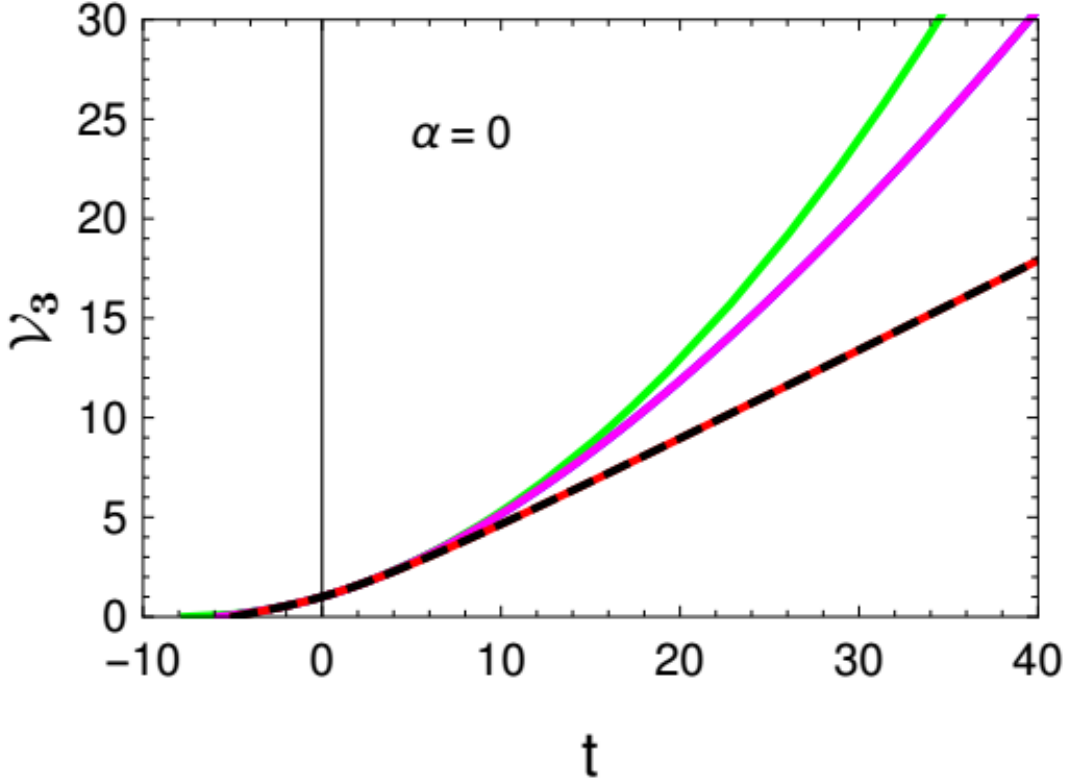}}
\subfigure{\includegraphics[scale=0.19]{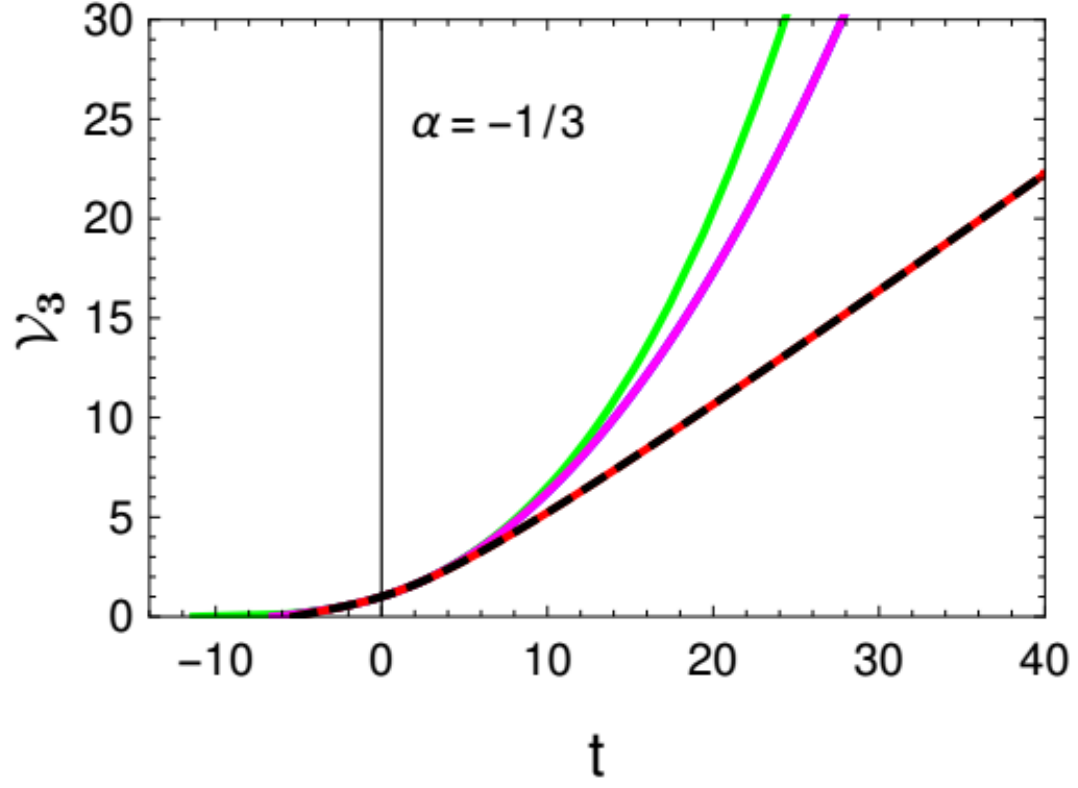}}
\subfigure{\includegraphics[scale=0.19]{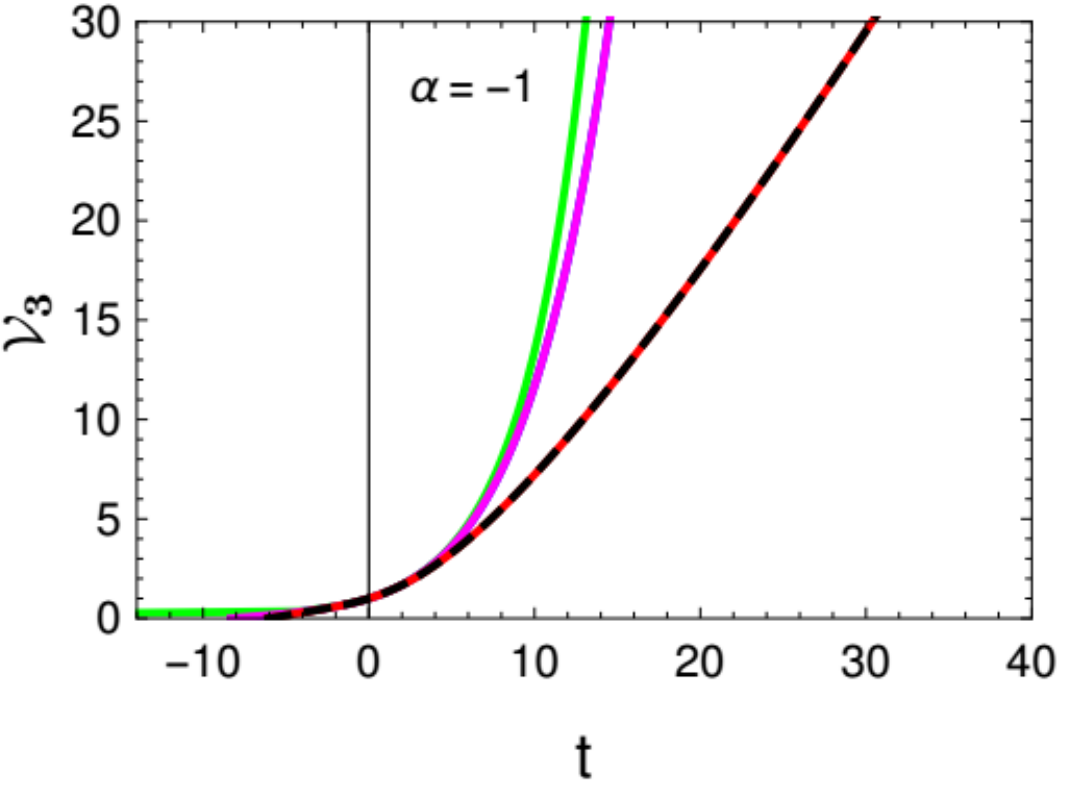}}
\subfigure{\includegraphics[scale=0.465]{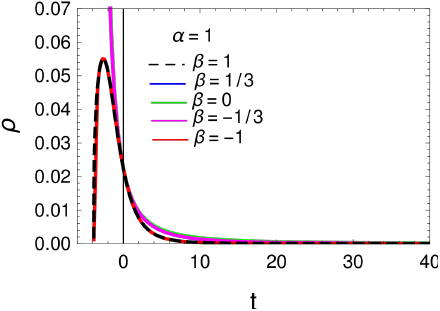}}
\subfigure{\includegraphics[scale=0.465]{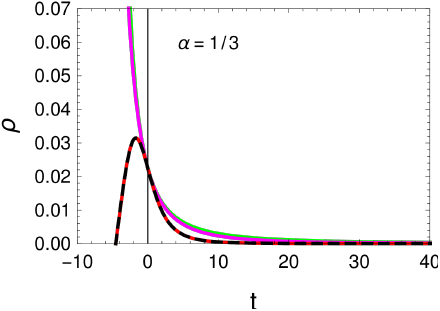}}
\subfigure{\includegraphics[scale=0.465]{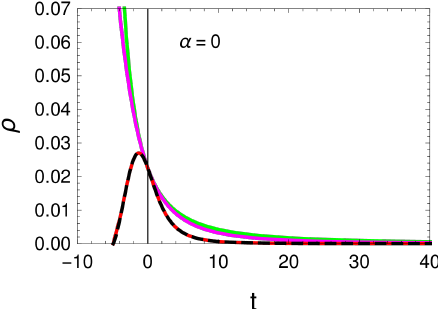}}
\subfigure{\includegraphics[scale=0.465]{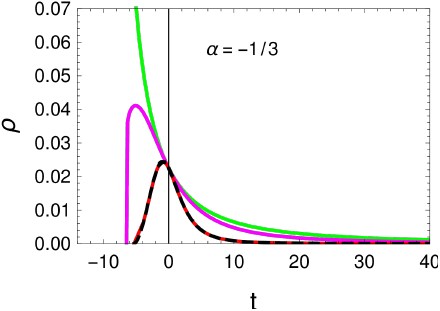}}
\subfigure{\includegraphics[scale=0.465]{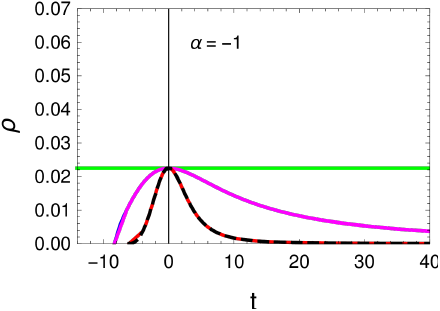}}
\subfigure{\includegraphics[scale=0.19]{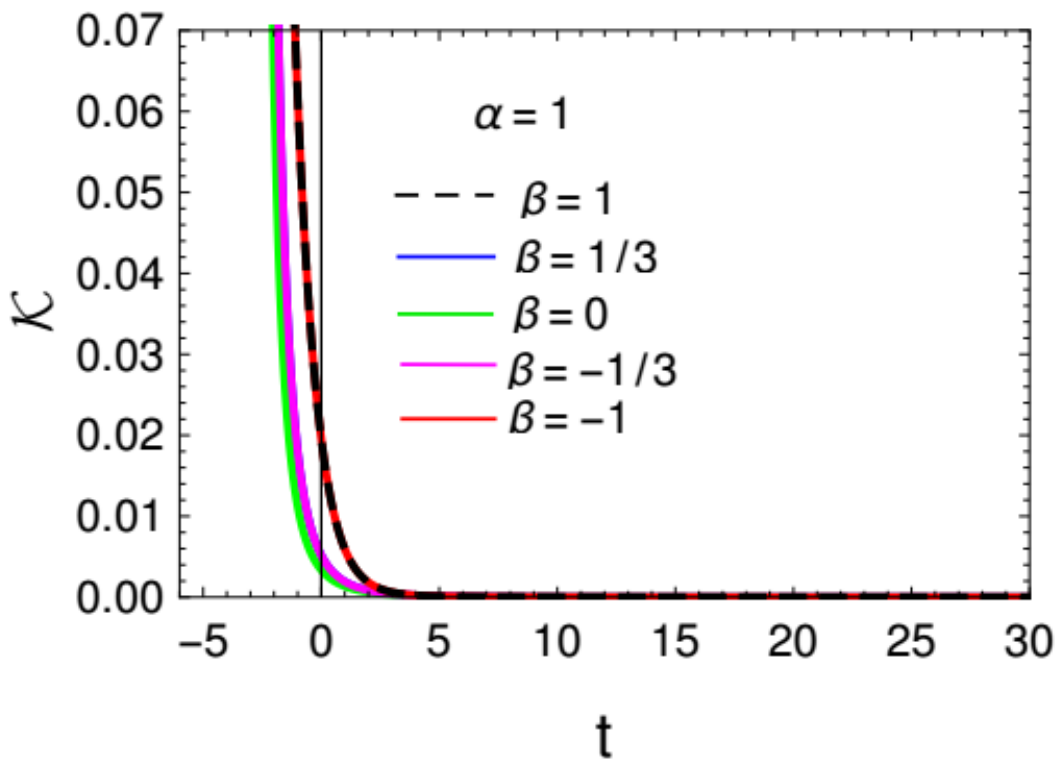}}
\subfigure{\includegraphics[scale=0.19]{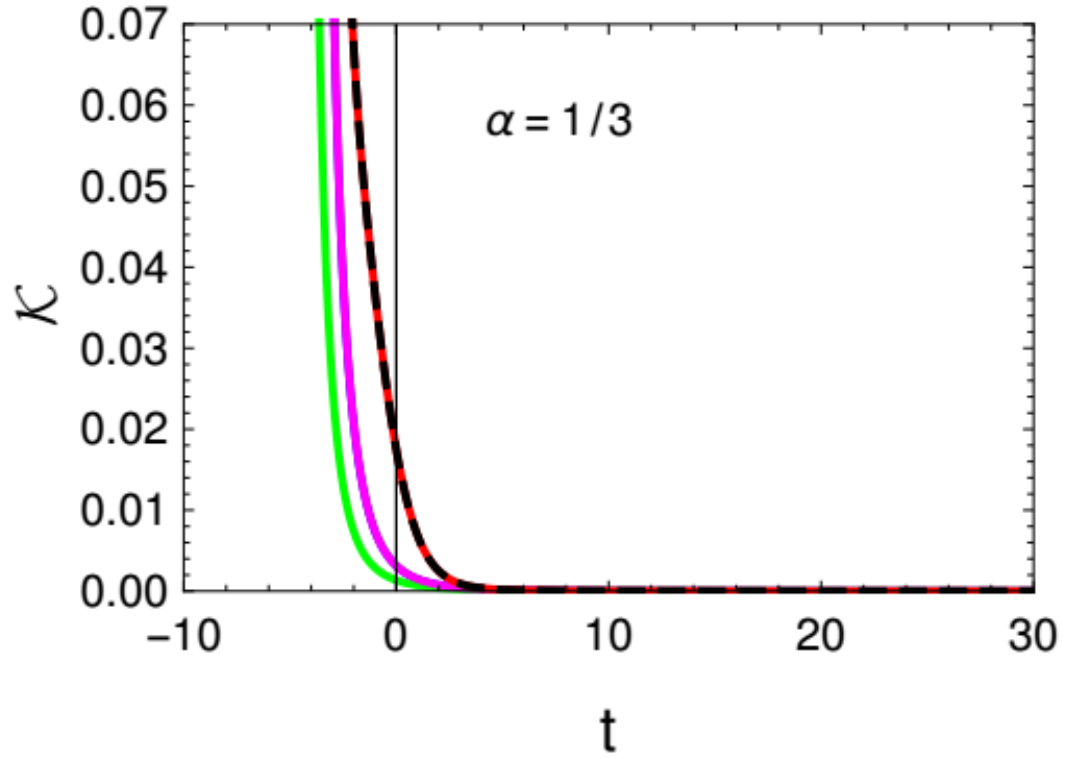}}
\subfigure{\includegraphics[scale=0.19]{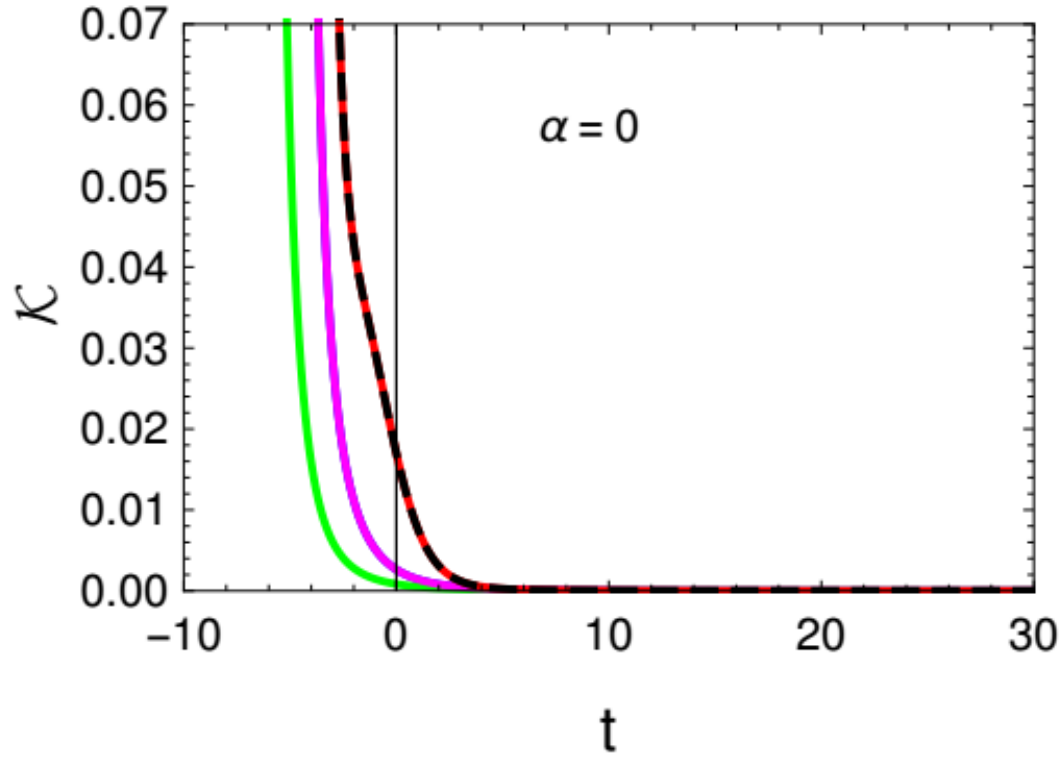}}
\subfigure{\includegraphics[scale=0.19]{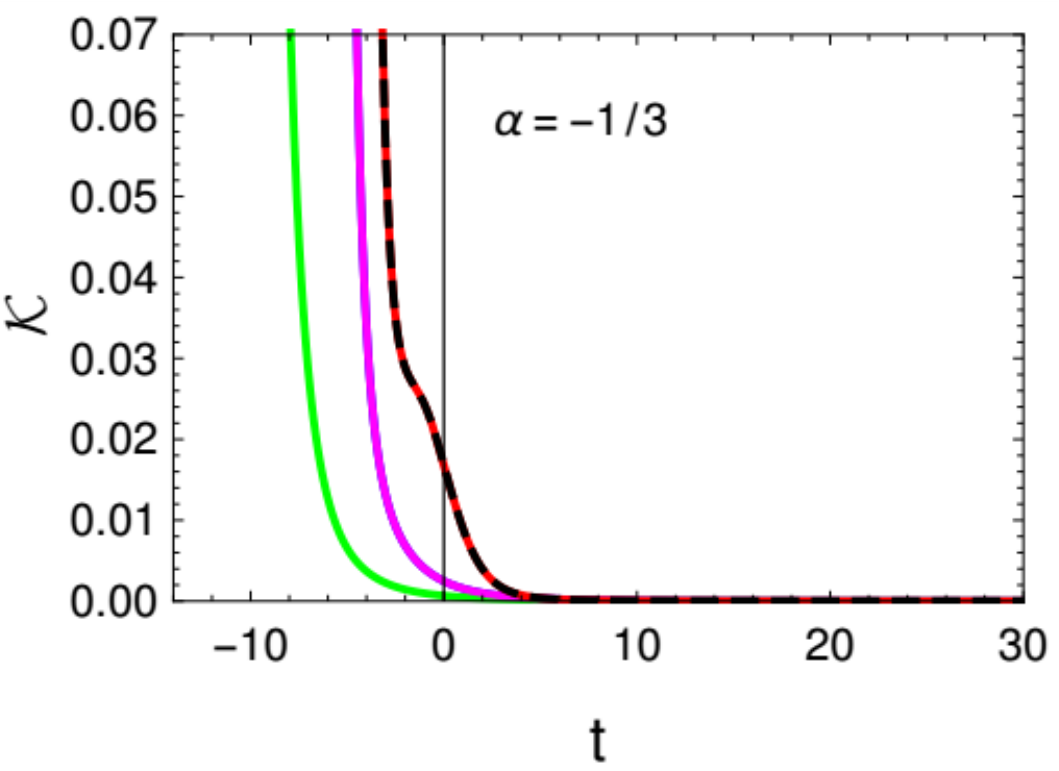}}
\subfigure{\includegraphics[scale=0.19]{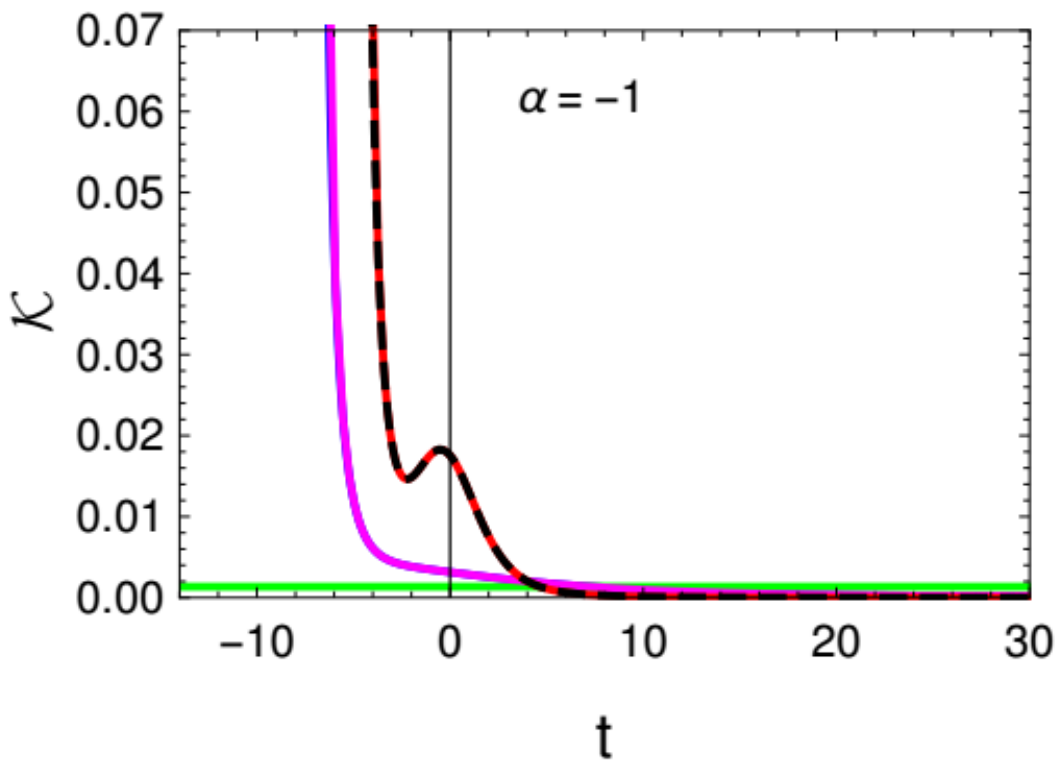}}
\caption{Plots of $c$, $d$, $e$, ${\cal V}_3$, $\rho$ and $\mathcal{K}$ 
for $\alpha, \beta = 0, \pm 1/3,\pm 1$ and $n=1$.
Each column is for the same value of $\alpha$. 
Values of $\beta$ are distinguished by color.}
\label{fig1_n1}
\end{figure}

\begin{figure}[]
\centering
\subfigure{\includegraphics[scale=0.45]{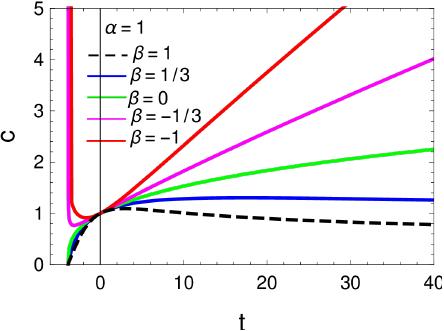}}
\subfigure{\includegraphics[scale=0.45]{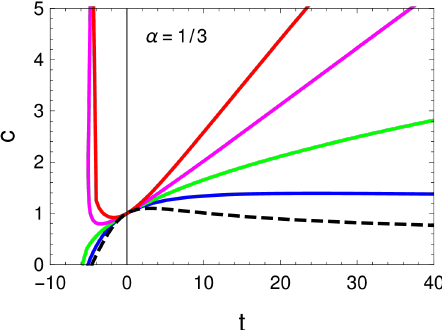}}
\subfigure{\includegraphics[scale=0.45]{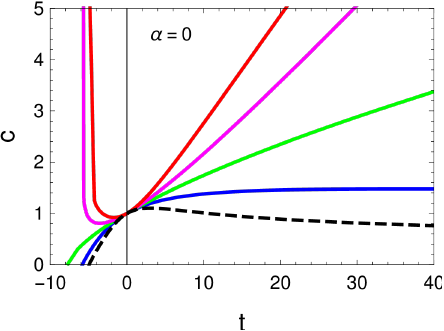}}
\subfigure{\includegraphics[scale=0.45]{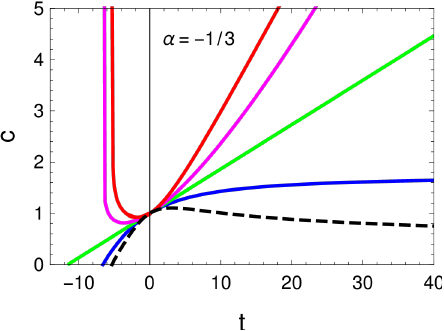}}
\subfigure{\includegraphics[scale=0.45]{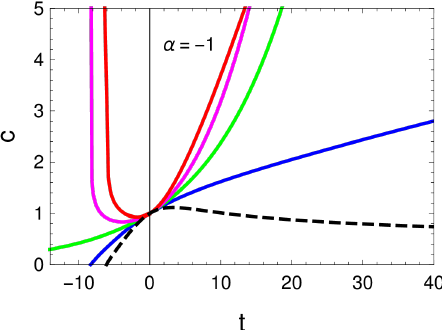}}
\subfigure{\includegraphics[scale=0.45]{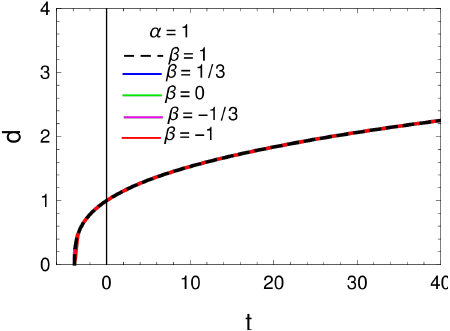}}
\subfigure{\includegraphics[scale=0.45]{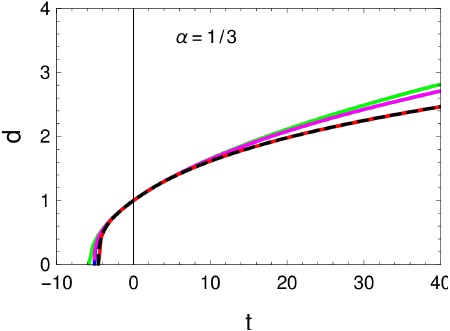}}
\subfigure{\includegraphics[scale=0.45]{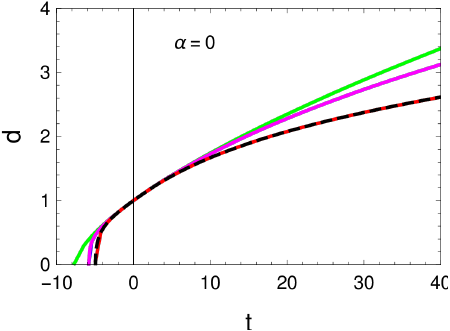}}
\subfigure{\includegraphics[scale=0.45]{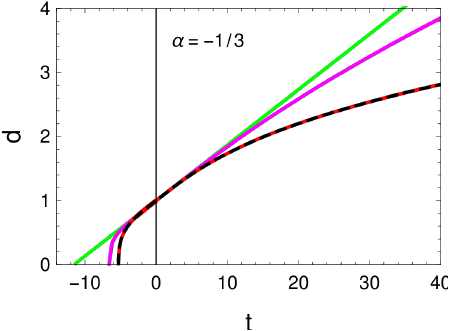}}
\subfigure{\includegraphics[scale=0.45]{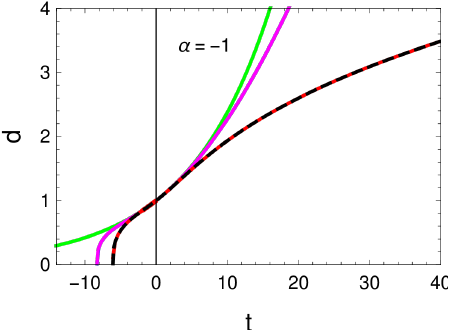}}
\subfigure{\includegraphics[scale=0.46]{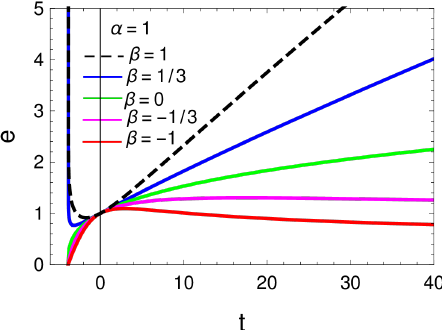}}
\subfigure{\includegraphics[scale=0.46]{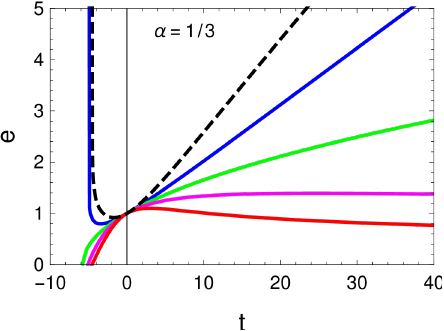}}
\subfigure{\includegraphics[scale=0.46]{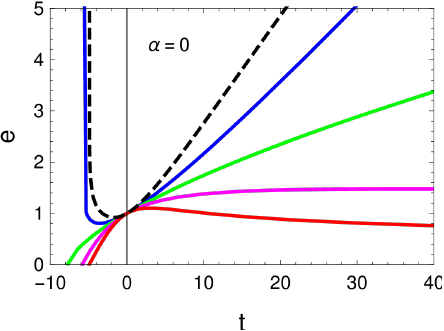}}
\subfigure{\includegraphics[scale=0.46]{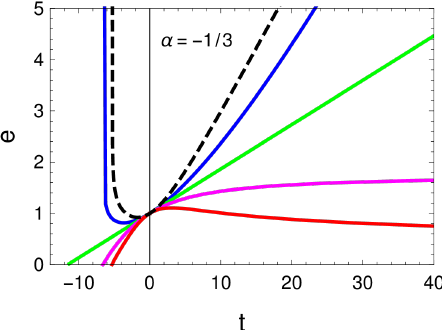}}
\subfigure{\includegraphics[scale=0.46]{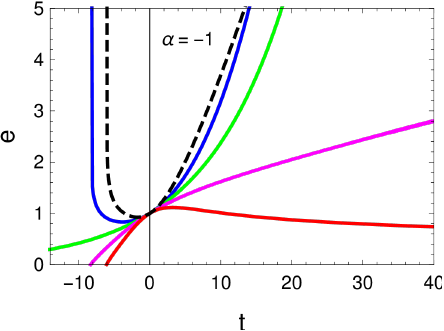}}
\subfigure{\includegraphics[scale=0.19]{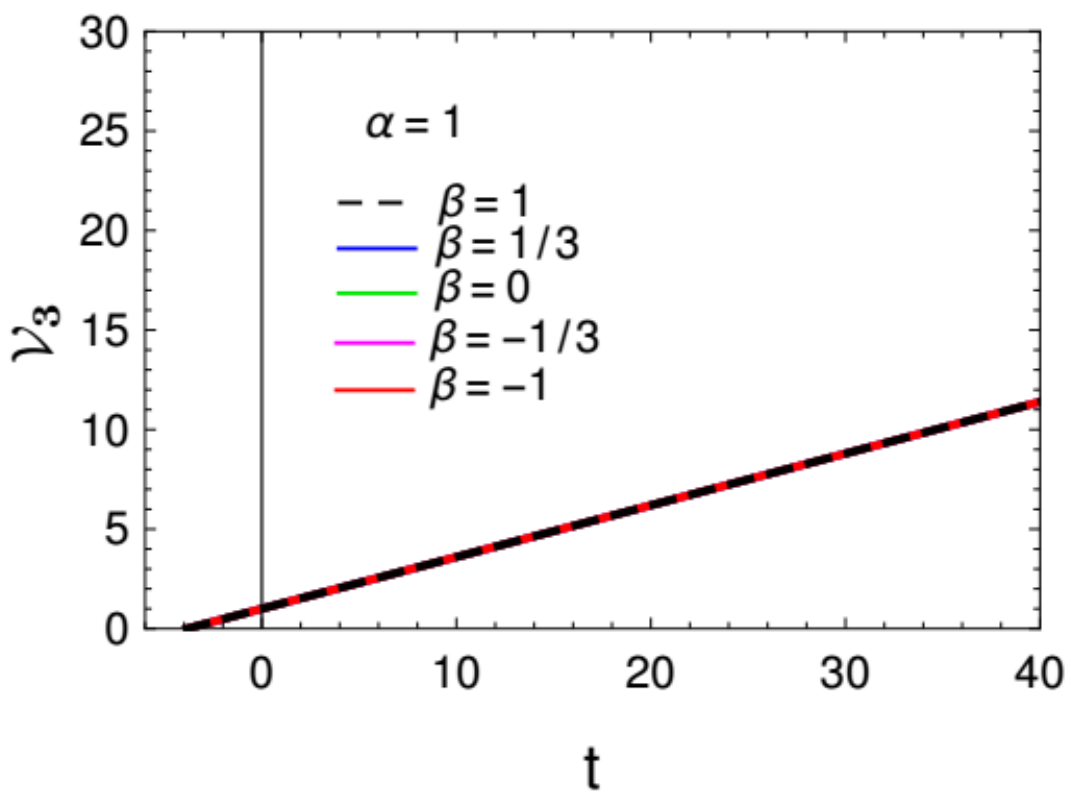}}
\subfigure{\includegraphics[scale=0.19]{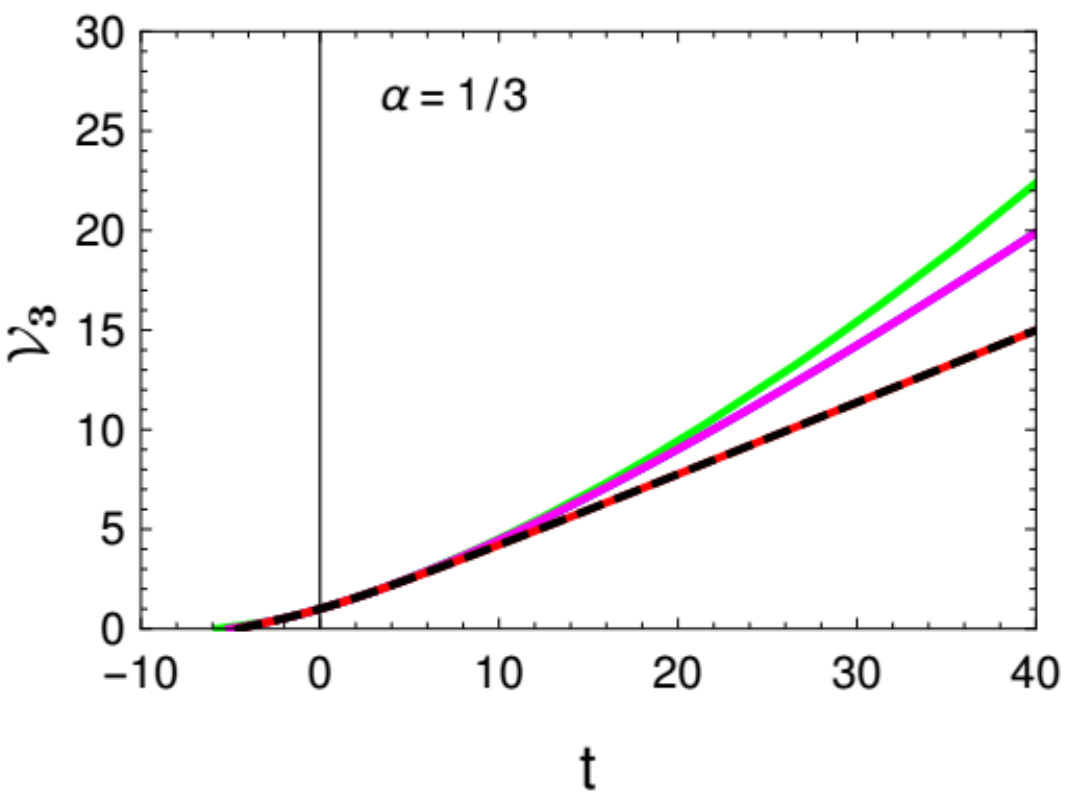}}
\subfigure{\includegraphics[scale=0.19]{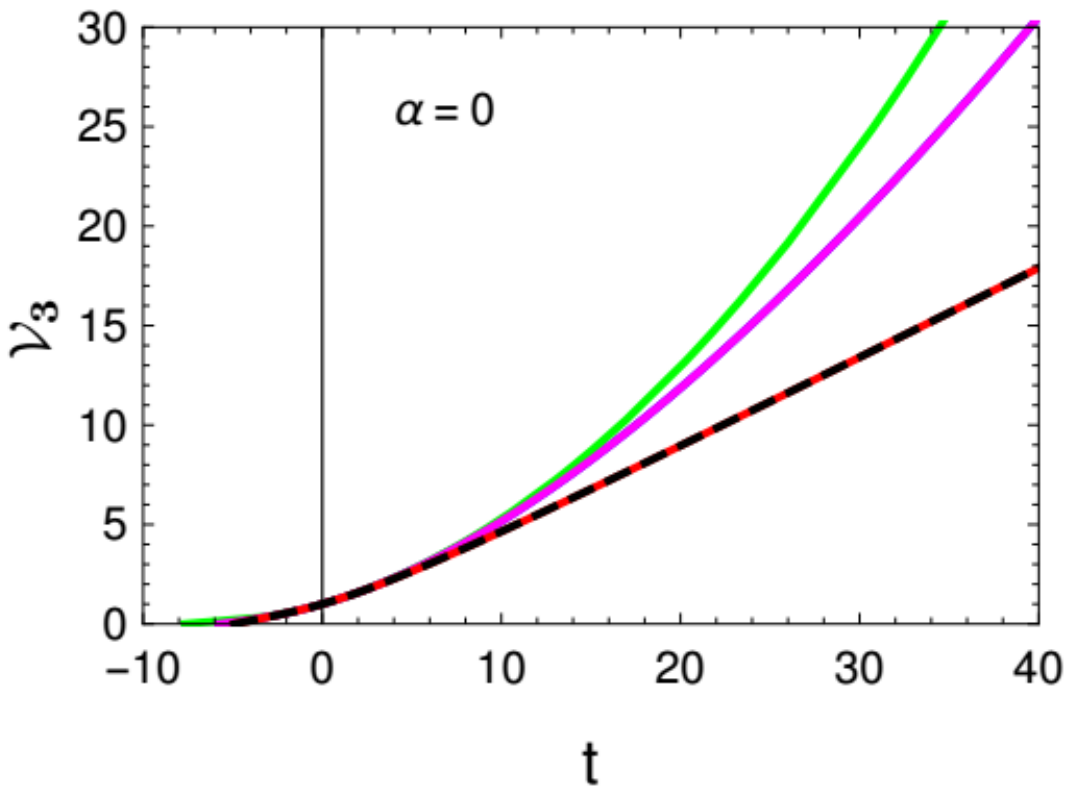}}
\subfigure{\includegraphics[scale=0.19]{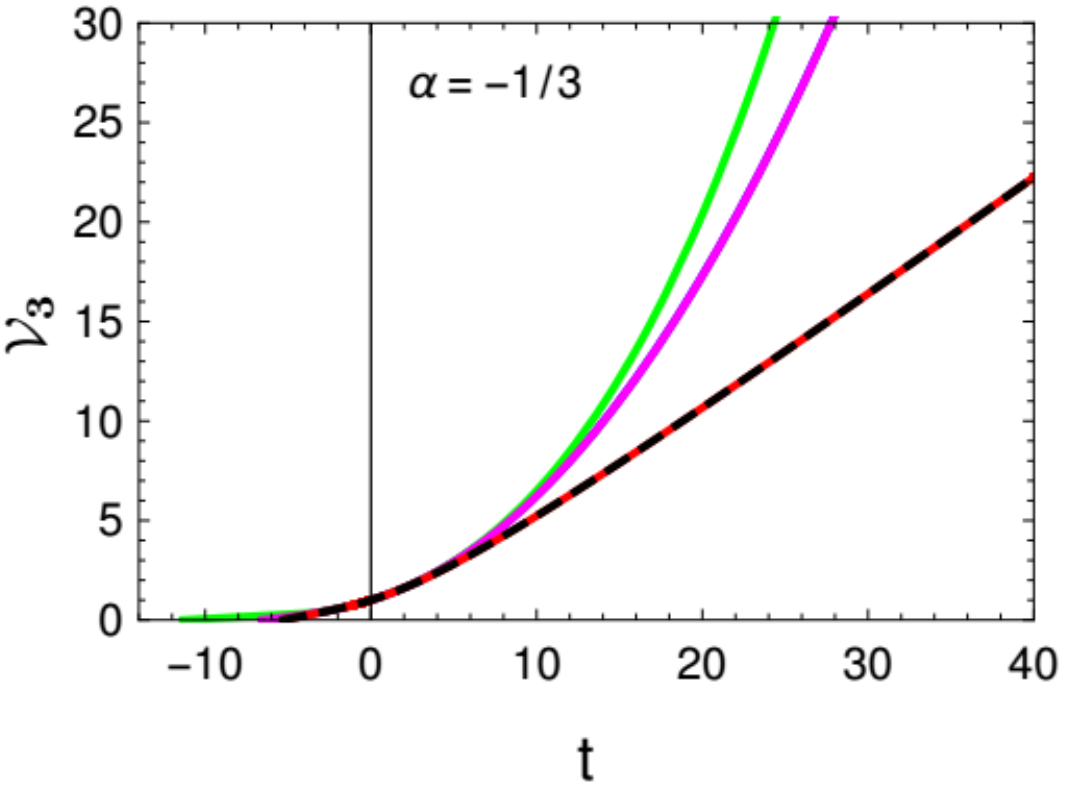}}
\subfigure{\includegraphics[scale=0.19]{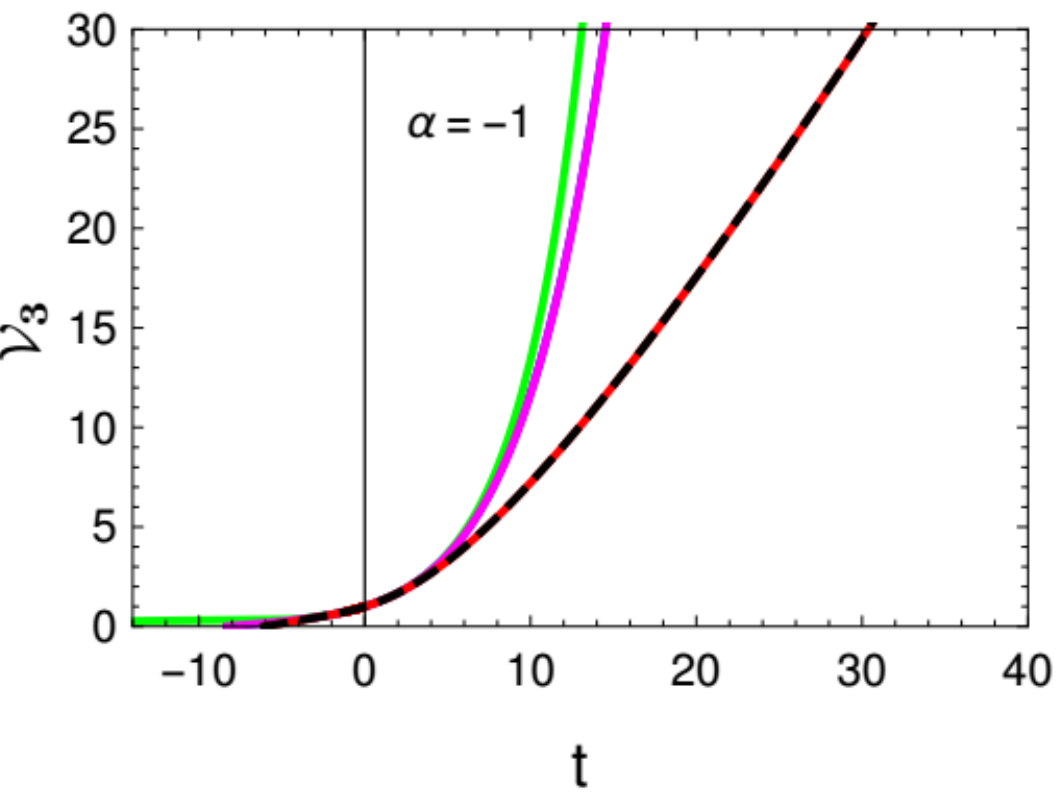}}
\subfigure{\includegraphics[scale=0.465]{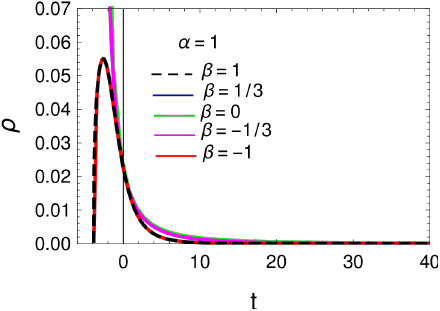}}
\subfigure{\includegraphics[scale=0.465]{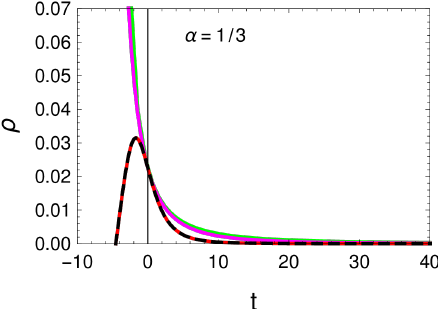}}
\subfigure{\includegraphics[scale=0.465]{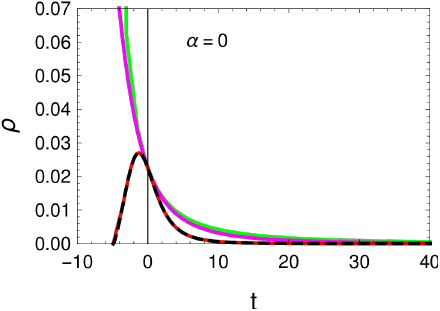}}
\subfigure{\includegraphics[scale=0.465]{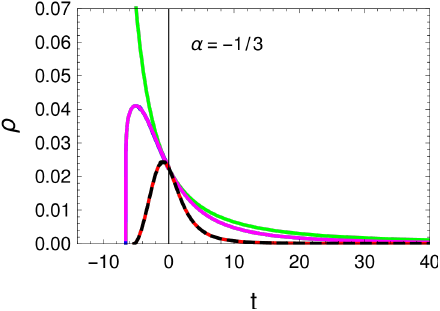}}
\subfigure{\includegraphics[scale=0.465]{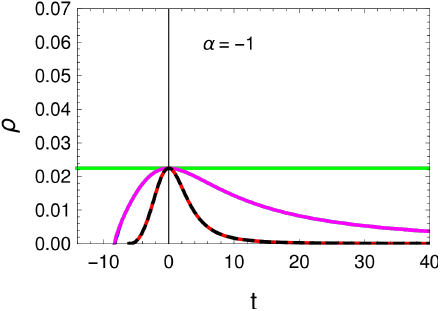}}
\subfigure{\includegraphics[scale=0.19]{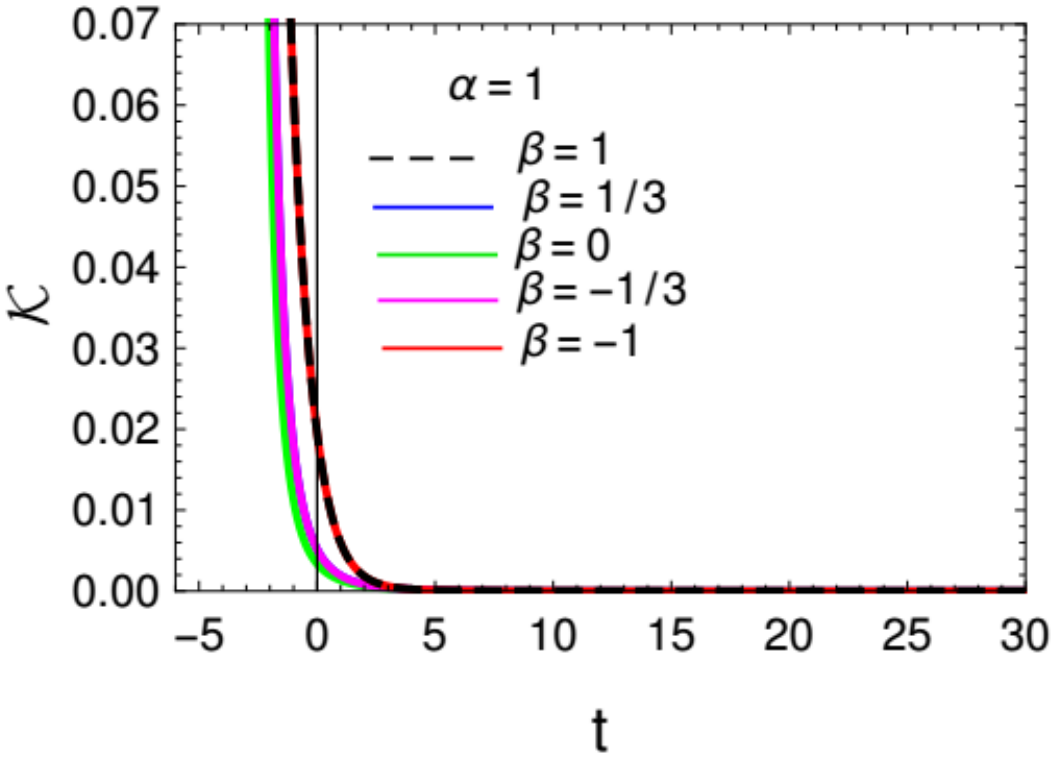}}
\subfigure{\includegraphics[scale=0.19]{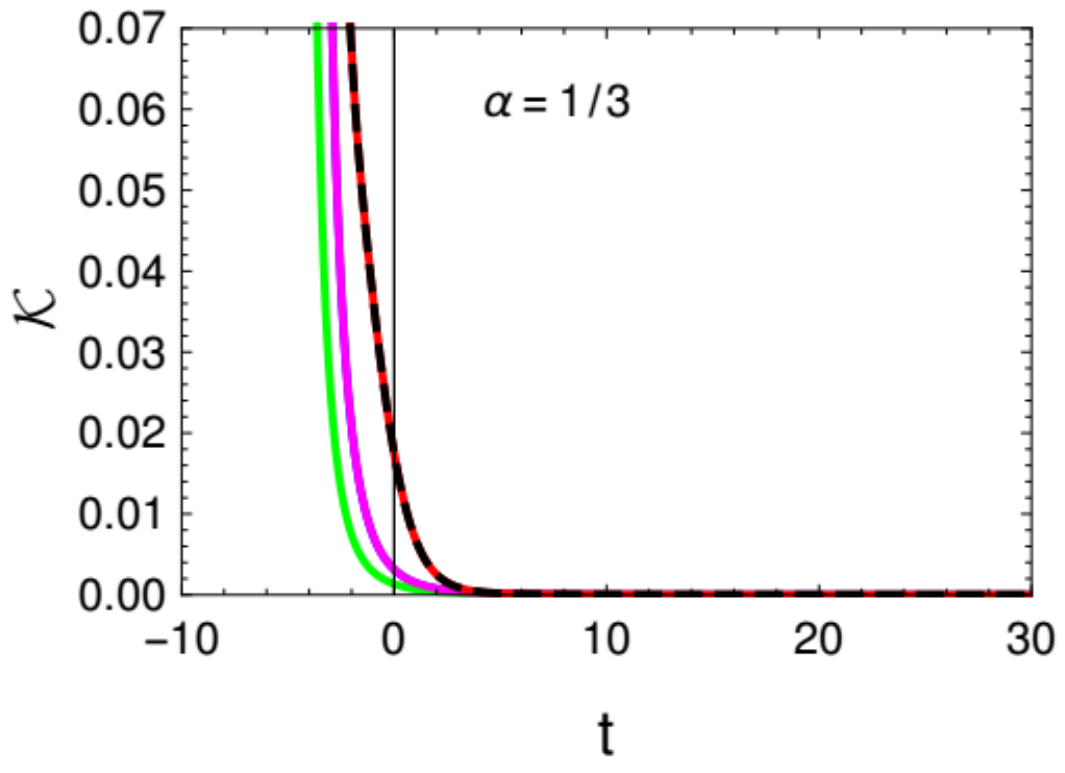}}
\subfigure{\includegraphics[scale=0.19]{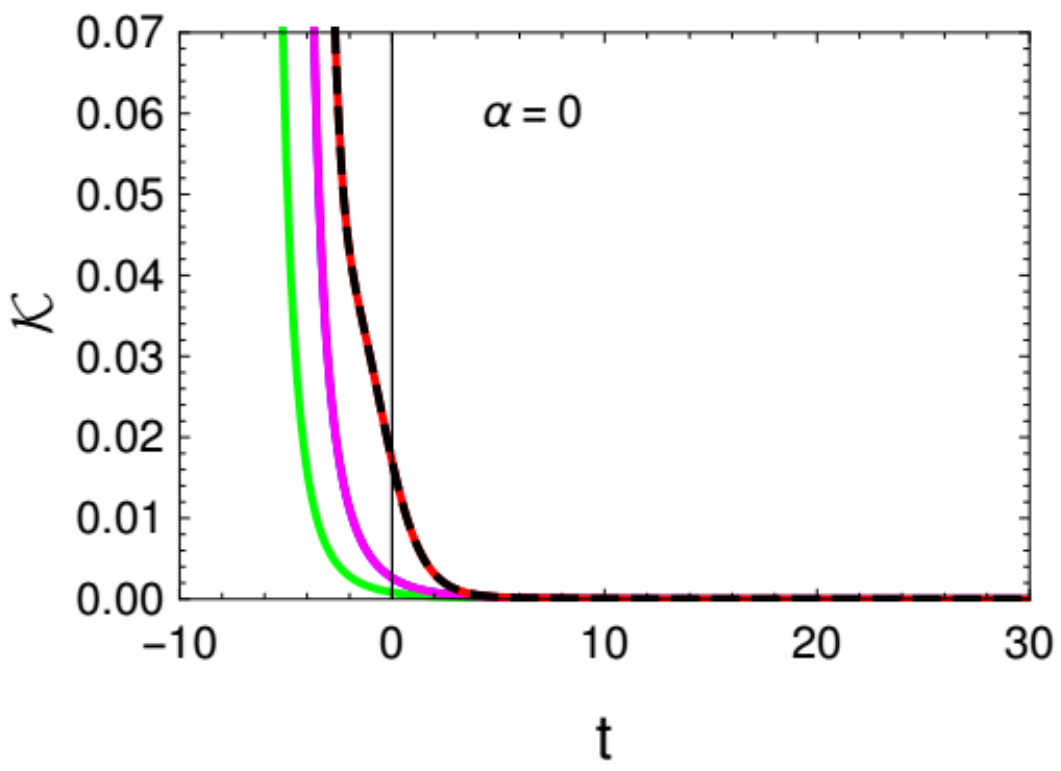}}
\subfigure{\includegraphics[scale=0.19]{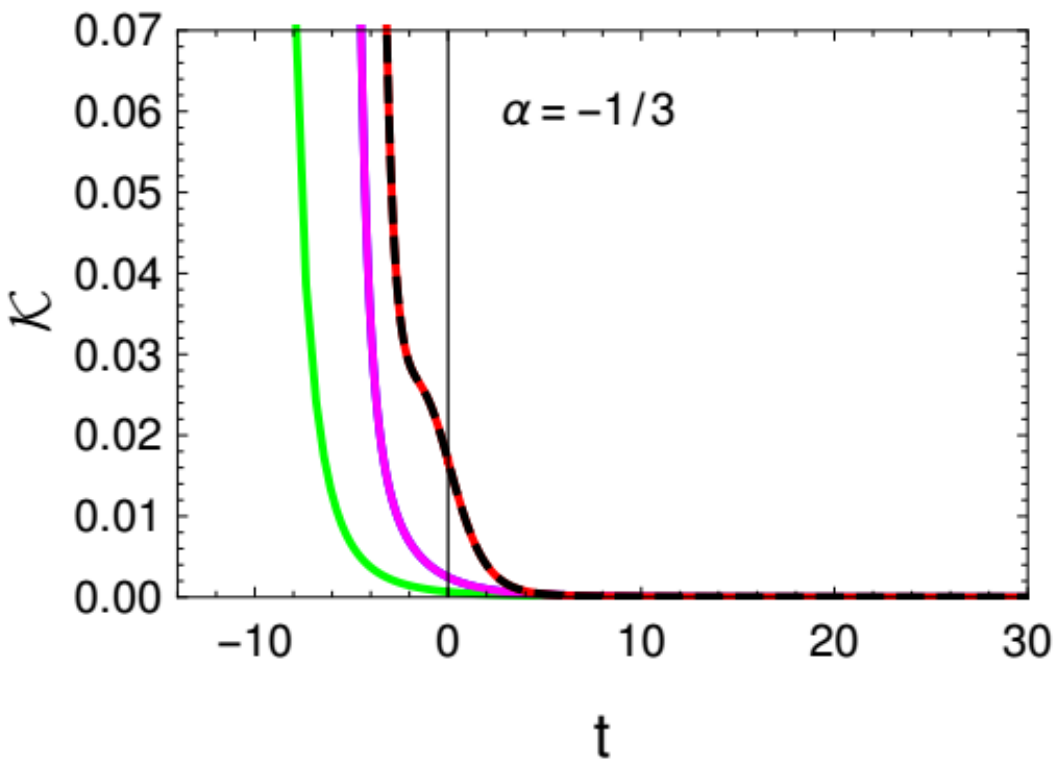}}
\subfigure{\includegraphics[scale=0.19]{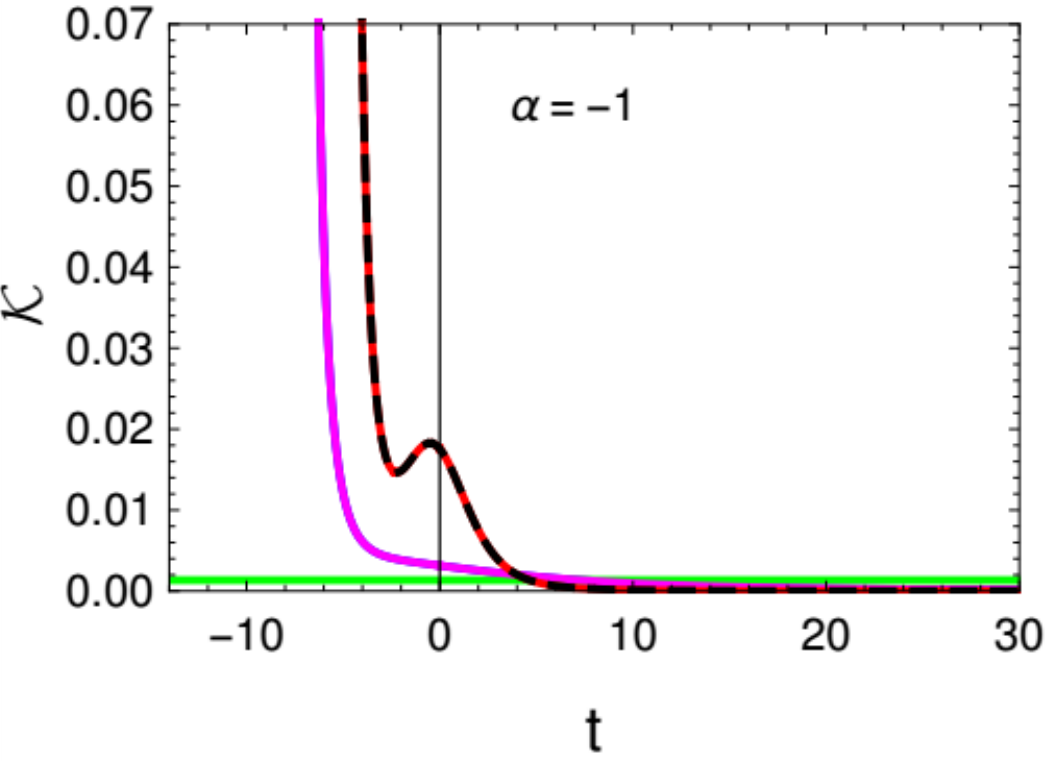}}
\caption{Plots of $c$, $d$, $e$, ${\cal V}_3$, $\rho$ and $\mathcal{K}$ 
for $\alpha, \beta = 0, \pm 1/3,\pm 1$ and $n=-1$.
Each column is for the same value of $\alpha$. 
Values of $\beta$ are distinguished by color.}
\label{fig1_nm1}
\end{figure}

The time-evolution of the metric functions $c$, $d$ and $e$ for $n=0,\pm1$ 
are shown in Figs.~\ref{fig1}-\ref{fig1_nm1}. If $\beta \neq 0$, they evolve in 
a different way, which exhibits an anisotropy. Depending on the signatures 
of $\alpha$, $\beta$ and $n$, they may diverge, or go to zero as $t \to t_s$. 
This means that they can exhibit initially either an expansion, or a contraction. 
For $n=0$, $c$ goes to zero and $d$ diverges as $t \to t_s$ if $\beta>0$, whereas 
they are opposite if $\beta<0$. On the other hand, irrespective of $\beta$, 
$e$ always goes to zero as $t \to t_s$. For $n=1$, the behaviour of 
$d$ as $t \to t_s$ remains the same as that for $n=0$, whereas those of 
$c$ and $e$ get interchanged. For $n=-1$, the behaviour 
of $c$ as $t \to t_s$ remains the same as that for $n=0$, whereas those 
of $d$ and $e$ get interchanged. Although, 
the metric functions may either diverge, or vanish at $t \to t_s$, the 
three-volume density ${\cal V}_3$ always vanishes there [except for the de Sitter 
$(\alpha,\beta)=(-1,0)$ case], which exhibits the big bang at the 
initial moment. ${\cal V}_3$ increases afterwards in time monotonically.

%%%%%%%%%%%%%%%%%%%%%%%%%%%%%%%%%%%%%%%%%%%%%%%%%%%%%%%%%%%%%%%%%%%%%%%%%%
\afterpage{\clearpage}
%%%%%%%%%%%%%%%%%%%%%%%%%%%%%%%%%%%%%%%%%%%%%%%%%%%%%%%%%%%%%%%%%%%%%%%%%%

%%%%%%%%%%%%%%%%%%%%%%%%%%%%%%%%%%%%%%%%%%%%%%%%%%%%%%%%%%%%%%%%%%%%%%%%%%
%%%%%%%%%%%%%%%%%%%%%%%%%%%%%%%%%%%%%%%%%%%%%%%%%%%%%%%%%%%%%%%%%%%%%%%%%%

%%%%%%%%%%%%%%%%%%%%%%%%%%%%%%%%%%%%%%%%%%%%%%%%%%%%%%%%%%%%%%%%%%%%%%%%%%
%%%%%%%%%%%%%%%%%%%%%%%%%%%%%%%%%%%%%%%%%%%%%%%%%%%%%%%%%%%%%%%%%%%%%%%%%%

\subsection{Energy density $\rho$}

The energy density $\rho$ in the Friedmann universe always diverges at the 
moment of big  bang ($t \to t_s$) and decreases afterwards monotonically 
for $\alpha > -1$. It remains constant in time for the de Sitter case 
$\alpha = -1$.

For $\beta \neq 0$, however, we observe some cases where, unlike in the 
Friedmann universe, $\rho$ does not diverge as $t\to t_s$; $\rho$ 
increases from zero at the early stage, reaches a maximum value 
and then starts to decrease in time (see Figs.~\ref{fig1}-\ref{fig1_nm1}).

\subsection{Kretschmann scalar $\mathcal{K}$}
\label{SecKR}

In the previous subsections, we found that the energy density $\rho$ 
does not necessarily diverge even though the three-volume density 
${\cal V}_3$ vanishes at $t=t_s$. It is therefore useful to evaluate 
the curvature invariants to lookup the nature of the 
spacetime at $t=t_s$. The Kretschmann scalar is given by
\begin{equation}
\mathcal{K} = R^{\mu\nu\rho\sigma} R_{\mu\nu\rho\sigma}
=\frac{(1/9216)}{(1+3n^2)^2 u^{\sqrt{3}}v^{\sqrt{3}}}\left(Y_1 u^{4s_1}+Y_2 u^{3s_1}v^{s_2}+Y_3 u^{2s_1}v^{2s_2}+Y_4 u^{s_1} v^{3s_2}+Y_5 v^{4s_2} \right),
\label{eq:kretschmann}
\end{equation}
where
\begin{eqnarray}
Y_1 &=& A_2^4 \left[\bar{n}_2^4 X_1- \bar{n}_1\bar{n}_2^3 X_2+ \bar{n}_1^2\bar{n}_2^2 X_3-\bar{n}_1^3 \bar{n}_2 X_4+ \bar{n}_1^4 X_5 \right] \nonumber \\
Y_2 &=& -A_1 A_2^3 \left[4n_2\bar{n}_2^3 X_1- (n_1\bar{n}_2+3\bar{n}_1 n_2)\bar{n}_2^2 X_2+ 2(n_1\bar{n}_2+\bar{n}_1 n_2)\bar{n}_1\bar{n}_2 X_3- (3n_1\bar{n}_2+\bar{n}_1 n_2)\bar{n}_1^2 X_4+ 4n_1\bar{n}_1^3 X_5 \right] \nonumber \\
Y_3 &=& A_1^2 A_2^2 \left[6n_2^2\bar{n}_2^2 X_1- 3(n_1\bar{n}_2+\bar{n}_1 n_2)n_2\bar{n}_2 X_2+ (n_1^2\bar{n}_2^2+4n_1\bar{n}_1 n_2\bar{n}_2+n_2^2\bar{n}_1^2) X_3- 3(n_1\bar{n}_2+\bar{n}_1 n_2)n_1\bar{n}_1 X_4\right. \nonumber \\
& & \left. + 6n_1^2\bar{n}_1^2 X_5 \right] \nonumber \\
Y_4 &=& -A_1^2 A_2 \left[4n_2^3\bar{n}_2 X_1- (3n_1\bar{n}_2+\bar{n}_1 n_2)n_2^2 X_2+ 2(n_1\bar{n}_2+\bar{n}_1 n_2)n_1 n_2 X_3- (n_1\bar{n}_2+3\bar{n}_1 n_2)n_1^2 X_4+ 4n_1^3\bar{n}_1 X_5 \right] \nonumber \\
Y_5 &=& A_1^4 \left[n_2^4 X_1- n_1 n_2^3 X_2+ n_1^2 n_2^2 X_3- n_1^3 n_2 X_4+ n_1^4 X_5 \right] \nonumber
\end{eqnarray}
and
\begin{eqnarray}
X_1 &=& (1+3n)^2\left[7+2\alpha+3\alpha^2+8(1+n)\beta+8(1+3n^2)\beta^2\right] \nonumber \\
X_2 &=& 8(1+3n)\left[1-6n+2\alpha+3\alpha^2+4(1-n)\beta+8(1+3n^2)\beta^2\right] \nonumber \\
X_3 &=& 6\left[5+27n^2+6(1-n^2)\alpha+9(1-n^2)\alpha^2-8n(5+3n^2)\beta+24(1-n^2)(1+3n^2)\beta^2\right] \nonumber \\
X_4 &=& 8(1-3n)\left[1+6n+2\alpha+3\alpha^2-4(1+n)\beta+8(1+3n^2)\beta^2\right] \nonumber \\
X_5 &=& (1-3n)^2\left[7+2\alpha+3\alpha^2-8(1-n)\beta+8(1+3n^2)\beta^2\right] \nonumber
\end{eqnarray}
In the limit $t\to t_s$, one of the terms in Eq. \eqref{eq:kretschmann} 
dominates. The nature of $\mathcal{K}$ in this limit 
depends on this dominating term. Note that, for $n=-1/3$, $Y_1=0$ and 
$Y_2=0$ as $X_1=0$, $X_2=0$ and $\bar{n}_1=0$. Therefore, the dominating 
term in this case is one of $Y_3$, $Y_4$ and $Y_5$. On the other 
hand, for $n=1/3$, $Y_4=0$ and $Y_5=0$ as $X_4=0$, $X_5=0$ and $n_2=0$. 
Therefore, the dominating term in this case is one of $Y_1$, 
$Y_2$ and $Y_3$. As discussed in the previous section, for $n=\pm 1/3$, 
two of the effective pressures in the diagonal energy-momentum tensor 
become equal and the solution becomes equivalent to the 
one obtained in Ref. \cite{Cho:2022rgs}. As shown in Ref. \cite{Cho:2022rgs}, 
$\mathcal{K}$ in these cases does not always diverge, but 
is finite as $t\to t_s$ for some values of $\alpha$ and $\beta$.

On the other hand, for $n\neq \pm 1/3$, all the effective pressures are 
different, and none of the coefficients $Y_1$, $Y_2$, 
$Y_4$ and $Y_5$ is zero. Therefore, the dominating 
term and hence the nature of $\mathcal{K}$ as $t\to t_s$ is different from 
the ones for $n=\pm 1/3$. For $n=0,\pm 1$, we have plotted ${\cal K}$ 
in Figs.~\ref{fig1}-\ref{fig1_nm1}. Note that, unlike the results presented 
in Ref. \cite{Cho:2022rgs} for $n=\pm 1/3$, $\mathcal{K}$ for $n=0,\pm 1$ always 
diverges and hence the solution is always singular. We will see this also 
later when we present the nature of $\mathcal{K}$ as $t\to t_s$ for arbitrary 
values of $n$, $\alpha$ and $\beta$.

\subsection{$\rho(t_s)$ \& $\mathcal{K}(t_s)$}

In order to examine the nature of the spacetime at $t\to t_s$ 
for arbitrary $n$, $\alpha$ and $\beta$, we investigate 
the behaviour of $\rho$ and $\mathcal{K}$ in the limit 
$t\to t_s$. Considering the signatures of the terms in the integrand, 
the $v$-solution for Class G in Eq.~\eqref{eq:t_gen} is classified into four cases, \\
(i) G1: $\beta<0\; \& \; s_2=\sqrt{3}(1-\alpha)+2\sqrt{1+3n^2}\beta > 0$, \\
(ii) G2: $\beta<0\; \& \; s_2=\sqrt{3}(1-\alpha)+2\sqrt{1+3n^2}\beta < 0$, \\
(iii) G3: $\beta>0\; \& \; s_1=\sqrt{3}(1-\alpha)-2\sqrt{1+3n^2}\beta > 0$, \\
(iv) G4: $\beta>0\; \& \; s_1=\sqrt{3}(1-\alpha)-2\sqrt{1+3n^2}\beta < 0$. \\
The behaviour of ${\cal V}_3$ as well as $\rho$ and $\mathcal{K}$ in the 
limit $t\to t_s$ is obtained in Appendix \ref{appendixB}. In this limit, 
$\rho$ and $\mathcal{K}$ behave as 
[see Eqs. \eqref{eq:early_rho} and \eqref{eq:early_K}]
\begin{equation}
\rho \sim \left\{
  \begin{array}{ll}
  \frac{1}{(t-t_s)^{[\sqrt{3}(1+\alpha)+2\sqrt{1+3n^2}\beta]/\sqrt{3}}} 
  & (\mbox{G1 and G2}) \\
  \frac{1}{(t-t_s)^{[\sqrt{3}(1+\alpha)-2\sqrt{1+3n^2}\beta]/\sqrt{3}}} 
  & (\mbox{G3 and G4}) 
  \end{array}
\right.,
\end{equation}

\begin{equation}
\mathcal{K} \sim \left\{
  \begin{array}{lllll}
  \frac{1}{(t-t_s)^4} 
  & \mbox{for $n\neq\pm 1/3$} \quad (\mbox{G1-G4}) \\
  \frac{1}{(t-t_s)^4} 
  & \mbox{for $n=-1/3$} \quad (\mbox{G1 and G2}) \\
  \frac{1}{(t-t_s)^{2(1+\alpha-4\beta/3)}} 
  & \mbox{for $n=-1/3$} \quad (\mbox{G3 and G4}) \\
  \frac{1}{(t-t_s)^{2(1+\alpha+4\beta/3)}} 
  & \mbox{for $n=1/3$} \quad (\mbox{G1 and G2}) \\
  \frac{1}{(t-t_s)^4} 
  & \mbox{for $n=1/3$} \quad (\mbox{G3 and G4}) 
  \end{array}
\right..
\end{equation}

\begin{figure}[]
\centering
\subfigure[$n=0$]{\includegraphics[scale=0.25]{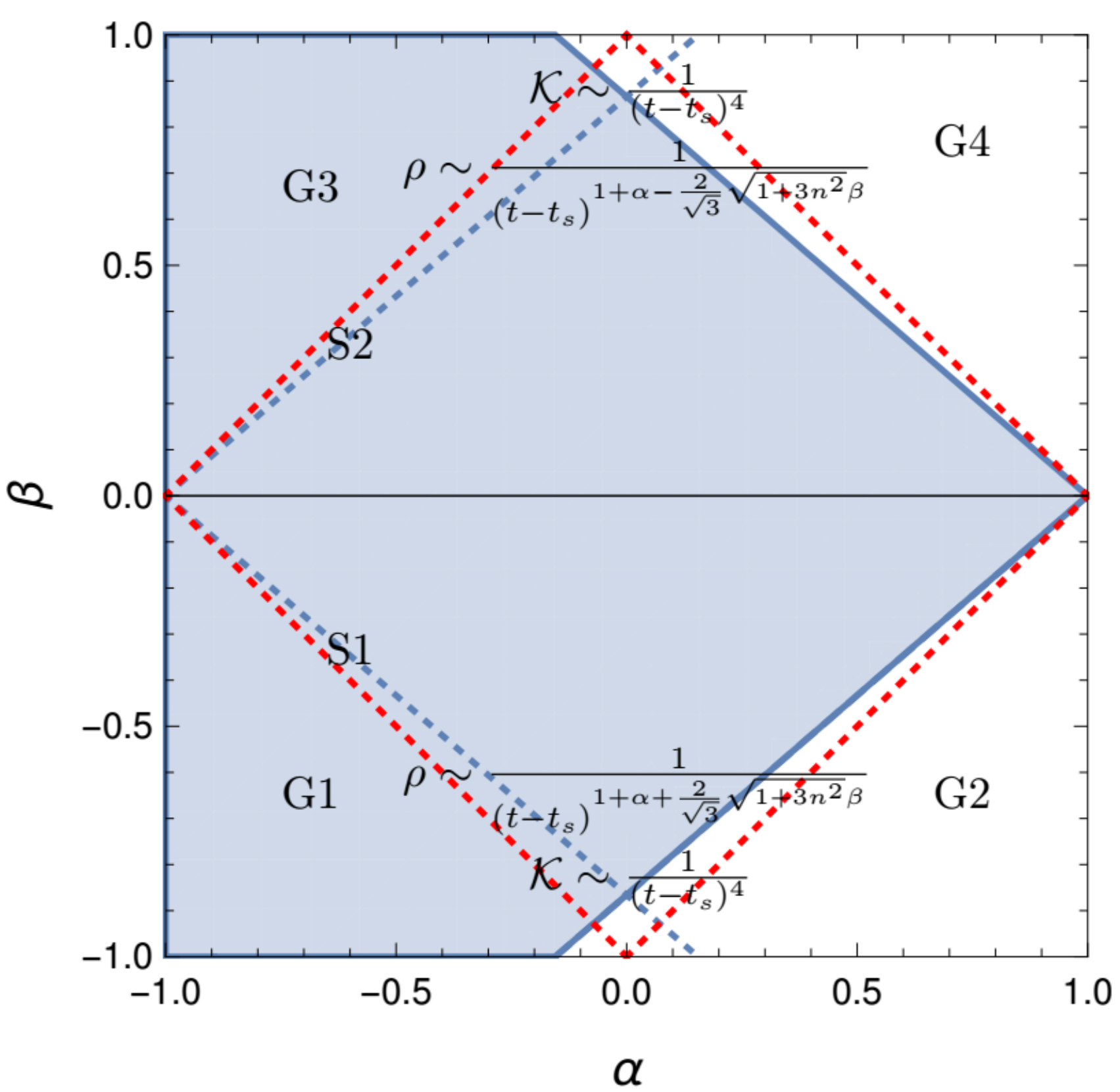}}
\subfigure[$n=-1/3$]{\includegraphics[scale=0.25]{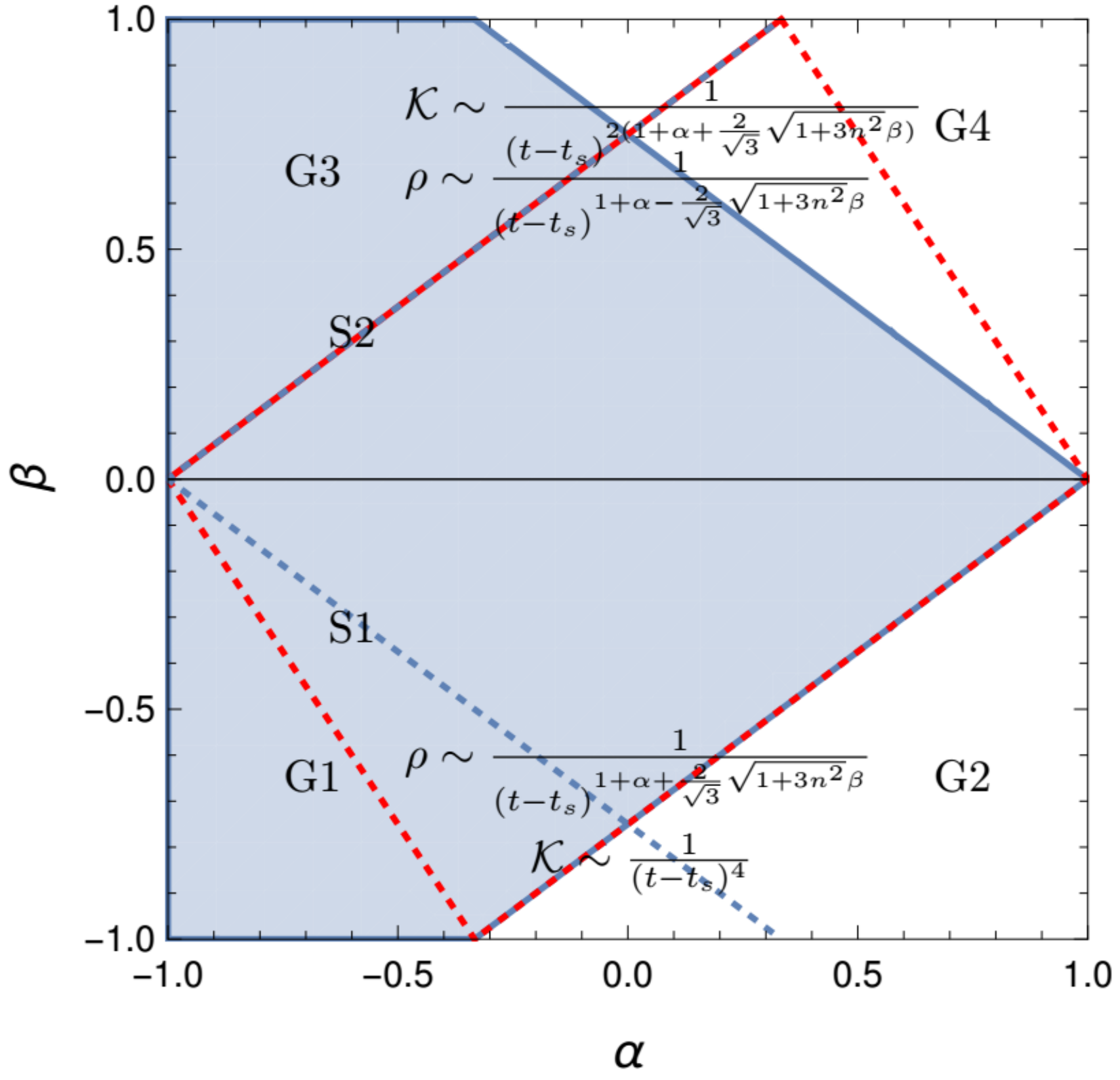}}
\subfigure[$n=1/3$]{\includegraphics[scale=0.25]{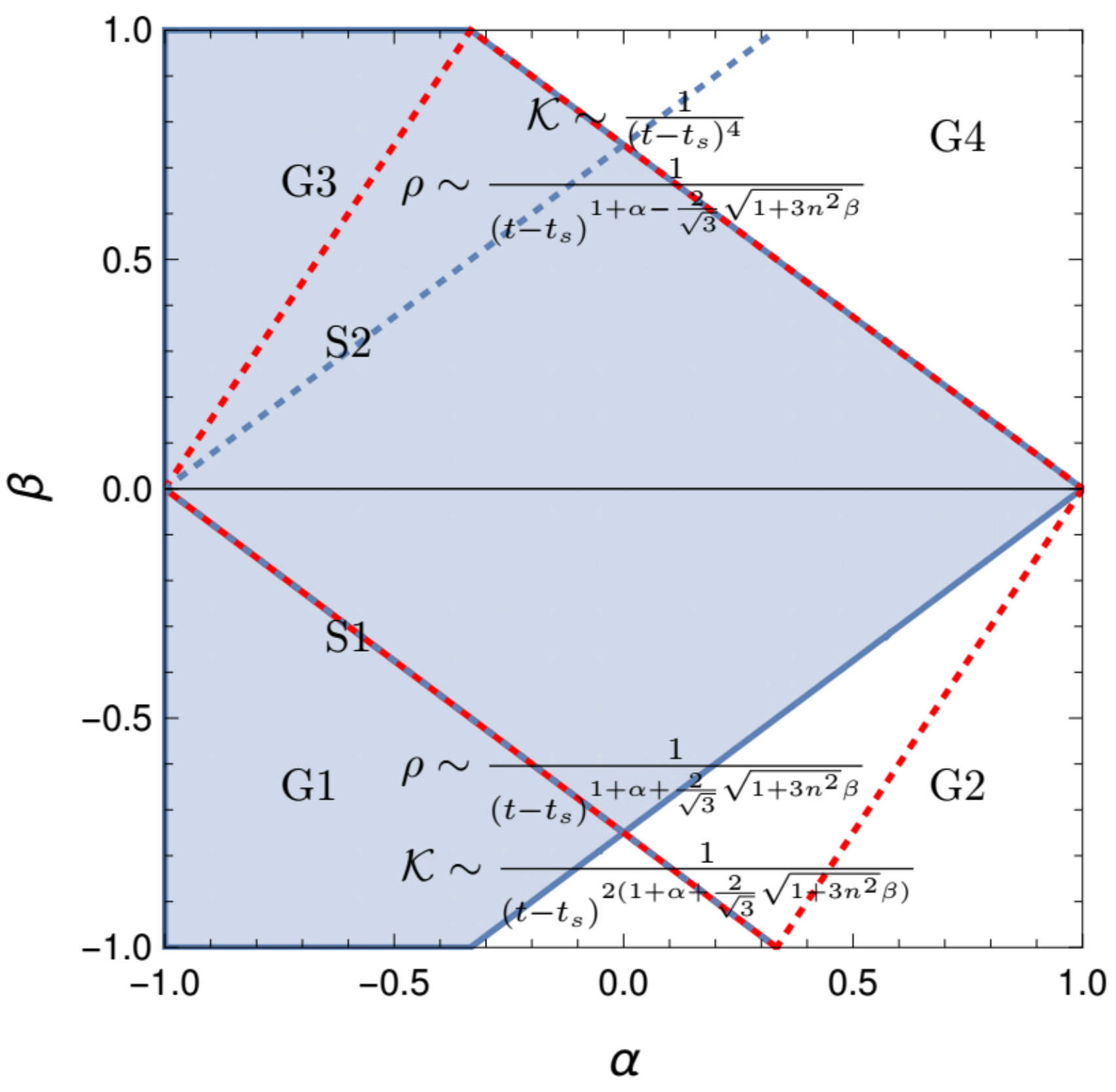}}
\subfigure[$n=\pm 1$]{\includegraphics[scale=0.30]{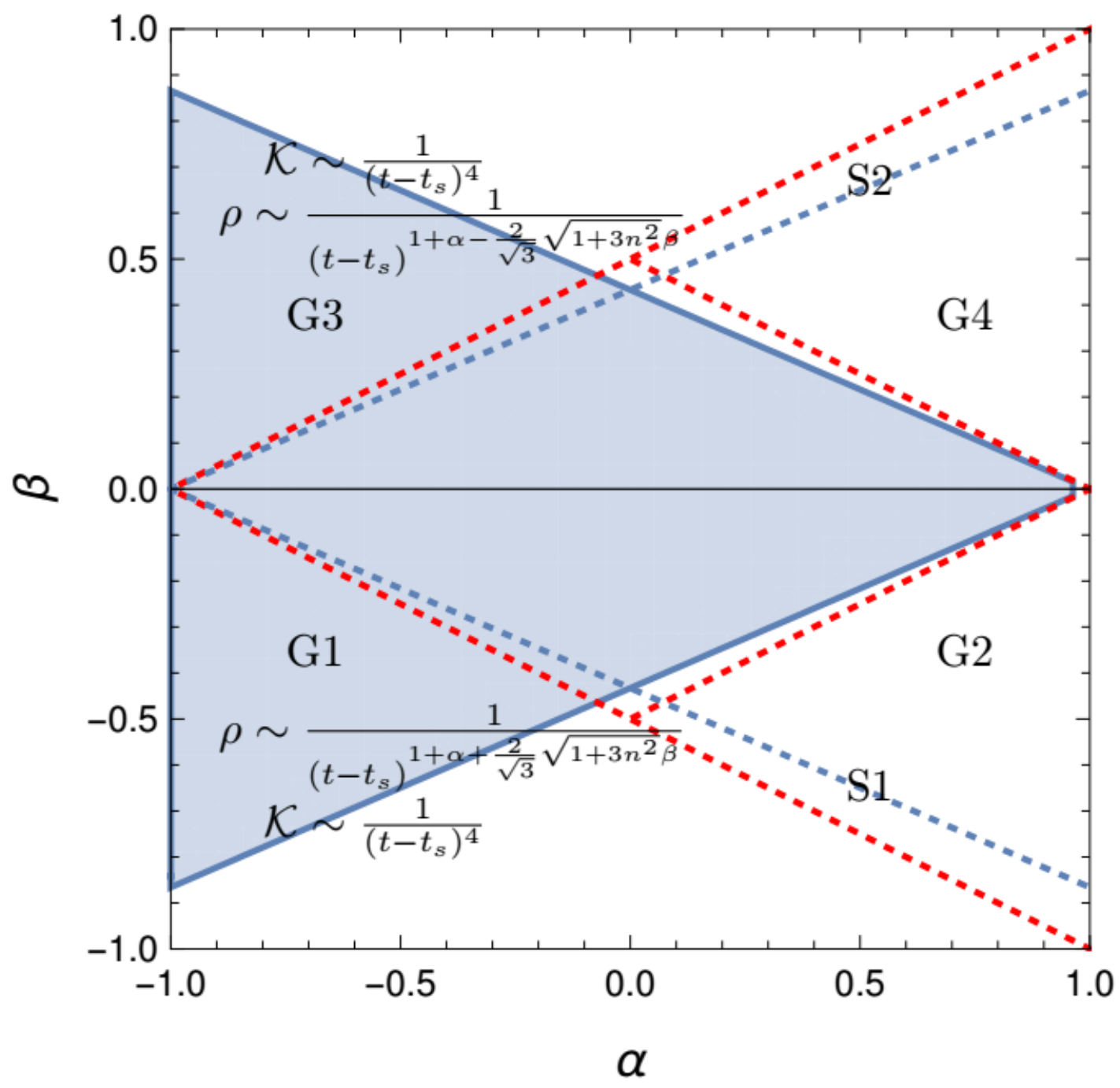}}
\caption{Diagram for $\rho$ and $\mathcal{K}$ at $t\to t_s$
in $\alpha$-$\beta$ plane.
Class G is split into four regions by G1-G4.
The shaded region ($\sqrt{3}(1-\alpha)-2\sqrt{1+3n^2}|\beta| > 0$) consists  of G1 and G3 divided by $\beta=0$,
and the unshaded region ($\sqrt{3}(1-\alpha)-2\sqrt{1+3n^2}|\beta| < 0$) consists of G2 and G4. 
The blue dotted lines S1 and S2 represent $\sqrt{3}(1+\alpha)+2\sqrt{1+3n^2}\beta=0$ and $\sqrt{3}(1+\alpha)-2\sqrt{1+3n^2}\beta=0$, respectively.}
\label{fig2}
\end{figure}

The behaviour of $\rho(t_s)$ and ${\cal K}(t_s)$ can be characterized by the regions 
in the $\alpha$-$\beta$ plane in Fig.~\ref{fig2}.
The regions are split by two lines S1 ($\sqrt{3}(1+\alpha)+2\sqrt{1+3n^2}\beta=0$) and S2 ($\sqrt{3}(1+\alpha)-2\sqrt{1+3n^2}\beta=0$).

1) In the region below S1: (i) $\rho(t_s)={\cal K}(t_s)=0$  for $n=1/3$, (ii) $\rho(t_s)=0$ but ${\cal K}(t_s)$ diverges for all other values of $n$. 

2) On S1: (i) $\rho(t_s)$ and ${\cal K}(t_s)$ are finite for $n=1/3$, (ii) $\rho(t_s)$ is finite but ${\cal K}(t_s)$ diverges for all other values of $n$.

3) In the region between S1 and S2: both $\rho(t_s)$ and ${\cal K}(t_s)$ diverge for all values of $n$.

4) On S2:  (i) $\rho(t_s)$ and ${\cal K}(t_s)$ are finite for $n=-1/3$, (ii) $\rho(t_s)$ is finite but ${\cal K}(t_s)$ diverges for all other values of $n$.

5) In the region above S2: (i) $\rho(t_s)={\cal K}(t_s)=0$  for $n=-1/3$, (ii) $\rho(t_s)=0$ but ${\cal K}(t_s)$ diverges for all other values of $n$.

\noindent
Therefore, in the region $\sqrt{3}(1+\alpha)-2\sqrt{1+3n^2}|\beta|\leq 0$, 
the energy density is finite for all the values of $n$, whereas the Kretschmann scalar 
is finite only for $n=\pm 1/3$. Therefore, for $n\neq \pm 1/3$, the 
solution is always singular. The initial singularity can only be removed 
on S1 and in the region below for $n=1/3$, and on S2 and in the region 
above for $n=-1/3$.

Let us discuss the energy conditions. The standard energy conditions \citep{poisson} for the diagonal energy-momentum tensor in Eq. \eqref{EMTdiag} turn out to be\\
(a) Weak Energy Condition (WEC):
\begin{eqnarray}
& & \rho\geq 0, \;\; \rho+p+(n+1)\sigma>0, \;\; \rho+p+(n-1)\sigma>0, \;\; \rho+p-2n\sigma>0 \nonumber\\
&\Rightarrow & \rho\geq 0, \;\; 1+\alpha+(n+1)\beta>0, \;\; 1+\alpha+(n-1)\beta>0 \;\; 1+\alpha-2n\beta>0.
\nonumber
\end{eqnarray}
(b) Null Energy Condition (NEC):
\begin{eqnarray}
& & \rho+p+(n+1)\sigma\geq 0, \;\; \rho+p+(n-1)\sigma\geq 0, \;\; \rho+p-2n\sigma\geq 0 \nonumber\\
&\Rightarrow & 1+\alpha+(n+1)\beta\geq 0, \;\; 1+\alpha+(n-1)\beta\geq 0 \;\; 1+\alpha-2n\beta\geq 0.
\nonumber
\end{eqnarray}
(c) Strong Energy Condition (SEC):
\begin{eqnarray}
& & \rho+p+(n+1)\sigma\geq 0, \;\; \rho+p+(n-1)\sigma\geq 0, \;\; \rho+p-2n\sigma\geq 0, \;\; \rho+3p\geq 0 \nonumber\\
&\Rightarrow & 1+\alpha+(n+1)\beta\geq 0, \;\; 1+\alpha+(n-1)\beta\geq 0 \;\; 1+\alpha-2n\beta\geq 0, \;\; 1+3\alpha\geq 0.
\nonumber
\end{eqnarray}
(d) Dominant Energy Condition (DEC):
\begin{eqnarray}
& & \rho\geq 0, \;\; \rho\geq |p+(n+1)\sigma|, \;\; \rho\geq |p+(n-1)\sigma|, \;\; \rho\geq |p-2n\sigma| \nonumber\\
&\Rightarrow & \rho\geq 0, \;\; 1-|\alpha+(n+1)\beta|\geq 0, \;\; 1-|\alpha+(n-1)\beta|\geq 0 \;\; 1-|\alpha-2n\beta|\geq 0.
\nonumber
\end{eqnarray}
Here, we have used the equations of state given in Eq. \eqref{eoss}. In the region enclosed by the red dotted lines in Fig. \ref{fig2} (rhombus for $n=0$ and parallelogram for $n\neq 0$), the WEC, NEC and DEC are satisfied. The SEC is satisfied also in this region if $\alpha\geq -1/3$. Outside of this enclosed region, at least one energy condition is violated. It is to be noted that this energy condition violating region consists of four triangles. Inside the two left triangles, all the energy conditions are violated, whereas only the DEC is violated inside the two right triangles. It is to be noted also that, for $n=\pm 1/3$, the region where the singularity can be removed belongs to the region where all the energy conditions are violated ($n=-1/3$: upper triangle, $n=+1/3$: lower triangle).

%%%%%%%%%%%%%%%%%%%%%%%%%%%%%%%%%%%%%%%%%%%%%%%%%%%%%%%%%%%%%%%%%%%%%%%%%%
%%%%%%%%%%%%%%%%%%%%%%%%%%%%%%%%%%%%%%%%%%%%%%%%%%%%%%%%%%%%%%%%%%%%%%%%%%

\section{Late time behaviour}
\label{sec:latetime}

\begin{figure}[]
\centering
\subfigure[$n=0$]{\includegraphics[scale=0.3]{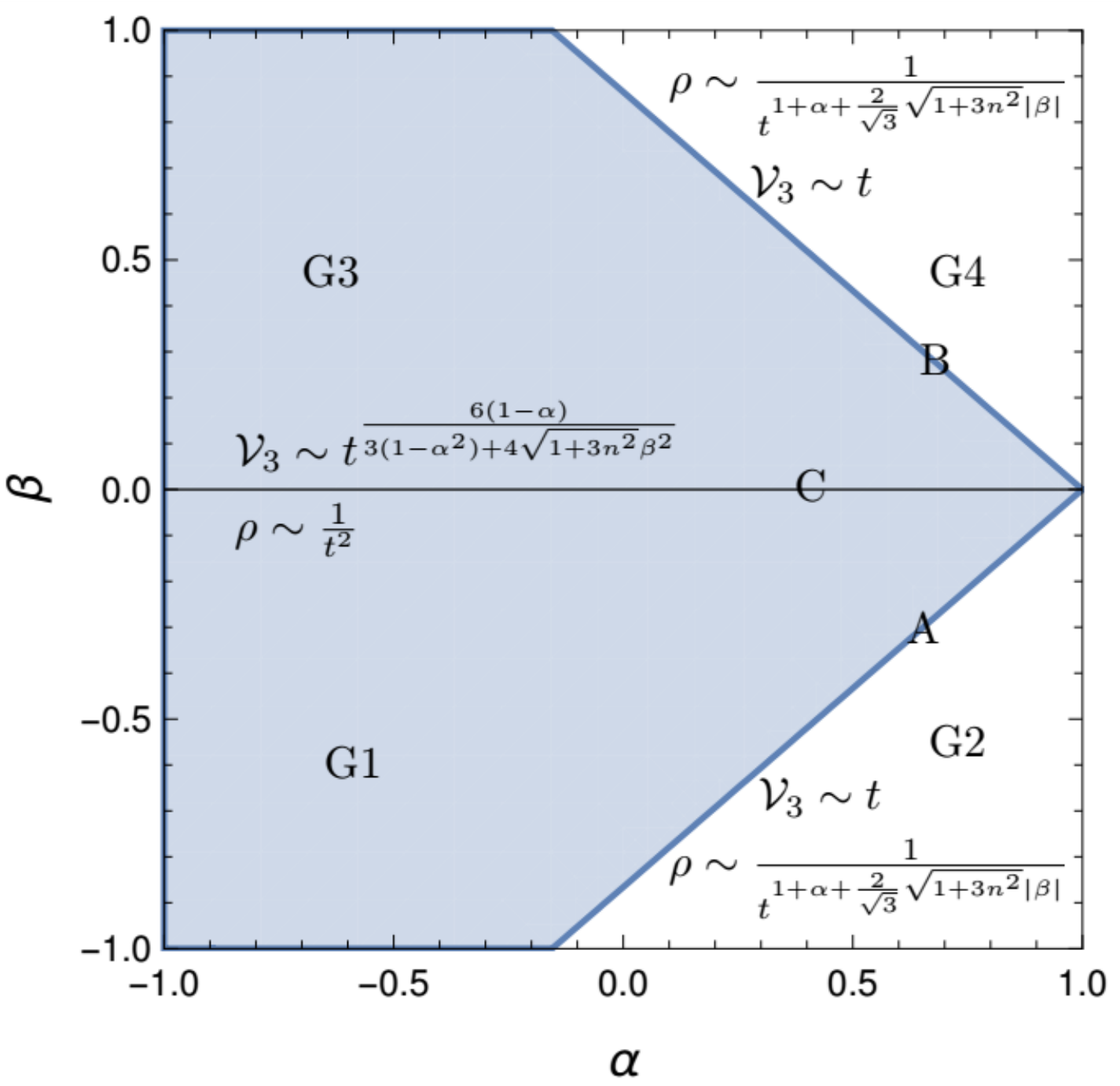}}
\subfigure[$n=\pm 1$]{\includegraphics[scale=0.3]{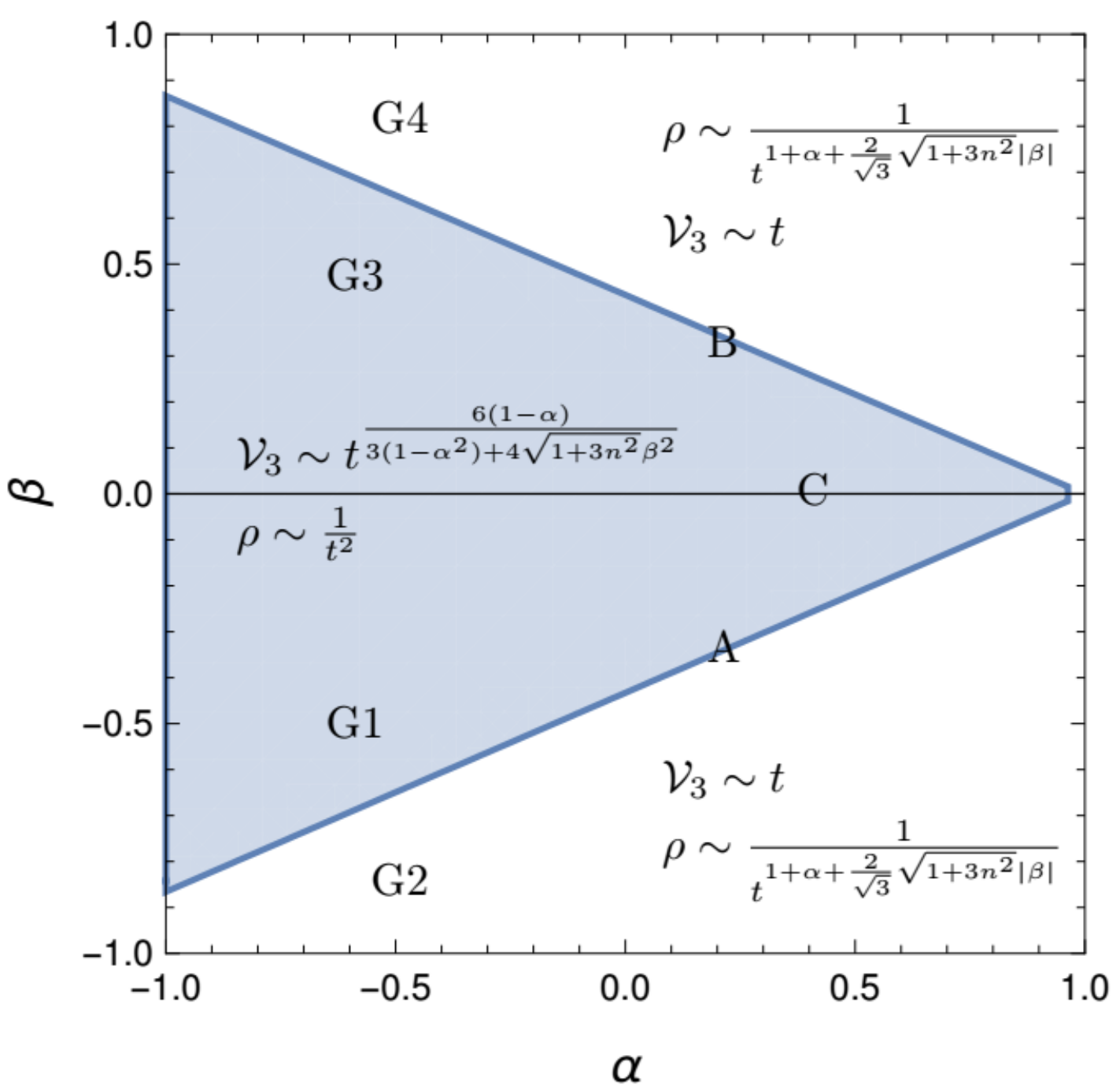}}
\caption{Diagram for ${\cal V}_3$ and $\rho$ at large $t$
in $\alpha$-$\beta$ plane.
The lower border between the shaded and the unshaded regions
corresponds to Class A,
and the upper border to Class B.
The solid horizontal line between G1 and G3 to Class C ($\beta=0$: FRW). 
Considering the power in $t$,
$\rho$ evolves in the same manner with that of FRW in the shaded region, 
and decreases faster than that of FRW in the unshaded region.
${\cal V}_3$ increases slower than that of FRW in all regions.
}
\label{fig3}
\end{figure}

We now analyse the late-time (large $t$) behaviour of the solutions for 
different classes separately. The results are summarized in Fig.~\ref{fig3}.

\vspace{12pt}
\noindent
\underline{{\bf Class G}: General Class}
\vspace{12pt}

The late-time behaviour of the solution in this class is obtained 
in Appendix \ref{appendixC}. As $t\to \infty$, the three-volume 
and the energy density behave as (see Eqs. \eqref{GV3LT_app} and \eqref{Grho3LT_app})
\begin{equation}
{\cal V}_3 \sim \left\{
  \begin{array}{ll}
  t^{\frac{6(1-\alpha)}{3(1-\alpha^2)+4(1+3n^2)\beta^2}} 
  & (\mbox{G1 and G3}) \\
  t 
  & (\mbox{G2 and G4}) 
  \end{array}
\right.,
\label{GV3LT}
\end{equation}
and
\begin{equation}
\rho\sim \left\{
  \begin{array}{ll}
  \frac{1}{t^2} 
  & (\mbox{G1 and G3}) \\
  \frac{1}{t^{[\sqrt{3}(1+\alpha)+2\sqrt{1+3n^2}|\beta|]/\sqrt{3}}}
  & (\mbox{G2 and G4}) 
  \end{array}
\right..
\label{Grho3LT}
\end{equation}
It is to be noted that ${\cal V}_3$ diverges and $\rho$ vanishes as $t\to \infty$.

\vspace{12pt}
\noindent
\underline{{\bf Class A} ($[\sqrt{3}(1-\alpha)+2\sqrt{1+3n^2}\beta]=0$) and {\bf Class B} ($[\sqrt{3}(1-\alpha)-2\sqrt{1+3n^2}\beta]=0$)}
\vspace{12pt}

For these classes, it is difficult to obtain the late-time behaviour of 
the solution directly from Eqs.~\eqref{eq:t_SC1} and \eqref{eq:t_SC2}. 
However, since Class A is the border of G1 and G2, and Class B is the 
same of G3 and G4 (see Fig.~\ref{fig3}), we can find out the late-time 
behaviour of ${\cal V}_3$ and $\rho$ by putting
$\left[\sqrt{3}(1-\alpha)-2\sqrt{1+3n^2}|\beta|\right]=0$ in 
Eqs.~\eqref{GV3LT} and \eqref{Grho3LT}. This gives
\begin{equation}
{\cal V}_3 \sim t, \qquad \rho\sim \frac{1}{t^2}.
\end{equation}

\vspace{12pt}
\noindent
\underline{{\bf Class C}: $\beta=0$}
\vspace{12pt}

{\bf (i) FRW universe}: 
For this class, at large $t$, we have, from Eq.~\eqref{eq:c_SC3},
\begin{equation}
c=d\sim \left\{
  \begin{array}{lr}
    e^{\frac{A_2}{2\sqrt{3}} t} &(\alpha =-1) \\
  t^{\frac{2}{3(1+\alpha)}} &(\alpha \neq -1)
  \end{array}
\right..
\end{equation}
Therefore, ${\cal V}_3$ and $\rho$ at large $t$ are given by
\begin{equation}
{\cal V}_3 \sim \left\{
  \begin{array}{lr}
    e^{\frac{\sqrt{3}A_2}{2} t} & (\alpha =-1)\\
  t^{\frac{2}{(1+\alpha)}} & (\alpha \neq -1)
  \end{array}
\right., \qquad 
\rho\sim \left\{
  \begin{array}{lr}
    {\rm const.} & (\alpha =-1) \\
  \frac{1}{t^2} & (\alpha \neq -1)
  \end{array}
\right..
\end{equation}
Class C corresponds to the horizontal axis in Fig.~\ref{fig3}.

{\bf (ii) Non-FRW universe}: 
The solution for this class is given by Eqs. \eqref{eq:U_SC3} and 
\eqref{eq:t_SC3}. However, we find that the late-time behaviours of 
${\cal V}_3$ and $\rho$ in this class remains the same as that in 
Class C-(i).

\vspace{12pt}
The late-time behaviours of ${\cal  V}_3$ and $\rho$ are summarized 
in Fig.~\ref{fig3} for $n=0,\pm 1$. 
In comparison to those in FRW, $\rho$ evolves in the same manner and 
${\cal  V}_3$ exhibits slower expansion in the shaded region 
(regions G1 and G3), whereas, in the unshaded region 
(regions G2 and G4), $\rho$ decreases faster and
${\cal  V}_3$ again exhibits slower expansion. Therefore, with 
off-diagonal stress, the expansion of ${\cal V}_3$ is slower, 
and $\rho$ decreases more rapidly (at most equally) compared 
with the FRW universe.

%In the shaded region (regions G1 and G3),
%$\rho$ evolves in the same manner with that of FRW (Class C),
%while ${\cal  V}_3$ increases slower than that of FRW.
%In the unshaded region (regions G2 and G4),
%$\rho$ decreases faster than that of FRW,
%while ${\cal  V}_3$ again increases slower than that of FRW.

%%%%%%%%%%%%%%%%%%%%%%%%%%%%%%%%%%%%%%%%%%%%%%%%%%%%%%%%%%%%%%%%%%%%%%%%%%

\section{Deceleration and anisotropic parameters}
\label{sec:decceleration}

\subsection{Deceleration parameter}

\begin{figure}[]
\centering
\subfigure{\includegraphics[scale=0.7]{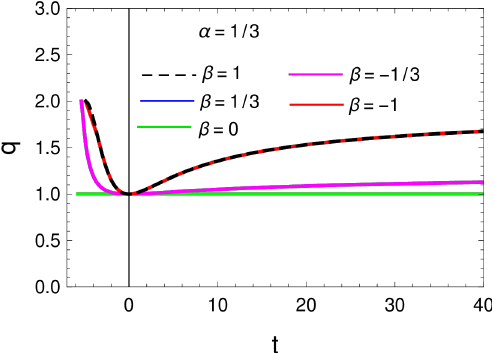}}
\subfigure{\includegraphics[scale=0.7]{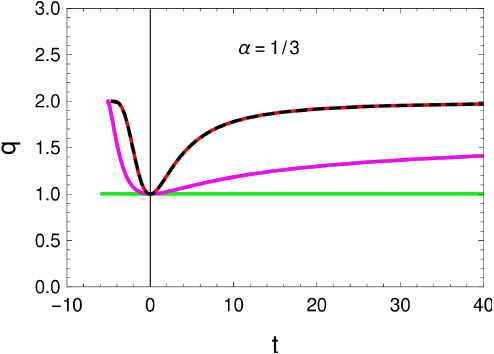}}
\subfigure{\includegraphics[scale=0.7]{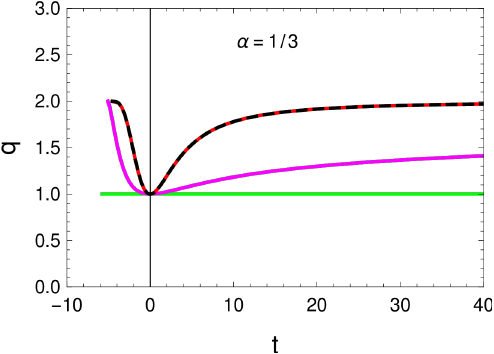}}
\subfigure{\includegraphics[scale=0.7]{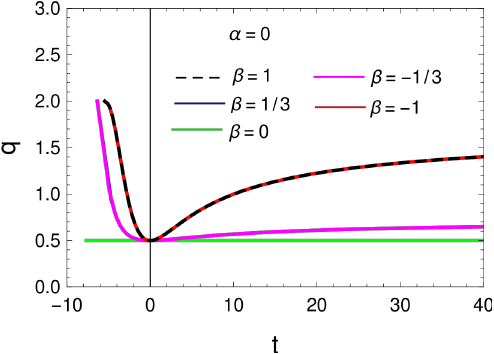}}
\subfigure{\includegraphics[scale=0.7]{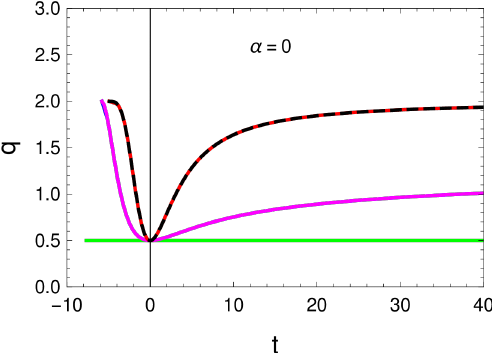}}
\subfigure{\includegraphics[scale=0.7]{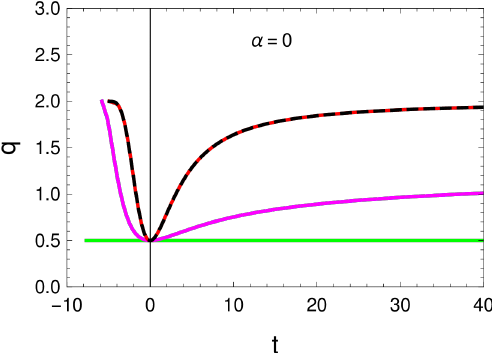}}
\caption{Plots of the deceleration parameter $q$ for $\alpha=1/3$ (radiation), 
$0$ (dust), $\beta = 0, \pm 1/3,\pm 1$, and $n=0$ (first column), $1$ (second 
column), $-1$ (third column). Values of $\beta$ are distinguished by color.}
\label{fig4}
\end{figure}

In FRW cosmology, the deceleration parameter $q$ is defined as 
\begin{equation}
q=-\frac{\ddot{\mathfrak{a}}}{\mathfrak{a}H^2}=-1-\frac{\dot{H}}{H^2},
\end{equation}
where $\mathfrak{a}$ is the scale factor, and $H=\dot{\mathfrak{a}}/\mathfrak{a}$ is the Hubble parameter. 
In our anisotropic case, we take $\mathfrak{a}$ to be the volume scale factor defined 
by $\mathfrak{a}={\cal V}_3^{1/3}=(cde)^{1/3}=c^{(1+n)/2}d^{(1-n)/2}$. Therefore, in 
our anisotropic case, the generalized Hubble parameter $H=\dot{\mathfrak{a}}/\mathfrak{a}$ turns 
out to be
\begin{equation}
H=\frac{1}{3}\left(\frac{\dot{c}}{c}+\frac{\dot{d}}{d}+\frac{\dot{e}}{e}\right)=\frac{1}{3}(H_c+H_d+H_e)=\frac{(1+n)}{2}\frac{\dot{c}}{c}+\frac{(1-n)}{2}\frac{\dot{d}}{d},
\end{equation}
where $H_c=\dot{c}/c$, $H_d=\dot{d}/d$ and $H_e=\dot{e}/e$. Using Eqs. 
\eqref{eq:cd_UV}, \eqref{eq:Udot2} and \eqref{eq:Vdot2}, we obtain
\begin{equation}
H=\frac{1}{4\sqrt{3}}\left(\frac{\dot{u}}{u}+\frac{\dot{v}}{v}\right)=\frac{1}{4\sqrt{3}}\left(\frac{A_1}{u^{\sqrt{3}/4}v^{\sqrt{3}/4-s_2}}+\frac{A_2}{u^{\sqrt{3}/4-s_1}v^{\sqrt{3}/4}}\right).
\end{equation}
We differentiate above equation and use Eqs. \eqref{eq:Udot2} and 
\eqref{eq:Vdot2} to obtain
\begin{equation}
q=2-\frac{6(1-\alpha)A_1 A_2 u^{s_1} v^{s_2}}{(A_1 v^{s_2} + A_2 u^{s_1})^2}.
\end{equation}
Note that, for the FRW case ($\beta=0$), $u=v$, $A_1=A_2$ and $s_1=s_2$. 
This gives $q=(1+3\alpha)/2$ which matches with the FRW results. Figure 
\ref{fig4} shows the time-evolution of $q$ for the dust and the radiation 
cases. It is to be noted that $q\to 2$ always as $t\to t_s$. In the 
late-time ($t\to\infty$), $q$ approaches a constant value which is 
closer to the FRW value for small $|\beta|$. Moreover, 
$q$ in our anisotropic case is always positive and greater than that of 
the FRW case. This also explains our finding that the late-time expansion 
of ${\cal V}_3$ in the anisotropic case is always slower than that of the 
FRW. 

The early- and late-time behaviour of $q$ can be obtained either by using 
the early- and late-time behaviour of $u$ and $v$ obtained in Appendices 
\ref{appendixB} and \ref{appendixC}, or from the early- and late-time 
behaviour of the volume scale factor $\mathfrak{a}={\cal V}_3^{1/3}$. The early-time 
behaviour is given by
\begin{equation}
q \sim \left\{
  \begin{array}{ll}
  2-{\rm const.}\times (t-t_s)^{4s_1/\sqrt{3}} & (\mbox{G1 and G2}) \\
  2-{\rm const.}\times (t-t_s)^{4s_2/\sqrt{3}}
  & (\mbox{G3 and G4})
  \end{array}
\right..
\end{equation}
On the other hand, the late-time behaviour is given by
\begin{equation}
q \sim \left\{
  \begin{array}{lll}
  \frac{1+3\alpha}{2}+\frac{2(1+3n^2)\beta^2}{1-\alpha} & (\mbox{G1 and G3}) \\
  2-{\rm const.}\times t^{-4|s_2|/\sqrt{3}} & (\mbox{G2}) \\
  2-{\rm const.}\times t^{-4|s_1|/\sqrt{3}}
  & (\mbox{G4})
  \end{array}
\right..
\end{equation}
It is to be noted that $q$ is always much larger than the FRW value at the early-time. In the late-time, it can be made closer to the FRW value for small $\beta$ only in the regions G1 and G3.

%%%%%%%%%%%%%%%%%%%%%%%%%%%%%%%%%%%%%%%%%%%%%%%%%%%%%%%%%%%%%%%%%%%%%%%%%%
%%%%%%%%%%%%%%%%%%%%%%%%%%%%%%%%%%%%%%%%%%%%%%%%%%%%%%%%%%%%%%%%%%%%%%%%%%

\subsection{Anisotropy parameter} 

\begin{figure}[]
\centering
\subfigure{\includegraphics[scale=0.7]{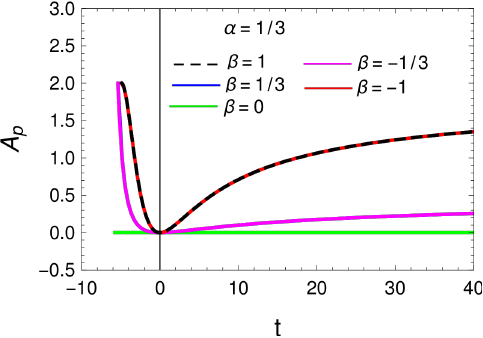}}
\subfigure{\includegraphics[scale=0.7]{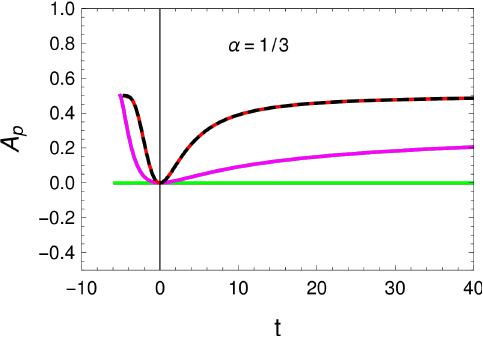}}
\subfigure{\includegraphics[scale=0.7]{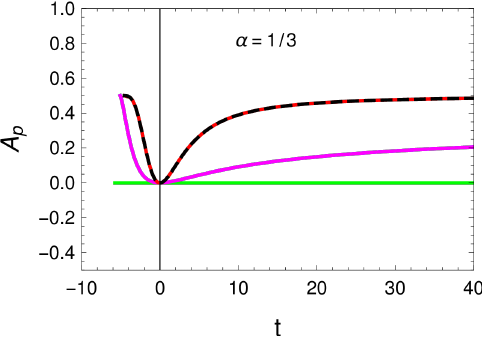}}
\subfigure{\includegraphics[scale=0.7]{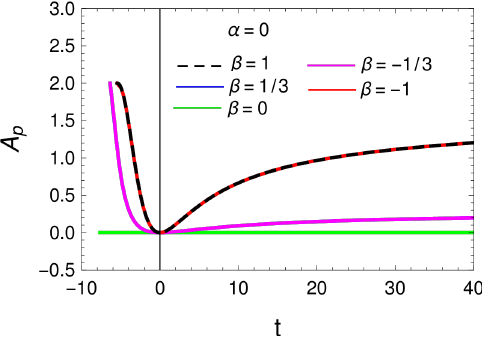}}
\subfigure{\includegraphics[scale=0.7]{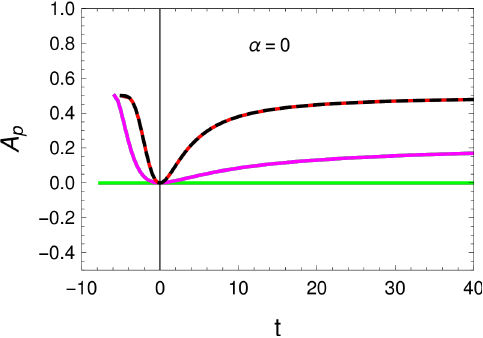}}
\subfigure{\includegraphics[scale=0.7]{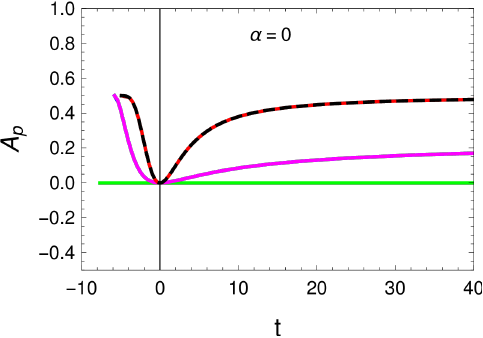}}
\caption{Plots of the anisotropy parameter $A_p$ for $\alpha=1/3$ (radiation), 
$0$ (dust), $\beta = 0, \pm 1/3,\pm 1$, and $n=0$ (first column), $1$ (second 
column), $-1$ (third column). Values of $\beta$ are distinguished by color.}
\label{fig5}
\end{figure}

The anisotropy parameter $A_p$ is defined as (see Ref. \citep{VC7})
\begin{eqnarray}
A_p &=& \frac{1}{3}\left[\left(\frac{H_c-H}{H}\right)^2+\left(\frac{H_d-H}{H}\right)^2+\left(\frac{H_e-H}{H}\right)^2\right]\\
&=& \frac{1}{6H^2}\left(\frac{\dot{c}}{c}-\frac{\dot{d}}{d}\right)^2,
\end{eqnarray}
where we have used the expressions for $H$, $H_c$, $H_d$ and $H_e$ in the last step. Using Eqs. \eqref{eq:cd_UV}, \eqref{eq:Udot2} and \eqref{eq:Vdot2}, we obtain
\begin{equation}
A_p = \frac{2}{1+3n^2}\left(\frac{\dot{u}/u-\dot{v}/v}{\dot{u}/u+\dot{v}/v}\right)^2 = \frac{2}{1+3n^2} \left(\frac{A_1 v^{s_2} - A_2 u^{s_1}}{A_1 v^{s_2} + A_2 u^{s_1}}\right)^2.
\end{equation}
Figure \ref{fig5} shows the time-evolution of $A_p$ for the dust and 
the radiation cases. For a given $n$, $A_p$ at $t= t_s$ is same for all 
$\alpha$ and $\beta$. In the late-time ($t\to\infty$), $A_p$ approaches 
a constant value which, for a given $n$ and $\alpha$, is smaller/larger 
for smaller/larger $|\beta|$. 

Following the same procedure as for $q$, the early-time behaviour of 
$A_p$ turns out to be
\begin{equation}
A_p \sim \left\{
  \begin{array}{ll}
  \frac{2}{1+3n^2}\left[1-{\rm const.}\times (t-t_s)^{4s_1/\sqrt{3}}\right] & (\mbox{G1 and G2}) \\
  \frac{2}{1+3n^2}\left[1-{\rm const.}\times (t-t_s)^{4s_2/\sqrt{3}}\right]
  & (\mbox{G3 and G4})
  \end{array}
\right..
\end{equation}
On the other hand, the late-time behaviour is given by
\begin{equation}
A_p \sim \left\{
  \begin{array}{lll}
  \frac{8\beta^2}{3(1-\alpha)^2} & (\mbox{G1 and G3}) \\
  \frac{2}{1+3n^2}\left(1-{\rm const.}\times t^{-4|s_2|/\sqrt{3}}\right) & (\mbox{G2}) \\
  \frac{2}{1+3n^2}\left(1-{\rm const.}\times t^{-4|s_1|/\sqrt{3}}\right)
  & (\mbox{G4})
  \end{array}
\right..
\end{equation}
It is to be noted that $\beta=0$ line is the intersection of the regions G1 and G3 (see Fig. \ref{fig2}). In these regions, $\sqrt{3}(1-\alpha)-2\sqrt{1+3n^2}|\beta| > 0$, i.e., $\alpha<1-2\sqrt{1+3n^2}|\beta|/\sqrt{3}$. Therefore, for a given $\alpha<1-2\sqrt{1+3n^2}|\beta|/\sqrt{3}$, the anisotropy in the late-time can be made small by choosing small values of $|\beta|$.

%%%%%%%%%%%%%%%%%%%%%%%%%%%%%%%%%%%%%%%%%%%%%%%%%%%%%%%%%%%%%%%%%%%%%%%%%%
%%%%%%%%%%%%%%%%%%%%%%%%%%%%%%%%%%%%%%%%%%%%%%%%%%%%%%%%%%%%%%%%%%%%%%%%%%

\section{Conclusions}
\label{sec:conclusion}

We have investigated in detail the effect of the off-diagonal stress tensor 
(shear viscosity) on the evolution of spacetime. We have solved the 
Eintein's equation and have obtained exact solutions. The shear viscosity 
part has diagonal as well as off-diagonal components. Under a certain 
condition ($p_1=p_2=p_3$), we find that the diagonal component $\sigma_1$ 
is proportional to the off-diagonal component $\sigma$, i.e., 
$\sigma_1=n\sigma$, $n$ being a constant. For $n=\pm 1/3$, the solution 
reduces to our earlier results obtained in Ref. \cite{Cho:2022rgs}. 
(The energy-momentum tensor as well as the metric can be put into diagonal 
forms through a suitable coordinate transformation).

We considered two equation of states, $p =\alpha\rho$ and 
$\sigma =\beta\rho$, and solved the Einstein's equation for the range 
$-1 \leq \alpha, \beta \leq 1$. Depending on the values of the 
parameters $\alpha$ and $\beta$, the solution was classified 
into several classes. We analysed the early-time and the late-time 
behaviour of the solution.

The three-volume density ${\cal V}_3$ vanishes at the initial moment $t_s$ 
for all the classes except for the pure de Sitter ($\alpha=-1, \beta=0$). 
${\cal V}_3$ increases monotonically afterwards. At late times, ${\cal V}_3$ 
increases slower than that of FRW. The evolution pattern of $\rho$, on 
the other hand, depends on the parameters. In the parameter region 
$\sqrt{3}(1+\alpha)-2\sqrt{1+3n^2}|\beta|\leq 0$, $\rho$ is finite at $t_s$. 
Outside of this region, $\rho$ diverges at $t_s$ and drops afterwards as 
the usual FRW universe ($\beta =0$). At late times, $\rho$ drops faster than, 
or at most equal to that in the Friedmann universe considering the power 
of $t$-dependence. The deceleration parameter $q$ approaches $2$ always as $t\to t_s$. 
At late times, $q$ approaches a constant value which is closer to the FRW 
value for small $|\beta|$. Moreover, $q$ in our anisotropic case is always 
positive and greater than that of the FRW case. This also explains our 
finding that the late-time expansion of ${\cal V}_3$ in the anisotropic 
case is always slower than that of the FRW.

For $n\neq \pm 1/3$, the solution is always singular as the Kretshmann 
scalar (${\cal K}$) always diverges at the initial moment $t_s$ in these 
cases. For $n=\pm 1/3$, however, the initial big-bang singularity can be 
removed for some parameter ranges (see Ref. \cite{Cho:2022rgs} for more 
details of these cases). For this case, the evolution along two spatial 
directions are equal which is Bianchi type-VII. For $n\neq \pm 1/3$, the 
evolution along all three directions are different which is Bianchi type-I. 
Therefore, the latter is more anisotropic (less symmetry) than the former, 
and the effect of this is reflected in the early-time behaviour of the 
solution; the solution is always singular for $n\neq \pm 1/3$, whereas it 
can be either singular or non-singular for $n= \pm 1/3$. 

For a given $n$, the anisotropy parameter $A_p$ at $t= t_s$ is same for all 
$\alpha$ and $\beta$. At late times, $A_p$ approaches 
a constant value which, for a given $n$ and $\alpha$, is smaller/larger 
for smaller/larger $|\beta|$. Therefore, the anisotropy can be made small 
by choosing small values of $|\beta|$.

\appendix

\section{Integrating the field equations}
\label{appendixA}

With  Eqs.~\eqref{eoss} and \eqref{eq:FE1}, 
Eqs.~\eqref{eq:FE2} and \eqref{eq:FE3} can be written as
\begin{equation}
2(1+n)\frac{\ddot{c}}{c} +(1+3n)(\alpha+n)\frac{\dot{c}^2}{c^2}+2(1+2\alpha-3n^2)\frac{\dot{c}}{c}\frac{\dot{d}}{d}+(1-3n)(\alpha-n)\frac{\dot{d}^2}{d^2}+2(1-n)\frac{\ddot{d}}{d}=0,
\label{eq:FE4}
\end{equation}
\begin{equation}
2\frac{\ddot{c}}{c} +(1+3n)(1+2\beta)\frac{\dot{c}^2}{c^2}-(6n-8\beta)\frac{\dot{c}}{c}\frac{\dot{d}}{d}-(1-3n)(1-2\beta)\frac{\dot{d}^2}{d^2}-2\frac{\ddot{d}}{d} =0. 
\label{eq:FE5}
\end{equation}
Now multiplying Eq. \eqref{eq:FE5} by $\sqrt{1+3n^2}/\sqrt{3}$ and then adding with Eq. \eqref{eq:FE4}, we obtain
\begin{equation}
\frac{d}{dt}\left[c^{q_1}d^{q_2}\left(n_1 d\dot{c}+n_2 c\dot{d}\right)\right]=0,
\end{equation}
where
\begin{equation}
q_1=\frac{1+3n}{2n_1}\left[\sqrt{3}(\alpha+n)+\sqrt{1+3n^2}(1+2\beta)\right],
\end{equation}
\begin{equation}
q_2=\frac{1-3n}{2n_2}\left[\sqrt{3}(\alpha-n)-\sqrt{1+3n^2}(1-2\beta)\right],
\end{equation}
\begin{equation}
n_1=\sqrt{3}(1+n)+\sqrt{1+3n^2}, \quad n_2=\sqrt{3}(1-n)-\sqrt{1+3n^2}.
\end{equation}
Integrating this equation, we get
\begin{equation}
n_1 d\dot{c}+n_2 c\dot{d}=\frac{A_2}{c^{q_1}d^{q_2}},
\label{eq:FE6}
\end{equation}
where $A_2$ is an integration constant. The conservation equation turns out to be
\begin{equation}
\dot{\rho}+\frac{3}{2}\left[(1+n)\frac{\dot{c}}{c}+(1-n)\frac{\dot{d}}{d} \right](\rho+p)-(1+3n^2)\left(\frac{\dot{c}}{c}-\frac{\dot{d}}{d} \right)\sigma=0,
\end{equation}
which, after using Eq. \eqref{eoss}, can be integrated to obtain
\begin{equation}
\rho=\frac{A_0}{c^{\frac{3}{2}(1+n)(1+\alpha)-(1+3n^2)\beta}d^{\frac{3}{2}(1-n)(1+\alpha)+(1+3n^2)\beta}},
\end{equation}
where $A_0$ is an integration constant. Equation \eqref{eq:FE1}, then, can be written as
\begin{equation}
\frac{1}{4c^2d^2}\left(n_1 d\dot{c}+n_2 c\dot{d}\right)\left(\bar{n}_1 d\dot{c}+\bar{n}_2 c\dot{d}\right)=\frac{A_0}{c^{\frac{3}{2}(1+n)(1+\alpha)-(1+3n^2)\beta}d^{\frac{3}{2}(1-n)(1+\alpha)+(1+3n^2)\beta}},
\label{eq:FE7}
\end{equation}
where
\begin{equation}
\bar{n}_1=\sqrt{3}(1+n)-\sqrt{1+3n^2}, \quad \bar{n}_2=\sqrt{3}(1-n)+\sqrt{1+3n^2}.
\end{equation}
Using Eq. \eqref{eq:FE6}, Eq. \eqref{eq:FE7} can be rewritten as
\begin{equation}
\bar{n}_1 d\dot{c}+\bar{n}_2 c\dot{d}=\frac{A_1}{c^{\bar{q}_1}d^{\bar{q}_2}},
\label{eq:FE8}
\end{equation}
where $A_0=A_1A_2/4$ and
\begin{equation}
\bar{q}_1=\frac{3}{2}(1+n)(1+\alpha)-(1+3n^2)\beta-2-q_1,
\end{equation}
\begin{equation}
\bar{q}_2=\frac{3}{2}(1-n)(1+\alpha)+(1+3n^2)\beta-2-q_2.
\end{equation}
Introducing new variables, $u=c^{\bar{n}_1}d^{\bar{n}_2}$ and $v=c^{n_1} d^{n_2}$, Eqs. \eqref{eq:FE8} and \eqref{eq:FE6} can be rewritten as
\begin{equation}
\frac{\dot{u}}{u}=\frac{A_1}{u^{-\frac{n_2(\bar{q}_1+1)}{n_1\bar{n}_2-\bar{n}_1 n_2}+\frac{n_1(\bar{q}_2+1)}{n_1\bar{n}_2-\bar{n}_1 n_2}}v^{\frac{\bar{n}_2(\bar{q}_1+1)}{n_1\bar{n}_2-\bar{n}_1 n_2}-\frac{\bar{n}_1(\bar{q}_2+1)}{n_1\bar{n}_2-\bar{n}_1 n_2}}},
\label{eq:Udot1}
\end{equation}
\begin{equation}
\frac{\dot{v}}{v}=\frac{A_2}{u^{-\frac{n_2(q_1+1)}{n_1\bar{n}_2-\bar{n}_1 n_2}+\frac{n_1(q_2+1)}{n_1\bar{n}_2-\bar{n}_1 n_2}}v^{\frac{\bar{n}_2(q_1+1)}{n_1\bar{n}_2-\bar{n}_1 n_2}-\frac{\bar{n}_1(q_2+1)}{n_1\bar{n}_2-\bar{n}_1 n_2}}}.
\label{eq:Vdot1}
\end{equation}
Introducing
\begin{equation}
s_1=-\frac{n_2(\bar{q}_1-q_1)}{n_1\bar{n}_2-\bar{n}_1 n_2}+\frac{n_1(\bar{q}_2-q_2)}{n_1\bar{n}_2-\bar{n}_1 n_2}=\frac{1}{4}\left[\sqrt{3}(1-\alpha)-2\sqrt{1+3n^2}\beta\right],
\end{equation}
\begin{equation}
s_2=-\frac{\bar{n}_2(\bar{q}_1-q_1)}{n_1\bar{n}_2-\bar{n}_1 n_2}+\frac{\bar{n}_1(\bar{q}_2-q_2)}{n_1\bar{n}_2-\bar{n}_1 n_2}=\frac{1}{4}\left[\sqrt{3}(1-\alpha)+2\sqrt{1+3n^2}\beta\right],
\end{equation}
Eqs. \eqref{eq:Udot1} and \eqref{eq:Vdot1} can be written as
\begin{equation}
\frac{\dot{u}}{u}=\frac{A_1}{u^{\sqrt{3}/4}v^{\sqrt{3}/4-s_2}},
\label{eq:Udot2_app}
\end{equation}
\begin{equation}
\frac{\dot{v}}{v}=\frac{A_2}{u^{\sqrt{3}/4-s_1}v^{\sqrt{3}/4}}.
\label{eq:Vdot2_app}
\end{equation}
In terms of $u$ and $v$, $\rho$ can be written as
\begin{equation}
\rho=\frac{(A_1A_2/4)}{u^{\sqrt{3}/2-s_1}v^{\sqrt{3}/2-s_2}}.
\label{eq:rho_app}
\end{equation}
So, all the field equations reduces to Eqs. \eqref{eq:Udot2_app}, 
\eqref{eq:Vdot2_app} and \eqref{eq:rho_app}.

\section{Early-time behaviour}
\label{appendixB}
Here, we investigate the behaviour of the solutions in the limit of $t\to t_s$. 
At $t=0$, we imposed the initial conditions, $c(0)=d(0)=1$ and $\dot{c}(0)=\dot{d}(0)$, 
which are equivalent to $u(0)=v(0)=1$ and $\dot{u}(0)=\dot{v}(0)$. 
Using these conditions, we obtain $A_2=A_1$ from Eq.~\eqref{eq:UVdot} and 
$A_3=(1-{s_1}/{s_2})=\sqrt{1+3n^2}\beta/{s_2}$ from Eq.~\eqref{eq:U_gen}. 
Now depending on the signatures of the terms in the integrand, 
the $v$-solution for Class G in Eq.~\eqref{eq:t_gen} can be classified into 
four cases,
\begin{equation}
t=\left\{
  \begin{array}{lr}
  \frac{1}{A_2}\int_{1}^v v^{\sqrt{3}/4-1} \left[\frac{s_1}{s_2}\left(v^{s_2}-v_*^{s_2}\right)\right]^{\frac{\sqrt{3}}{4s_1}-1} dv 
  & [\mbox{G1: } \beta<0\; \& \; s_2=\sqrt{3}(1-\alpha)+2\sqrt{1+3n^2}\beta > 0] \\
  \frac{1}{A_2}\int_{1}^v v^{\sqrt{3}/4-1} \left[\frac{s_1}{|s_2|}\left(-v^{s_2}+v_*^{s_2}\right)\right]^{\frac{\sqrt{3}}{4s_1}-1} dv 
  & [\mbox{G2: } \beta<0\; \& \; s_2=\sqrt{3}(1-\alpha)+2\sqrt{1+3n^2}\beta < 0] \\
  \frac{1}{A_2}\int_{1}^v v^{\sqrt{3}/4-1} \left[\frac{s_1}{s_2}\left(v^{s_2}+v_*^{s_2}\right)\right]^{\frac{\sqrt{3}}{4s_1}-1} dv 
  & [\mbox{G3: } \beta>0\; \& \; s_1=\sqrt{3}(1-\alpha)-2\sqrt{1+3n^2}\beta > 0] \\
  \frac{1}{A_2}\int_{1}^v v^{\sqrt{3}/4-1} \left[\frac{|s_1|}{s_2}\left(-v^{s_2}+v_*^{s_2}\right)\right]^{\frac{\sqrt{3}}{4s_1}-1} dv 
  & [\mbox{G4: } \beta>0\; \& \; s_1=\sqrt{3}(1-\alpha)-2\sqrt{1+3n^2}\beta < 0] 
  \end{array}
\right.,
\label{t_all_app}
\end{equation}
where $v_{*}=\left(\sqrt{1+3n^2}|\beta|/|s_1|\right)^{1/s_2}$. 
Note that, for $-1\leq \alpha\leq 1$, we have $\sqrt{3}(1-\alpha)-2\sqrt{1+3n^2}\beta > 0$ ($s_1>0$) if $\beta<0$, and $\sqrt{3}(1-\alpha)+2\sqrt{1+3n^2}\beta > 0$ ($s_2>0$) if $\beta>0$. 
As $t\to t_s$, $v$ approaches to its minimum value; 
$v\to v_{*}$ for G1 and G2, and $v\to 0$ for G3 and G4. 
We split the integration range into two parts as
\begin{equation}
\int_1^v[\cdots]dv = \left( \int_1^{v_{s}} + \int_{v_{s}}^v \right)[\cdots]dv, 
\quad \mbox{where }
\int_1^{v_{s}}[\cdots]dv\equiv t_s 
\mbox{ and } v_s=\left\{
  \begin{array}{l}
  \mbox{$v_{*}$ (for G1 and G2)} \\
  \mbox{$0$ (for G3 and G4)}
  \end{array}
  \right..
\end{equation}
Therefore, Eq.~\eqref{t_all_app} can be rewritten as
\begin{equation}
t-t_s=\left\{
  \begin{array}{lr}
  \frac{1}{A_2}\int_{0}^v v^{\sqrt{3}/4-1} \left[\frac{s_1}{s_2}\left(v^{s_2}-v_*^{s_2}\right)\right]^{\frac{\sqrt{3}}{4s_1}-1} dv 
  & (\mbox{G1}) \\
  \frac{1}{A_2}\int_{0}^v v^{\sqrt{3}/4-1} \left[\frac{s_1}{|s_2|}\left(-v^{s_2}+v_*^{s_2}\right)\right]^{\frac{\sqrt{3}}{4s_1}-1} dv 
  & (\mbox{G2})\\
  \frac{1}{A_2}\int_{v_{*}}^v v^{\sqrt{3}/4-1} \left[\frac{s_1}{s_2}\left(v^{s_2}+v_*^{s_2}\right)\right]^{\frac{\sqrt{3}}{4s_1}-1} dv 
  & (\mbox{G3})\\
  \frac{1}{A_2}\int_{v_{*}}^v v^{\sqrt{3}/4-1} \left[\frac{|s_1|}{s_2}\left(-v^{s_2}+v_*^{s_2}\right)\right]^{\frac{\sqrt{3}}{4s_1}-1} dv 
  & (\mbox{G4})
  \end{array}
\right..
\end{equation}
We now perform the Taylor expansion about $v=v_{*}$ and keep the leading 
order of the term in the parentheses of the integrand for G1 and G2. For 
G3 and G4, we set $v=0$ in the parentheses of the integrand and get the 
leading order. After doing so, the integration gives, in the limit of $t \to t_s$,
\begin{equation}
v \sim \left\{
  \begin{array}{ll}
  v_{*}+{\rm const.}\times (t-t_s)^{4s_1/\sqrt{3}} & (\mbox{G1 and G2}) \\
  (t-t_s)^{4/\sqrt{3}}
  & (\mbox{G3 and G4})
  \end{array}
\right..
\end{equation}
Using $v$, we obtain $u$ from Eq.~\eqref{eq:U_gen},
\begin{equation}
u\sim \left\{
  \begin{array}{ll}
  (t-t_s)^{4/\sqrt{3}}  & (\mbox{G1 and G2}) \\
  u_{*}+{\rm const.}\times (t-t_s)^{4s_2/\sqrt{3}}  & (\mbox{G3 and G4})
  \end{array}
\right.,
\end{equation}
where $u_{*}=\left(|s_1|v_{*}^{s_2}/|s_2|\right)^{1/{s_1}}$. Note that the dependence of $c$ and $d$ on $(t-t_s)$ can be obtained from the expressions
\begin{equation}
c=u^{-\frac{n_2}{n_1\bar{n}_2-\bar{n}_1 n_2}} v^{\frac{\bar{n}_2}{n_1\bar{n}_2-\bar{n}_1 n_2}}, \quad d=u^{\frac{n_1}{n_1\bar{n}_2-\bar{n}_1 n_2}} v^{-\frac{\bar{n}_1}{n_1\bar{n}_2-\bar{n}_1 n_2}}.
\label{eq:cd_UV}
\end{equation}
We obtain, in the limit of $t \to t_s$,
\begin{equation}
c \sim \left\{
  \begin{array}{ll}
  (t-t_s)^{-\frac{\sqrt{3}(1-n)-\sqrt{1+3n^2}}{3\sqrt{1+3n^2}}} & (\mbox{G1 and G2}) \\
  (t-t_s)^{\frac{\sqrt{3}(1-n)+\sqrt{1+3n^2}}{3\sqrt{1+3n^2}}} 
  & (\mbox{G3 and G4})
  \end{array}
\right.,
\end{equation}
\begin{equation}
d \sim \left\{
  \begin{array}{ll}
  (t-t_s)^{\frac{\sqrt{3}(1+n)+\sqrt{1+3n^2}}{3\sqrt{1+3n^2}}} & (\mbox{G1 and G2}) \\
  (t-t_s)^{-\frac{\sqrt{3}(1+n)-\sqrt{1+3n^2}}{3\sqrt{1+3n^2}}} 
  & (\mbox{G3 and G4})
  \end{array}
\right..
\end{equation}
Therefore, the three-volume density behaves as 
\begin{equation}
{\cal V}_3 \sim t-t_s \quad\mbox{(G1-G4)}, 
\end{equation}
the energy density becomes
\begin{equation}
\rho \sim \left\{
  \begin{array}{ll}
  \frac{1}{(t-t_s)^{[\sqrt{3}(1+\alpha)+2\sqrt{1+3n^2}\beta]/\sqrt{3}}} 
  & (\mbox{G1 and G2}) \\
  \frac{1}{(t-t_s)^{[\sqrt{3}(1+\alpha)-2\sqrt{1+3n^2}\beta]/\sqrt{3}}} 
  & (\mbox{G3 and G4}) 
  \end{array}
\right.,
\label{eq:early_rho}
\end{equation}
and the Kretschmann scalar becomes
\begin{equation}
\mathcal{K} \sim \left\{
  \begin{array}{lllll}
  \frac{1}{(t-t_s)^4} 
  & \mbox{for $n\neq\pm 1/3$} \quad (\mbox{G1-G4}) \\
  \frac{1}{(t-t_s)^4} 
  & \mbox{for $n=-1/3$} \quad (\mbox{G1 and G2}) \\
  \frac{1}{(t-t_s)^{2(1+\alpha-4\beta/3)}} 
  & \mbox{for $n=-1/3$} \quad (\mbox{G3 and G4}) \\
  \frac{1}{(t-t_s)^{2(1+\alpha+4\beta/3)}} 
  & \mbox{for $n=1/3$} \quad (\mbox{G1 and G2}) \\
  \frac{1}{(t-t_s)^4} 
  & \mbox{for $n=1/3$} \quad (\mbox{G3 and G4}) 
  \end{array}
\right..
\label{eq:early_K}
\end{equation}

\section{Late time behaviour}
\label{appendixC}

Here we obtain the late time behaviour of the Class G solution. 
We expect the integration in Eq.~\eqref{t_all_app} to diverge 
as $t\to \infty$. 
For Class G1, G2 and G3, 
we have observed from our calculations 
that the integration diverges as $v\to\infty$. 
For the class G4, however,
the integration diverges as $v\to v_{*}$.

At large $t$, the integrations in $v$ yield
\begin{equation}
v \sim \left\{
  \begin{array}{ll}
  t^{\frac{4[\sqrt{3}(1-\alpha)-2\sqrt{1+3n^2}\beta]}{3(1-\alpha^2)+4(1+3n^2)\beta^2}} 
  & (\mbox{G1 and G3}) \\
  t^{4/\sqrt{3}}
  & (\mbox{G2}) \\
  v_{*}-{\rm const.}\times t^{-|\sqrt{3}(1-\alpha)-2\sqrt{1+3n^2}\beta|/\sqrt{3}} 
  & (\mbox{G4}) 
  \end{array}
\right..
\label{eq:Vlate_app}
\end{equation}
Using $v$, we obtain $u$ from Eq.~\eqref{eq:U_gen} at large $t$,
\begin{equation}\label{dlate_app}
u \sim \left\{
  \begin{array}{ll}
  t^{\frac{4[\sqrt{3}(1-\alpha)+2\sqrt{1+3n^2}\beta]}{3(1-\alpha^2)+4(1+3n^2)\beta^2}}
   & (\mbox{G1 and G3}) \\
  u_{*}-{\rm const.}\times t^{-|\sqrt{3}(1-\alpha)+2\sqrt{1+3n^2}\beta|/\sqrt{3}}
   & (\mbox{G2}) \\
  t^{4/\sqrt{3}} 
  & (\mbox{G4}) 
  \end{array}
\right..
\end{equation}
The late-time behaviour of $c$ and $d$, therefore, become
\begin{equation}
c \sim \left\{
  \begin{array}{ll}
  t^{\frac{2[1-\alpha-2(1-n)\beta]}{3(1-\alpha^2)+4(1+3n^2)\beta^2}} 
  & (\mbox{G1 and G3}) \\
  t^{\frac{\sqrt{3}(1-n)+\sqrt{1+3n^2}}{3\sqrt{1+3n^2}}}
  & (\mbox{G2}) \\
  t^{-\frac{\sqrt{3}(1-n)-\sqrt{1+3n^2}}{3\sqrt{1+3n^2}}}
  & (\mbox{G4}) 
  \end{array}
\right.,
\end{equation}
\begin{equation}
d \sim \left\{
  \begin{array}{ll}
  t^{\frac{2[1-\alpha+2(1+n)\beta]}{3(1-\alpha^2)+4(1+3n^2)\beta^2}} 
  & (\mbox{G1 and G3}) \\
  t^{-\frac{\sqrt{3}(1+n)-\sqrt{1+3n^2}}{3\sqrt{1+3n^2}}}
  & (\mbox{G2}) \\
  t^{\frac{\sqrt{3}(1+n)+\sqrt{1+3n^2}}{3\sqrt{1+3n^2}}}
  & (\mbox{G4}) 
  \end{array}
\right.,
\end{equation}
The three-volume density becomes
\begin{equation}
{\cal V}_3 \sim \left\{
  \begin{array}{ll}
  t^{\frac{6(1-\alpha)}{3(1-\alpha^2)+4(1+3n^2)\beta^2}} 
  & (\mbox{G1 and G3}) \\
  t 
  & (\mbox{G2 and G4}) 
  \end{array}
\right..
\label{GV3LT_app}
\end{equation}
The energy density becomes
\begin{equation}
\rho\sim \left\{
  \begin{array}{ll}
  \frac{1}{t^2} 
  & (\mbox{G1 and G3}) \\
  \frac{1}{t^{[\sqrt{3}(1+\alpha)+2\sqrt{1+3n^2}|\beta|]/\sqrt{3}}}
  & (\mbox{G2 and G4}) 
  \end{array}
\right..
\label{Grho3LT_app}
\end{equation}
Note that ${\cal V}_3$ diverges and $\rho$ vanishes as $t\to \infty$.

\section*{Acknowledgements}
This work was supported by the grant from the National Research Foundation
funded by the Korean government, No. NRF-2020R1A2C1013266.


\begin{thebibliography}{99}

\bibitem{MTW}
W. Misner, K. S. Thorne, and J. A. Wheeler, 
``Gravitation,"
(Princeton University Press, USA, 2017).

\bibitem{Aluri} Pavan Kumar Aluri et al, ``Is the observable Universe consistent with the cosmological principle?,'' Class. Quantum Grav. {\bf 40}, 094001 (2023).

\bibitem{Misner:1967uu}
C.~W.~Misner,
``The Isotropy of the universe,''
Astrophys. J. \textbf{151}, 431-457 (1968)
doi:10.1086/149448
%542 citations counted in INSPIRE as of 01 Jun 2023

\bibitem{Stewart1968}
J.~M.~Stewart, 
``Neutrino Viscosity in Cosmological Models,'' 
Astrophysical Letters \textbf{2}, 133-135 (1968).

\bibitem{Stewart1969}
J.~M.~Stewart, 
``Non-Equilibrium Processes in the Early Universe,''
Monthly Notices of the Royal Astronomical Society \textbf{145}, 347-356 (1969). 
%https://doi.org/10.1093/mnras/145.3.347

\bibitem{Doro1968}
A.~G.~Doroshkevich, Ya.~B.~Zel'dovich, and I.~D.~Novikov, 
``Weakly Interacting Particles in the Anisotropic Cosmological Model ,'' 
Soviet Phys. JETP \textbf{26}, 408 (1968).

\bibitem{Noh:2004bc}
H.~Noh and J.~C.~Hwang,
``Second-order perturbations of the Friedmann world model,''
Phys. Rev. D \textbf{69}, 104011 (2004).
%doi:10.1103/PhysRevD.69.104011
%209 citations counted in INSPIRE as of 03 Nov 2023

\bibitem{Hwang:2007ni}
J.~C.~Hwang and H.~Noh,
``Second-order perturbations of cosmological fluids: Relativistic effects of pressure, multi-component, curvature, and rotation,''
Phys. Rev. D \textbf{76}, 103527 (2007).
%doi:10.1103/PhysRevD.76.103527
%[arXiv:0704.1927 [astro-ph]].
%35 citations counted in INSPIRE as of 03 Nov 2023

\bibitem{Hwang:2017oxa}
J.~C.~Hwang, D.~Jeong and H.~Noh,
``Gauge dependence of gravitational waves generated from scalar perturbations,''
Astrophys. J. \textbf{842}, no.1, 46 (2017).
%doi:10.3847/1538-4357/aa74be
%[arXiv:1704.03500 [astro-ph.CO]].
%64 citations counted in INSPIRE as of 03 Nov 2023

\bibitem{Eckart:1940te}
C.~Eckart,
%``The Thermodynamics of irreversible processes. 3.. Relativistic theory of the simple fluid,''
Phys. Rev. \textbf{58}, 919-924 (1940)
doi:10.1103/PhysRev.58.919
%894 citations counted in INSPIRE as of 01 Jun 2023

\bibitem{Pimentel}
O.~M.~Pimentel, G.~A.~Gonz\'alez and F.~D.~Lora-Clavijo,
``The Energy-Momentum Tensor for a Dissipative Fluid in General Relativity,''
Gen. Rel. Grav. \textbf{48}, no.10, 124 (2016)
%doi:10.1007/s10714-016-2121-7
[arXiv:1604.01318 [gr-qc]].
%23 citations counted in INSPIRE as of 01 Jun 2022

\bibitem{VC1}
S.~Bravo Medina, M.~Nowakowski and D.~Batic,
``Viscous Cosmologies,''
Class. Quant. Grav. \textbf{36}, no.21, 215002 (2019)
%doi:10.1088/1361-6382/ab45bb
[arXiv:1901.09787 [gr-qc]].
%9 citations counted in INSPIRE as of 01 Jun 2022

\bibitem{VC1-1}
V. A. Belinski and I. M. Khalatnikov, 
``Influence of viscosity on the character of cosmological evolution,''
JETP \textbf{69}, 401 (1975).

\bibitem{VC3}
A.~Banerjee, S.~B.~Duttachoudhury and A.~K.~Sanyal,
``Bianchi type I cosmological model with a viscous fluid,''
J. Math. Phys. \textbf{26}, 3010-3015 (1985)
%doi:10.1063/1.526676
[arXiv:2103.07342 [gr-qc]].
%23 citations counted in INSPIRE as of 01 Jun 2022

\bibitem{VC2}
A.~K.~Banerjee and N.~O.~Santos,
``Spatially Homogeneous Cosmological Models,''
Gen. Rel. Grav. \textbf{16}, no.03, 217-224 (1984)
%doi:10.1007/BF00762537

\bibitem{VC4}
O.~Gron,
``Viscous Inflationary Universe Models,''
Astrophys. Space Sci \textbf{173}, 191-225 (1990)
%doi:10.1007/BF00643930

\bibitem{VC5}
A.~Banerjee, A.~K.~Sanyal and S.~Chakrabarty,
``Bianchi II, VIII and IX viscous fluid cosmology,''
Astrophys. Space Sci. \textbf{166}, 259-268 (1990)
%doi:10.1007/BF01094897
[arXiv:2105.10696 [gr-qc]].
%8 citations counted in INSPIRE as of 01 Jun 2022

\bibitem{VC6}
T.~Singh and R.~Chaubey,
``Bianchi type-V universe with a viscous fluid and $\Lambda$-term,''
Pramana \textbf{68}, 721-734 (2007)
%doi:10.1007/s12043-007-0072-y
%17 citations counted in INSPIRE as of 01 Jun 2022

\bibitem{VC7} N. Mostafapoor and O. Gron, ``Bianchi type-I universe models with nonlinear viscosity,'' Astrophys Space Sci {\bf 343}, 423-434 (2013).

%\cite{Cho:2022rgs}
\bibitem{Cho:2022rgs}
I.~Cho and R.~Shaikh,
``Perfect fluid with shear viscosity and spacetime evolution,''
Chin. J. Phys. \textbf{87}, 452-464 (2024).
%doi:10.1016/j.cjph.2023.12.017
%[arXiv:2209.12544 [gr-qc]].
%1 citations counted in INSPIRE as of 15 Feb 2024

\bibitem{poisson}
E.~Poisson,
``A Relativist's Toolkit,''
Cambridge University Press, 2007.





\end{thebibliography}
\end{document}